\documentclass[a4paper,fleqn]{cas-dc}
\pdfoutput=1
\usepackage[numbers]{natbib}
\usepackage{graphicx}
\usepackage[switch,columnwise]{lineno}
\usepackage[caption=false]{subfig}
\usepackage[boxruled]{algorithm2e}
\usepackage[utf8]{inputenc} 
\usepackage{url}
\usepackage{amsmath}
\usepackage{amsthm}
\usepackage{amsfonts}
\usepackage{mathtools}
\usepackage{bm}
\usepackage{acronym}
\usepackage{flushend}
\usepackage{color}

\renewcommand{\printorcid}{}

\newcommand{\beq}{\begin{equation}}
\newcommand{\eeq}{\end{equation}}
\newcommand{\beqa}{\begin{eqnarray}}
\newcommand{\eeqa}{\end{eqnarray}}

\DeclareMathOperator*{\argmax}{arg\,max}
\DeclareMathOperator*{\argmin}{arg\,min}

\newenvironment{megaalgorithm}[1][htb]
{
	\begin{algorithm}%
	}{\end{algorithm}}

\begin{document}

	\sloppy
	\let\WriteBookmarks\relax
	\def\floatpagepagefraction{1}
	\def\textpagefraction{.001}
	\shorttitle{Supervised Feature Selection Techniques in Network Intrusion Detection: a Critical Review}
	\shortauthors{Di Mauro et~al.}
	
	\title [mode = title]{Supervised Feature Selection Techniques in \\ Network Intrusion Detection: a Critical Review}               
	
	\author[1]{M. Di~Mauro}
	\address[1]{Department of Information and Electrical Engineering and Applied Mathematics (DIEM), University of Salerno, 84084, Fisciano, Italy}
	\cortext[cor1]{Corresponding author}
	\cormark[1]
	\ead{mdimauro@unisa.it}
	 
	\author[2]{G. Galatro} 
	\address[2]{Amazon AWS, Belgard Retail Park, Tallaght, Dublin, Ireland}
	\ead{galatrogiovanni@gmail.com}

    \author[3]{G. Fortino}
    \address[3]{Department of Informatics, Modeling, Electronics and Systems, University of Calabria, Italy}
    \ead{g.fortino@unical.it}
    
    \author[4]{A. Liotta}
    \address[4]{Faculty of Computer Science, Free University of Bozen-Bolzano, Italy}
    \ead{antonio.liotta@unibz.it}

\begin{abstract}
Machine Learning (ML) techniques are becoming an invaluable support for network intrusion detection, especially in revealing anomalous flows, which often hide cyber-threats. Typically, ML algorithms are exploited to classify/recognize data traffic on the basis of statistical features such as inter-arrival times, packets length distribution, mean number of flows, etc. Dealing with the vast diversity and number of features that typically characterize data traffic is a hard problem. This results in the following issues: \textit{i)} the presence of so many features leads to lengthy training processes (particularly when features are highly correlated), while prediction accuracy does not proportionally improve; \textit{ii)} some of the features may introduce bias during the classification process, particularly those that have scarce relation with the data traffic to be classified. To this end, by reducing the feature space and retaining only the most significant features, Feature Selection (FS) becomes a crucial pre-processing step in network management and, specifically, for the purposes of network intrusion detection. In this review paper, we complement other surveys in multiple ways: \textit{i)} evaluating more recent datasets (updated w.r.t. obsolete KDD~$99$) by means of a designed-from-scratch Python-based procedure; \textit{ii)} providing a synopsis of most credited FS approaches in the field of intrusion detection, including Multi-Objective Evolutionary techniques; \textit{iii)} assessing various experimental analyses such as feature correlation, time complexity, and performance. Our comparisons offer useful guidelines to network/security managers who are considering the incorporation of ML concepts into network intrusion detection, where trade-offs between performance and resource consumption are crucial.
\end{abstract}

\begin{keywords}
Feature Selection \sep Machine Learning \sep Network Intrusion Detection \sep Network Performance 
\end{keywords}

\maketitle

\section{Introduction}

With the rapid growth of digital technology and communications, we are overwhelmed by network data traffic, which are diverse for media type (e.g. video, voice, text, sensory, etc.), and originate from (and are transported through) a broad range of sources (e.g. mobile networks, cloud infrastructures, Internet of Things, etc.). Consequently, we handle high-dimensionality data, calling for increasingly more sophisticated classification methods \cite{Camastra1,Camastra2}. 

Typically, we refer to high dimensionality when we deal with data whereby a large number of features may be extracted, to the point that the features may even exceed the number of observations. This leads to major issues, particularly the massive increase in training times. 

To this end, Feature Selection (FS) is a promising research direction, looking at ways to reduce the feature space in order to pinpoint only the most significant features. As a fundamental pre-processing step in machine learning, FS is gaining prominence in network management and, specifically, for the purposes of network intrusion detection and network traffic classification problems \cite{granville1,granville2,stadler,boutaba}. 

More generally, FS finds an even much broader applicability in field as diverse as bioinformatics \cite{bioinf0,bioinf1,bioinf2,bioinf3},image recognition/retrieval \cite{image_recog1,image_recog2,image_recog3,image_retr1,image_retr2,image_recog4,image_recog5}, fault diagnosis \cite{fault_diag1,fault_diag2}, text mining \cite{text_min1,text_min2,text_min3} and, interestingly, in network traffic analysis/classification, whose pertinent bibliography will be covered more closely in the following.

Machine learning engines can be easily embedded in network intrusion detection systems (IDS), which represent an essential part of network infrastructures to guarantee security \cite{nids1,nids3,nids4} and availability \cite{ava_1,ava_2,ava_3,ava_4,esrel15,esrel17,tnsmava,nomsmdm1,nomsmdm2}. 
Specifically, modern NIDS can be equipped with software probes in charge of analyzing network traffic on the basis of some characterizing features such as: distribution of inter-arrival times, distribution of packet sizes, presence of specific TCP/IP flags, percentage of forward/backward flows. Such statistical information is instrumental to revealing anomalous traffic, which is often behind Distributed Denial of Service (DDoS) attacks \cite{ddos1,ddos2}, covert Voice-over-IP sessions \cite{dimvoip}, threat diffusion \cite{dimkend} and Peer-2-Peer traffic \cite{cerroni1,cerroni2,cerroni3}. In many cases, these flows would pass unobserved through other conventional signature-based analyses. 

In principle, a larger number of features would allow to perform more granular analyses; yet, two main drawbacks emerge. First, a proliferation of correlated features leads to levels of redundancy that are useless, while resulting in increasingly longer training times. Also, not all features are valuable in characterizing traffic, incurring bias during the classification step.

If we consider, as an example, some novel probes such as ISCXFlowMeter \cite{cicflowmeter}, these can actually generate more than $80$ features to characterize data traffic, which makes it extremely difficult for the network analyst to deal with this vast amount of information. This is where an FS pre-processing step becomes invaluable. 

Across the scientific literature, many works are devoted to surveying machine learning algorithms with applications to internet traffic classification. Yet, few attempts have been made in the field of FS techniques, where two main shortcomings emerge:  \textit{i)} many works analyze and compare FS algorithms applied to non-specific datasets \cite{survey_general_1,survey_general_2,survey_general_3}; \textit{ii)} are either based on outdated datasets, or lack the experimental dimension. Overall, not much novel algorithms have been compared (refer to details provided in Table $1$, discussed further ahead).  
Aimed at filling these gaps, our paper provides the main following contributions: 
\begin{itemize}
\item We perform an experimental analysis of more recent datasets (classic literature focuses on the $20$-year old KDD$99$ dataset) by means of our designed-from-scratch Python based routine. This allows us to \textit{i)} perform cleaning, re-balancing, and data mixing operations, and \textit{ii)} automatically interact with specific ML-based engines.
\item We present a variety of experimental results (including: feature correlation, time complexity, performance) aimed at critically comparing selected families of FS algorithms, ranging from classic ones (rank search, linear forwarding selection) to modern bio-inspired algorithms (genetic search, ant colonies, multi-objective evolutionary), thus going well beyond a conventional literature survey.
\end{itemize}
\noindent
Our experimental-based, comparative assessment finds fruitful applications in the field of network/security management, where machine learning techniques are proved to offer a precious support to intrusion detection tasks, and where critical trade-offs between performance classification and resource consumption arise.

The paper is organized as follows: Section $2$ offers a general perspective on FS methods, in line with other accredited taxonomies. In Section $3$ we analyze how diverse FS techniques have been applied in the literature, often based on the old KDD$99$ dataset. Section $4$ proposes an {\em excursus} of popular FS algorithms (classic and modern) along with some necessary technical details. In Section $5$ we analyze the exploited novel datasets by grouping the most relevant features in families. In Section $6$ we perform an experimental-based comparative analysis of prominent FS algorithms, providing performance, feature correlation, and time-complexity figures. Finally, Section $7$ draws conclusions and provides future-direction indications. 

\section{Overview of Feature Selection}

Feature Selection refers to that set of techniques and strategies that allow to optimize the feature space, namely, an $n$-dimensional space where each sample is represented as a point. When dealing with large feature spaces, the analysis of data, which starts from their representative features, can be tremendously time and resource consuming. Hence, the need to devise suitable FS strategies aimed at eliminating $i)$ irrelevant features, namely, those features that are not actually needed to build an optimal feature subset; and $ii)$ redundant features, namely, those features that strongly depend on other features \cite{relred}. 

Feature selection can be considered as a special case of feature extraction methods \cite{Camastra2008}. The latter refer to a set of techniques (e.g. PCA, Single Value Decomposition, Linear Discriminant Analysis and others) useful to transform the original feature space in a new one aimed at alleviating the effects of the notorious {\em curse of dimensionality} problem \cite{blum,dash}. It is important to note that feature extraction comes with the critical risk of obtaining a transformed feature space that could lose its original physical meaning, whereas classic FS aims at preserving it \cite{sel_vs_extr,sel_vs_extr2}. 
 
 A more formal definition of the FS problem follows. Given a feature set $X=\{x_i: i=1,\dots, N \}$, find a subset $S_K=\{x_{i_1}, x_{i_2}, \dots, x_{i_K}\}$ with $K<N$, where an objective function Y($\cdot$) is optimized, namely:
 \beq
 \{x_{i_1}, x_{i_2}, \dots, x_{i_K}\}=\argmax_{K,i_k}[Y\{x_i: i=1,\dots, N \}].
 \eeq
 Unless otherwise stated, the problem is to select the optimal subset of features (according to a specific criterion) from the initial set, where two steps are typically performed: the first one involves a search strategy to pinpoint candidate subsets; the second one involves an objective function to evaluate the selected candidate subsets. The latter can be split in two types \cite{filters_wrappers,filters_wrappers_2,filters_wrappers_3,filters_wrappers_4}: $i)$ {\em filters}, referring to objective functions that rely on properties of the data by evaluating the information content (e.g. correlation measures, inter-class distance); $ii)$ {\em wrappers}, referring to objective functions that exploit training models by starting from a subset of features and, then, adding or removing features based on the previous model.  

A commonly accepted taxonomy of FS methods is presented in \cite{taxonomy}, where three classic approaches have been identified as \textit{supervised}, \textit{unsupervised}, and \textit{semi-supervised}. According to the supervised approach, labeled data are exploited to single out a feature subset, considering specific criteria for measuring the features importance.
Conversely, unsupervised techniques seek to unveil the intrinsic data structure to select the most significant features, without assuming any {\em a priori} knowledge \cite{tnsm1}. 

Finally, the semi-supervised approach is based on a mixed strategy, striving to enrich an unlabeled set with some labeled data, so as to improve the FS phase. 
Both the unsupervised and the semi-supervised approaches exhibit the drawback of neglecting potential correlations among features, resulting in the analysis of sub-optimal sets. This may prove  critical when dealing with traffic analysis, where we need to take into account statistical-based features (e.g. inter-arrival times variance, average packet length, etc.) and deterministic ones (e.g. IP addresses, port numbers, etc.). 
Let us consider, for example, a particular kind of traffic directed towards a fixed destination port for a certain period of time. An unsupervised approach could lead to a subset of features which does not include the destination port. Thus, a crucial (as well as deterministic) piece of information gets lost. 

On the other hand, a supervised approach can offer optimal results, provided that the data are correctly labeled. This case typically occurs in a controlled network environment, where, with the help of network analyzers, it is possible to automatically label the type of passing data traffic.

Because of the great potential of supervised methods, and thanks to the availability of suitable labeled datasets, we have decided to focus our experimental-based comparative evaluation on supervised FS methods.

\section{Related Work on Feature Selection applied to ML-based Intrusion Detection}

\begin{table*}[pos=ht]
\centering
	\caption{Prominent related work surveying FS techniques applied to Network Intrusion Detection.}
	\small
	\renewcommand{\arraystretch}{1.1}
	\begin{tabular}{|p{2.5cm}|p{4cm}|p{3.1cm}|p{6cm}|}
		\hline	
		\textbf{Authors} &  \textbf{Experiments} &  \textbf{Single/Multi Class} &  \textbf{Description} \\  \hline
		Wang et al. \cite{table_wang} & Performance analysis & Single Class & Feature Selection performed on the KDD99 dataset, by applying C4.5 and Bayesian Network algorithms \\ \hline
		
		Janarthanan et al. \cite{table_jana} & Performance analysis & Single/Multi Class & Empirical selection of features by performing tests on KDD99 and UNSW-NB15 datasets by means of Random Forest algorithm \\ \hline
		
		El-Khatib \cite{table_khatib} & Performance analysis & Single/Multi Class & Feature selection combining Filter/Wrapper methods for Wireless IDS. Tests have been performed on a WLAN environment \\ \hline
		
		Chen et al. \cite{table_chen} & Performance analysis, Time analysis & Single Class & Correlation-based Feature selection using the KDD99 dataset \\ \hline
		
		Nisioti et al. \cite{table_nisioti} & N/A & N/A & Classic survey on ML-based techniques (including FS) with pointers to detailed sources, but with no experiments  \\ \hline
		
		Iglesias et al. \cite{table_iglesias} & Performance analysis & Single/Multi Class & Comparison among $4$ FS algorithms on the NSL-KDD dataset \\ \hline
		
		Singh et al. \cite{table_singh} & Performance analysis, Time analysis & Single Class & Comparison among various FS algorithms on the KDD99 dataset \\ \hline
		
		Bahrololum et al. \cite{table_bahrololum} & Performance analysis & Single/Multi Class & Comparison among PSO, Decision Tree, Flexible neural tree algorithms on the KDD99 dataset \\ \hline
		
		Dhote et al. \cite{table_dhote} & N/A & N/A & Limited survey of FS  techniques applied to network traffic, with pointers to detailed sources, but with no experiments  \\ \hline
		
		This work & Performance analysis, Feature Correlation analysis, Time analysis & Single/Multi Class & Feature Selection performed on the CIC-IDS-2017/2018 dataset, by considering $9$ FS algorithms from classic (Rank, Scatter) to modern (Genetic, Multi-Objective Evolutionary) \\ \hline
	\end{tabular}
\end{table*}

Most scientific literature involving ML approaches is typically focused on the proposal of algorithms or techniques for classification/detection of specific network traffic (\cite{classif_buczak,classif_mishra,classif_dimauro, classif_dimauro2, classif_dimauro3,classif_liotta,classif_nn,classif_dimauro4}). Unfortunately, the straight application of machine learning to network traffic analysis is not always feasible, due to the broad variety of traffic types, which leads to unmanageable feature spaces. Intuitively, in fact, a highly diversified traffic (multimedia, asynchronous, bursty, etc.) requires a large set of features able to capture the variegated ``nature" of the different flows. That is why appropriate pre-processing steps (i.e. FS) play a crucial role, thus a significant portion of ML-based literature has shown interest in FS techniques.

For instance, to improve the performance of IDS frameworks, the authors of \cite{ids_18_1} propose a mixed strategy involving Principal Component Analysis and fuzzy clustering with KNN-based FS techniques. 
A correlation-based FS approach coupled with a Support Vector Machine (SVM) classifier is proposed in \cite{ids_18_2} to build a cloud-based IDS. 
An IDS based on deep learning methods, along with a filter-based FS algorithm, is introduced in \cite{ids_19}. Similarly, a Convolutional Neural Network (CNN) approach is exploited in \cite{ids_18_3} to select traffic features from raw data sets, improving the accuracy of an intrusion detector.  
Yu and Liu propose a mutual information-based algorithm that can analytically select optimal features for classification, by handling  linearly and non-linearly dependent data \cite{ids_16}. Again, FS is exploited jointly with Artificial Neural Networks \cite{ids_ann_19} and Deep Neural Networks \cite{ids_dnn_19}, respectively, to improve IDS performance. Furthermore, authors in \cite{ids_2011} embed in an IDS two FS algorithms that are compared against mutual information-based methods. 

A major shortcoming of these works is that methods are validated and compared through the outdated KDD$99$ dataset. This contains information about older network attacks (that have been mitigated by now) or, in some cases, adopts an updated version of KDD$99$, namely NSL-KDD \cite{nsl-kdd}. Yet, although NSL-KDD adds some improvements onto KDD$99$ (e.g. no redundant records, better balancing between training and test set, etc.), it does not take into account features that characterize novel cyber attacks. 
Other works in the field of machine learning applied to intrusion detection rely on more recent datasets such as  UNSW-NB15 \cite{unsw1,unsw2,unsw3,unsw4}. Although quite recent, such a dataset has two limitations: first,  the traffic has been collected in a reduced testbed; and secondly,  the number of features is limited to $49$, which is too small to appreciate the effectiveness of feature selection techniques.
In contrast, in our work we rely on an up-to-date dataset from the Canadian Institute for Cybersecurity \cite{cic} which has the following benefits: $i)$ it contains data traffic gathered over a vast network area; and $ii)$ it accounts for about $80$ features, allowing to extensively test the feature selection algorithms. Additional details about this dataset are provided in Section \ref{sec:dataset}.

Going more specifically into the set of works that share with this paper the aim to compare or survey FS methods for intrusion detection, we have collected the significant papers in Table $1$. In it, we have identified the material covered in the literature, which helps appreciating the contributions of our paper. The first column of Table $1$ points to the source; the second column highlights the type of experimental analysis (if any); the third column pinpoints the type of datasets utilized (single or multi-class); the last column provides a concise description of the surveyed material.  

\section{Review of Feature Selection Algorithms under Scrutiny}

In this section, we briefly describe the algorithms under scrutiny, which belong to different families of FS techniques, ranging from classic rank-guided (Rank, Linear Forward Selection) and meta-heuristic (Tabu, Scatter, Particle Swarm) ones, to nature-inspired algorithms (Ant, Cuckoo), and up to modern techniques (Genetic, Multi-Objective Evolutionary). 

For each algorithm, we provide a brief recap along with its pertinent application in network traffic analysis and security in literature. 

\subsection{Rank-based Feature Selection}

Algorithms belonging to this family follow an approach based on two macro-steps: in the first one, the features are \textit{ranked} according to a certain statistical measure, whereas in the second step the algorithm chooses the top ranked features (eventually partitioned in clusters). We investigate and put to test two representative algorithms of this family: Rank Search and Linear Forward Selection. 

\subsubsection{Rank Search}
The Rank Search technique refers not only to a single algorithm, but, to an umbrella of methods able to produce a list of attributes ranked with some criteria. One of the most applied rank-based techniques in FS relies on the Information Gain (IG) concept \cite{rank}. Given an attribute {\em A} and a class {\em C}, the {\em entropy} of the class without and with prior observation of the attribute are, respectively:

\beq
H(C)=-\sum_{c \in C}p(c)log_2p(c),
\label{eq:entr_noattr}
\eeq
and
\beq
H(C|A)=-\sum_{a \in A}p(a)\sum_{c \in C}p(c|a)log_2p(c|a).
\label{eq:entr_withattr}
\eeq

The IG is given by $H(C)-H(C|A_i)$, and represents the amount of the class entropy decreasing due to the {\em a priori} knowledge introduced by $i$-th attribute. 

Over network traffic analysis literature, the rank search method has been widely exploited in conjunction with standard machine learning algorithms typically used during the classification step. In \cite{rank_fuzzy}, having the NSL-KDD dataset as input, the proposed algorithm $i)$ evaluates the information gain value, $ii)$ applies fuzzy rules to remove unwanted features, $iii)$ calculates the mean value of IG across the new subset of features, $iv)$ refines the feature subsets by applying an algorithm based on conditional probability evaluation. 
Authors in \cite{rank_svm} propose a detection model able to cope with network attacks based on the feature IG ranking; once obtained a satisfying feature set, a triangle area based KNN, combining both SVM and greedy techniques, is exploited to single out even more discriminative and useful features.      
A combination between an IG-based feature selection method and a C4.5-based classifier is advanced in \cite{rank_c45}, where the resulting algorithm has been optimized (in terms of power consumption) to reveal DoS attacks in ad hoc networks. Similar ensemble has been exploited in \cite{rank_supervised}, where authors consider a broader set of classifiers (C4.5, Random Forest, Naive Bayes, etc.) and where the analysis focuses on generic malware detection. Again, a mixed Genetic/KNN approach for intrusion detection presented in \cite{rank_knn} benefits from a feature reduction procedure obtained applying ID3 algorithm to find higher IG.
\subsubsection{Linear Forward Selection}
Such a technique to reduce the feature space dimensionality has been presented in \cite{lfs}. Linear Forward Selection (LFS) can be considered as an improvement of Sequential Forward Selection (SFS) method which starts with an empty set of features and sequentially adds one feature at a time. In order to face the $\mathcal{O}(N^2)$ complexity of SFS, in LFS the number of considered features at each step does not exceed a certain user-specified constant, thus, the resulting performance is impressively ameliorated. Two methods for limiting the number of features have been implemented in LFS algorithm: $i)$ {\em Fixed Set}, where a score obtained by means of a wrapper evaluator is used to choose the top-$k$ ranked attributes, thus, the maximum number of evaluations reduces to $k/2(k+1)$ ; $ii)$ {\em Fixed Width}, where, at each forward selection step, the number of features is increased by one, such that the set of candidate expansions
consists of the best $k$ features not been selected so far during the search. In such case, the maximum number of evaluations amounts to $N \cdot k-k/2(k+1)$.
Forward Selection-based techniques have been exploited across the traffic classification domain for detection of malicious application on Android \cite{lfs_android}, detection of {\em zero day} attacks \cite{lfs_zero}, designing novel anomaly detection systems \cite{lfs_ids}.

\subsection{Meta-heuristic Feature Selection}

Techniques belonging to this family rely on the principle that it is possible to select \textit{heuristics}, namely,  approximate algorithms searching for a sufficiently good solution to an optimization problem, useful when some constraints arise (e.g. limited computational resources, incomplete information). We select three representative algorithms of this family: Tabu Search, Scatter Search, and Particle Swarm Optimization.

\subsubsection{Tabu Search}

Tabu Search (TS) algorithm has been originally proposed in \cite{glover_article}, and, then, it has been extended and exploited to solve practical optimization problems \cite{glover_book,rego_book,tabu}. TS is based on a {\em metaheuristic} method able to pilot a local heuristic search in exploring
the space of solutions beyond the local optimality. Two main features characterize TS: \textit{adaptive memory} and \textit{responsive exploration}. The former allows to perform local choices guided by information collected during the search, whereas, the latter allows to make strategic choices. In other words, TS exploits a local search procedure combined with memory-based strategies, thus, the issue of getting trapped in local optimal solutions is avoided. To implement the memory-based mechanism, TS builds a map of recently visited solutions called Tabu List (TL).
A simplified TS algorithm is illustrated below:

\begin{megaalgorithm}
	\SetAlgoLined
	\caption{Tabu Search}
	\label{alg_tabu}
	1. Given a function $f(x)$ to be optimized over a set $\mathcal{X}$, start from initial solution $x_0 \in \mathcal{X}$, initialize Tabu List (TL), and initialize a counter $i=0$. \\
	2. Given $N(x_i) \subset \mathcal{X}$ the {\em neighborhood} of $x_i$, $N(x_i)$ can be reached by $x_i$ by means of a {\em move} operation. Thus, generate a neighborhood move list $M(X_i$).\\
	3. Given $x_{i+1}$ the best solution in $M(X_i$), update TL. \\
	4. In case stopping conditions are met, terminate. Otherwise, repeat Step $2$.
\end{megaalgorithm}

\noindent It is interesting to notice that, the memory structures characterizing TS method operate according four {\em dimensions} \cite{glover_book}: 
\begin{itemize}
	\item {\em Recency}: concerns the ability of keeping track of solutions attributes that have changed across the recent past.
	\item {\em Frequency}: involves the mechanism exploited to broad the foundation for selecting preferred moves.
	\item {\em Quality}: pertains to the capacity of discriminating the merit of solutions visited during the search operation.
	\item {\em Influence}: takes into account the impact of choices performed during the search.
\end{itemize}

In the field of network traffic classification, TS has been exploited in \cite{tabu_fuzzy} jointly with fuzzy techniques aimed at optimizing the exploration of feature search space in intrusion detection problems. A combination of TS technique and KNN is presented in \cite{tabu_knn}, where KNN is initially exploited to generate a subset of non-redundant features, whereas, TS is used to refine the obtained subset. Again, authors in \cite{tabu_c45} propose a {\em wrapper} FS algorithm, dubbed GATS-C4.5, that embeds an hybrid Genetic and Tabu-based method as feature selection strategy and a supervised ML algorithm (C4.5) as the evaluation function. Similar combinations between TS and Genetic techniques have been used in \cite{tabu_genetic} where some tests (vs pure Genetic algorithms) across DARPA dataset have been carried out, and in \cite{tabu_genetic2} where SVM has been adopted as a classification criterion.

\subsubsection{Scatter Search}

Scatter Search is a metaheuristic algorithm involving memory-based mechanisms similar to those exploited in Tabu Search \cite{scatter_source}. The main strategy relies on an iterative process that organizes high-quality optimal solutions into subsets, and where five ``methods" emerge: $i)$ {\em Diversification}, to create a set of different trial solutions by exploiting a seed solution; $ii)$ {\em Improvement}, to convert a trial solution in one or more improved solutions; $iii)$ {\em Reference Set Update}, to create and keep update a reference set of best solutions; $iv)$ {\em Subset Generation}, to manipulate the reference set aimed at deriving a subset of solutions; $v)$ {\em Solution Combination}, to linearly re-combine the solutions on the basis of subsets obtained with the previous methods. 

As an effective FS procedure, Scatter Search has been exploited together with NLP solvers for global optimization \cite{scatter_nlp}, with rough sets for the implementation of credit scoring mechanisms \cite{scatter_roughset}, or in a parallelized fashion to improve the feature subset selection problem \cite{scatter_parallel}. Again, Scatter Search has been profitably exploited in issues involving credit cards fraud detection \cite{scatter_credit}, and software security characterization \cite{scatter_sec_sw}.

\subsubsection{PSO Search}

The Particle Swarm Optimization (PSO) algorithm has been originally discovered in \cite{pso_scource}, through simulations carried out across a simplified social model aimed at reproducing the behavior of birds flocking. 

Then, the population of agents became more similar to a swarm than flock, and single individuals were named {\em particles} that represent the candidate solutions.
In a mathematical form, given $\mathcal{A} \subset \mathbf{R}^n$ the search space, and $f:\mathcal{A}\xrightarrow{}Y\subseteq\mathbf{R}$, the swarm is defined by a set $\mathbf{S}=\{x_1,x_2,\dots,x_N\}$ of $N$ particles, where $x_i=(x_{i1},x_{i2},\dots,x_{in})^T \in \mathcal{A}$ with $i=1,2,\dots,N$. It is assumed that particles are able to iteratively move within search space $\mathcal{A}$ by means of a velocity parameter defined as $v_i=(v_{i1},v_{i2},\dots,v_{in})^T$ with $i=1,2,\dots,N$. 
In PSO method, it is also defined a memory set $\mathbf{P}=\{p_1,p_2,\dots,p_N\}$ containing the best positions $p_i=(p_{i1},p_{i2},\dots,p_{in})^T$ (with $i=1,2,\dots,N$) visited by each particle. Given $t$ the time counter, the current position and velocity for particle $i$ are, respectively, $x_i(t)$ and $v_i(t)$, whereas $p_i(t)=\argmin_{t} f_i(t)$.
Accordingly, PSO is defined by the following equations:

\beqa
v_{ij}(t+1)&=&v_{ij}(t)+\gamma_1 r_1 [p_{ij}(t)-x_{ij}(t)]
\\ \nonumber
&+&\gamma_2 r_2 [g(t)-x_{ij}(t)], 
\label{eq:pso_vel}
\\
x_{ij}(t+1)&=&x_{ij}(t)+v_{ij}(t+1),
\label{eq:pso_pos}
\eeqa
where: $r_1$ and $r_2$ denote random variables uniformly distributed in [0,1], $\gamma_1$ is the {\em cognitive} parameter which affects the step size that the particle takes towards its best candidate solution $p_{ij}(t)$, and $\gamma_2$ is the {\em social} parameter which affects the step size that the particle takes towards the swarm's best solution $g(t)$. At each iteration, best positions $p_{i}(t+1)$ are updated as well, namely

\begin{equation}
\label{eq:pso_eq}
\centering
p_{i}(t+1)=
\left\{
\begin{array}{l}
{\begin{array}{ll}
	\hspace{-0.2cm} x_i(t+1) \;\;\;\;\; \text{if } \; f(x_i(t+1))\leq f(p_i(t)),
	\end{array}}
\\
\\
p_{i}(t) \;\;\;\;\;\text{otherwise.}
\end{array}
\right.
\end{equation} 

In the field of FS applied to network traffic analysis, PSO has been profitably exploited jointly with classification techniques. It is the case of \cite{pso_svm,pso_svm2}, where a particle swarm selection method has been exploited with SVM-based classifiers. Again, a hybrid FS model based on PSO and Random Forest has been exploited in \cite{pso_rf}, where  independent measures and a learning algorithm are exploited to evaluate feature subsets.
Interesting is also an evolution of classic PSO advanced in \cite{pso_bigdata}, and dubbed Accelerated PSO, amenable to deal with FS on Big Data streams.

\subsection{Nature-inspired Feature Selection}

Although relying on meta-heuristic concepts, this family of algorithms takes inspiration from the nature ecosystem, where many animal species exhibit impressive behaviors aimed at optimizing their life cycle. We assess Ant Optimization and Cuckoo search as representative algorithms of such family.

\subsubsection{Ant Search}
This technique, originally proposed in \cite{ant}, is inspired by ants colonies behavior, where the optimization problem is solved by a ``colony" of cooperating agents. Analyses carried out by ethologists showed that, following pheromone trails, each ant is able to follow a preceding ant which releases such substance, thus, a whole ant colony is able to self-organize itself. The emerging collective behavior relies on a positive feedback loop: the probability which an ant chooses a certain path increases with the number of ants that choosing the same path, since the trail is continuously reinforced with new pheromone.
The Ant algorithm can be profitably exploited in feature selection problems by evaluating the probability $p_{ij}^k$ that the $k$-th ``ant" could arrive to feature $j$ by starting from feature $i$, namely,

\begin{equation}
\label{eq:ant_eq}
\centering
p_{ij}^k=
\left\{
\begin{array}{l}
{\begin{array}{ll}
	\hspace{-0.2cm} \frac{[\tau_{ij}]^\alpha \cdot [\eta_{ij}]^\beta}{\sum_{k \in \mathscr{U}_k} [\tau_{ik}]^\alpha \cdot [\eta_{ik}]^\beta } \;\;\;\;\; \text{if } \; j \in \mathscr{U}_k,
	\end{array}}
\\
\\
0 \;\;\;\;\;\text{otherwise,}
\end{array}
\right.
\end{equation}   
where, $\mathscr{U}_k$ is the set of feasible attributes (not visited yet), $\tau_{ij}$ is the amount of pheromone across the $ij$ path, $\eta_{ij}$ represents the heuristic information for the selected attribute $j$, $\alpha$ and $\beta$ are parameters in charge of controlling pheromone trials and heuristic information, respectively. 

As regards the FS problem in network traffic analysis, Ant-based methods have been used in: \cite{ant_svm} where a SVM classifier is adopted on KDD99 dataset, after a feature reduction obtained through ACO (Ant Colony Optimization) technique; in \cite{ant_stream} where an ACO-based feature selection method allows to deal with big streamed data; \cite{ant_faco} where an improved ACO-based algorithm (named FACO) has been designed and tested across the classic KDD99 dataset.  

\subsubsection{Cuckoo Search}

This technique is inspired to the {\em brood parasitism} strategy characterizing some cuckoos species. In particular, such species lay their eggs in the nests of other birds (hosts). Since host birds can engage a conflict as they recognize alien eggs, particular cuckoo species have evolved in such a way that females are able in mimicking colour and size of eggs of some host species, thus the hosts are cheated, and the probability of cuckoos reproductivity grows. Considering an egg in a nest as a solution, three idealized rules emerge in Cuckoo Search procedure \cite{cuckoo_book}: $i)$ each cuckoo lays one egg at time, in a randomly chosen nest; $ii)$ the bests nests (having high-quality eggs) are candidate to carry over next generations; $iii)$ the number of nests is fixed and, as a host bird discovers alien (cuckoo) eggs with a probability $p_d$, it gets rid of it. The aim of the algorithm is to exploit new (and eventually better) solutions in place of not-so-good solutions in the nest. 
New solutions $\mathbf{x}_i^{t+1}$ for cuckoo $i$ are obtained through the following expression which represents the stochastic equation for random walk, namely, 

\beq
\mathbf{x}_i^{t+1}=\mathbf{x}_i^t + \alpha \otimes \mathscr{L}(\lambda), 
\label{eq:cuckoo}
\eeq
where, $\alpha>0$ is a scale factor, the product $\otimes$ refers to entrywise multiplications, whereas $\mathscr{L}$ is the L\'evy distribution with ($1<\lambda\leq 3$). 

Cuckoo search method has been exploited in network traffic analysis paired with various techniques and technologies. In \cite{cuckoo_pca}, authors propose an algorithm that uses PCA and Cuckoo Search to reduce the feature space and to optimize the clustering center selection. A Cuckoo-based FS algorithm is proposed in \cite{cuckoo_ids} to preprocess network data aimed at improving the IDS detection accuracy in cloud environments. A Cuckoo search strategy has been also used in \cite{cuckoo_ann} to optimize Artificial Neural Networks when dealing with traffic anomaly detection issues. Again, coupled with SVM, Cuckoo search has been adopted in FS to deal with problem of phishing mail detection \cite{cuckoo_svm}.
Recently, extended versions of Cuckoo Search algorithm have been advanced to cope with classification of {\em tweets} in sentiment analysis \cite{cuckoo_sentiment}, or to defeat attacks in Software Defined Network infrastructures \cite{cuckoo_sdn}. 

\subsection{Evolutionary Feature Selection}
\vspace{0.2cm}
Such family of algorithms is inspired by natural selection theory, claiming that living organisms survived across millions of years thanks to an adaptation process. In a similar way, this aptitude can be translated in search for optimal solutions to a problem. Two exemplary tested algorithms are: Genetic search and Multi Objective Evolutionary search. 

\vspace{0.4cm}
\subsubsection{Genetic Search}

Genetic Algorithms (GAs) have been designed around the mid-1950s, when biologists started to perform computer-based simulations aimed at analyzing more in deep the evolution of genetic processes \cite{weise}. Then, GAs have been extended to face problems ranging from neural networks weight estimation \cite{genetic1} to inequalities-based problems \cite{genetic2}. A pioneering work in this field has been carried out by Holland \cite{holland_1962,holland_1992}, and, today, many variants of GAs exist \cite{genetic_survey} and are applied in economy, computer science, sociology.   

The basic skeleton of a GA includes three operators \cite{Goldberg:1989:GAS:534133}: {\em Reproduction, Crossover} and {\em Mutation}. 

Reproduction refers to a process in charge of evaluating the ability of an individual to be selected (among others) for reproduction, on the basis of a {\em fitness} score. 

Crossover concerns the capability of a genetic operator in recombining information to create new offspring. Typically, offspring is generated by exchanging genes of parents until a {\em crossover point} is reached. 

Mutation pertains to the probability that some offspring genes could be modified or altered.    

Genetic-based feature selection in network traffic analysis has been used in conjunction with many ML-based methods. Authors in \cite{ga_ann} exploit a GA-based FS approach to optimize network traffic data before applying an artificial neural network to perform attacks detection across cloud infrastructures. A combination of a genetic FS method and a supervised classifier based on J48 algorithm is proposed in \cite{ga_j48}. More frequent across the scientific literature is the coupling between genetic FS and SVM classifiers applied to network traffic classification problems (see \cite{ga_svm,ga_svm2,ga_svm3}).

When dealing with FS problems, GAs allow to explore the solution space by selecting the most promising regions, thus, avoiding a costly exhaustive search. In our domain, the initial population is represented by the whole feature space and the fitness function relies on the correlation among features and expressed by means of a {\em merit} indicator defined further ahead in eq. (\ref{eq:merit}). 

\begin{figure}[pos=ht]
	\centering
	\captionsetup{justification=centering}
	\includegraphics[scale=0.3,angle=90]{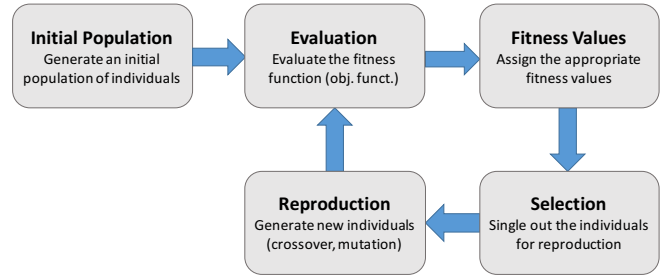}
	\caption{Genetic Algorithms life cycle.}
	\label{fig:GA}
\end{figure}

Once entered the cycle represented in Fig. \ref{fig:GA}, the algorithm calculates the fitness of each candidate solution per iteration, selects individuals to reproduce, and generates a new population by taking into account crossover (feature recombination with a certain probability), and mutation (one feature can be turned into another feature with a certain probability). 

\subsubsection{Multi-Objective Evolutionary Search}

The family of solutions concerning a multiobjective optimization problem (MO) includes all the elements of the search space whose objective vectors cannot be simultaneously improved (Pareto optimality concept) \cite{moea_source}. The set of such objective vectors is said non-dominated.

More formally, a MO problem can be formulated as follows: given a vector of $n$ objective functions $\mathbf{f}$ of a vector variable $\mathbf{x}$ in a domain $\mathcal{D}$ defined as
\beq
\mathbf{f}(\mathbf{x})=(f_1(\mathbf{x}), f_2(\mathbf{x}), \dots, f_n(\mathbf{x}),
\eeq
a decision vector $\mathbf{x}_h \in \mathcal{D}$ is Pareto-optimal {\em iff} there is no $\mathbf{x}_k \in \mathcal{D}$ such that:

\begin{align}
\label{eq:reward_het_col}
\left\{
{\begin{array}{ll}
	\hspace{-0.2cm} \;\;\ \forall i \in \{1,\dots,n\}, k_i \leq h_i  \\  \\
	\;\;\;\;\;\;\;\;\; \;\;\; \;\;\;    \wedge  \\    \\
	\exists i \in \{1,\dots,n\} : k_i < h_i. 
	\end{array}}
\right.
\end{align}

On the other hand, Evolutionary Algortihms (EAs) can be profitably exploited in MO-based problems since many  ``individuals" can search in parallel for multiple solutions, with the possibility of taking advantage of similarities among solutions belonging to the same family. Possible implementations of MO-EA techniques are ENORA (Evolutionary NOn-dominated Radial slots based Algorithm), and NSGA (Non-dominated Sorted Genetic Algorithm) compared in \cite{moea}. 

Applied to the FS problem, the purpose of a multi-objective search algorithm is to discover a subset of features (a family of solutions) being a good approximation of the Pareto front.  
The MO-EA approach has been exploited in network traffic classification jointly with ensemble ML-based methods \cite{moea_ensemb}, where some objectives such as maximizing true positive rate, maximizing classification accuracy, minimizing feature number, and minimizing false positive rate, are satisfied simultaneously and with no conflicts. 
An NSGA-based approach to ameliorate the performance classification of IDS platforms has been adopted in \cite{moea_nsga1} and in \cite{moea_nsga2}.
Again, a MO-EA technique jointly with fuzzy classifiers for coping with the traffic classification problem has been introduced in \cite{moea_fuzzy}. 

\section{The considered Datasets}
\label{sec:dataset}

One of the great issues when dealing with supervised FS approaches in traffic analysis, is finding training sets that are both recent and labeled.
As remarked in Sect. $3$, many works rely on the obsolete KDD$99$ dataset \cite{kdd99}. Created about $20$ years ago, KDD$99$ has been broadly employed to validate machine learning algorithms, particularly to differentiate malicious data traffic from benign one. However, the KDD$99$ dataset does no longer reflect the characteristics of modern data traffic, which has sensibly changed across time. This is very much the case of multimedia traffic (e.g. voice, video), being the principle revenue-making stream for service providers but also the vehicle of covert malicious data. 
\textit{Per contra}, in this work we consider more recent datasets obtained by means of CICFlowMeter \cite{cicflowmeter,icissp-dataset}, an open-source engine that can both gather and label network traffic in a controlled environment. 
Each dataset contains records labeled either as Benign or Malicious (malicious traffic is often split in sub-labeled traffic representing different kinds of attacks). 
Specifically, we consider the following datasets:
\begin{itemize}
	\item \textit{DDoS} which contains traffic relating to distributed denial of service attacks, aimed at saturating the network resources of specific targets;
	\item \textit{Portscan} which contains traffic relating to Portscan attacks, aimed at discovering open, network device ports;
	\item  \textit{Webattack}, including malicious traffic which implements various web-based attacks such as Brute Force, Cross-Site Scripting, and Sql Injection;
	\item \textit{TOR} which includes traffic passing over the TOR network, an anonymous and private data circuit often exploited to carry dangerous information or malicious encrypted traffic;
	\item \textit{Android} which embeds various families of Android-based threats (adwares, ransomwares, etc.). 
\end{itemize} 
We want to notice that the first four datasets are exploited into the single class analysis, whereas the \textit{Android} dataset is exploited in the forthcoming multi-class analysis. In this latter analysis, we build a new dataset called \textit{MultiAndroid} (see Sect. $6.2$), obtained by selecting the most relevant mobile threats from the \textit{Android} dataset mixed with some benign traffic.
In order to avoid the issue of bias that typically arises during classification in imbalanced datasets, we have first re-arranged the datasets. We have achieved well-balanced benign/malicious features, spanning across about $50k$ instances for each dataset. Each one contains up to $78$ features, except for the TOR dataset including 30 features.   
For the sake of convenience, we found it useful to group the features in $5$ macro-classes (the  complete list of features can be found in \cite{icissp-dataset2}): 
\begin{itemize}
	\item \textbf{Time-based features:} Forward/Backward inter-arrival times (IAT) between two flows, duration of active flow (min, max, mean, std), duration of idle flow (min, max, mean, std), etc.; 
	\item \textbf{Byte-based features:} Forward/Backward number of bytes in a flow, Forward/Backward number of bytes used for headers, etc.;  
	\item \textbf{Packet-based features:} Forward/Backward number of packets in a flow, Forward/Backward length of packets in a flow (min, max, mean, std), etc.;  
	\item \textbf{Flow-based features:} Length of a flow (mean, max, etc.);  
	\item \textbf{Flag-based features:} Number of packets with active TCP/IP flags (FIN, SYN, RST, PUSH, URG, etc.).
\end{itemize}

\section{Experimental Results}

\begin{figure*}[pos=htp] 
	\centering
	\begin{tabular}{cccc}
				\subfloat[Ant (21 fts)]{\includegraphics[width=2.3in]{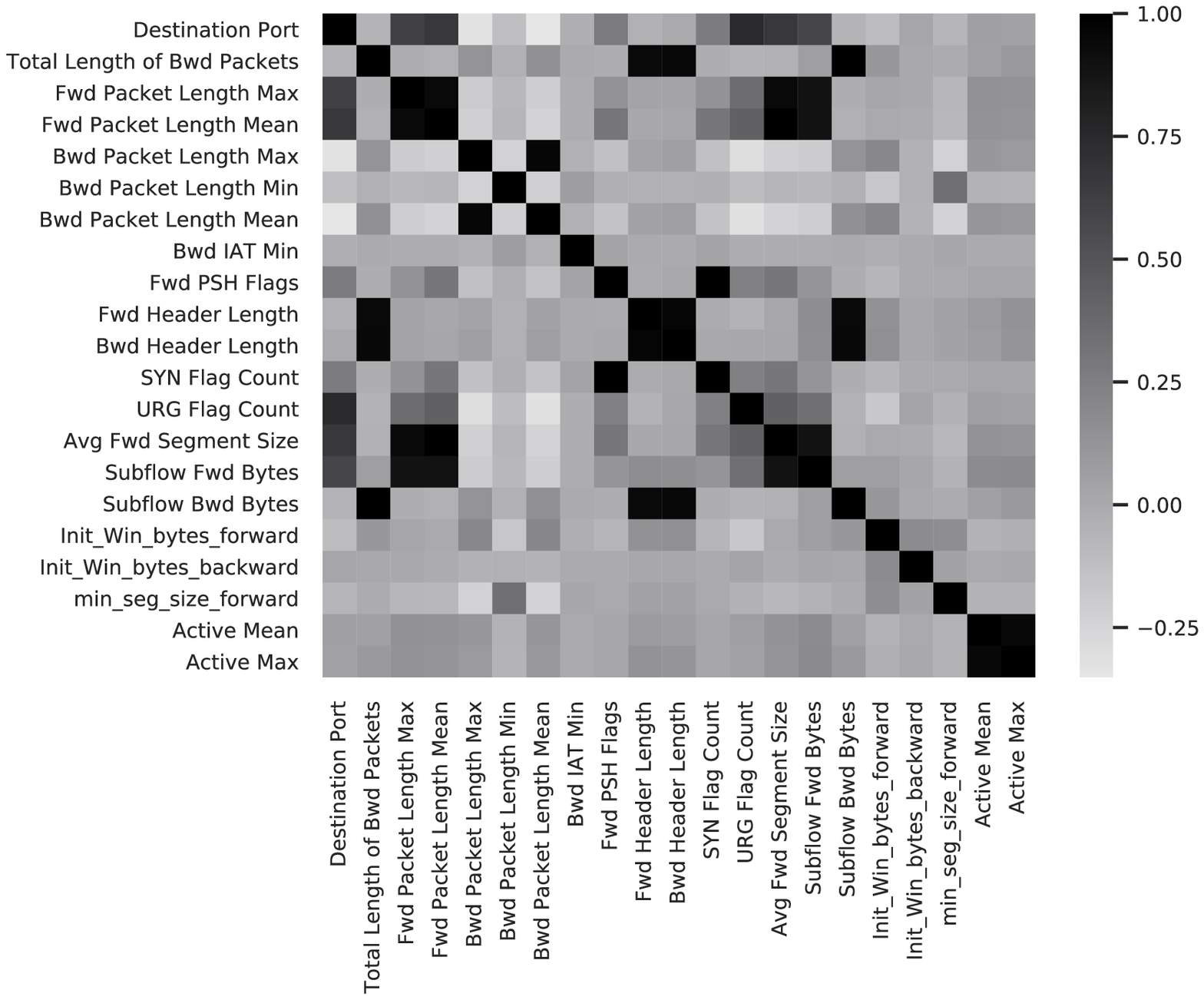}} \hfil
				\subfloat[Scatter (4 fts)]{\includegraphics[width=2.3in]{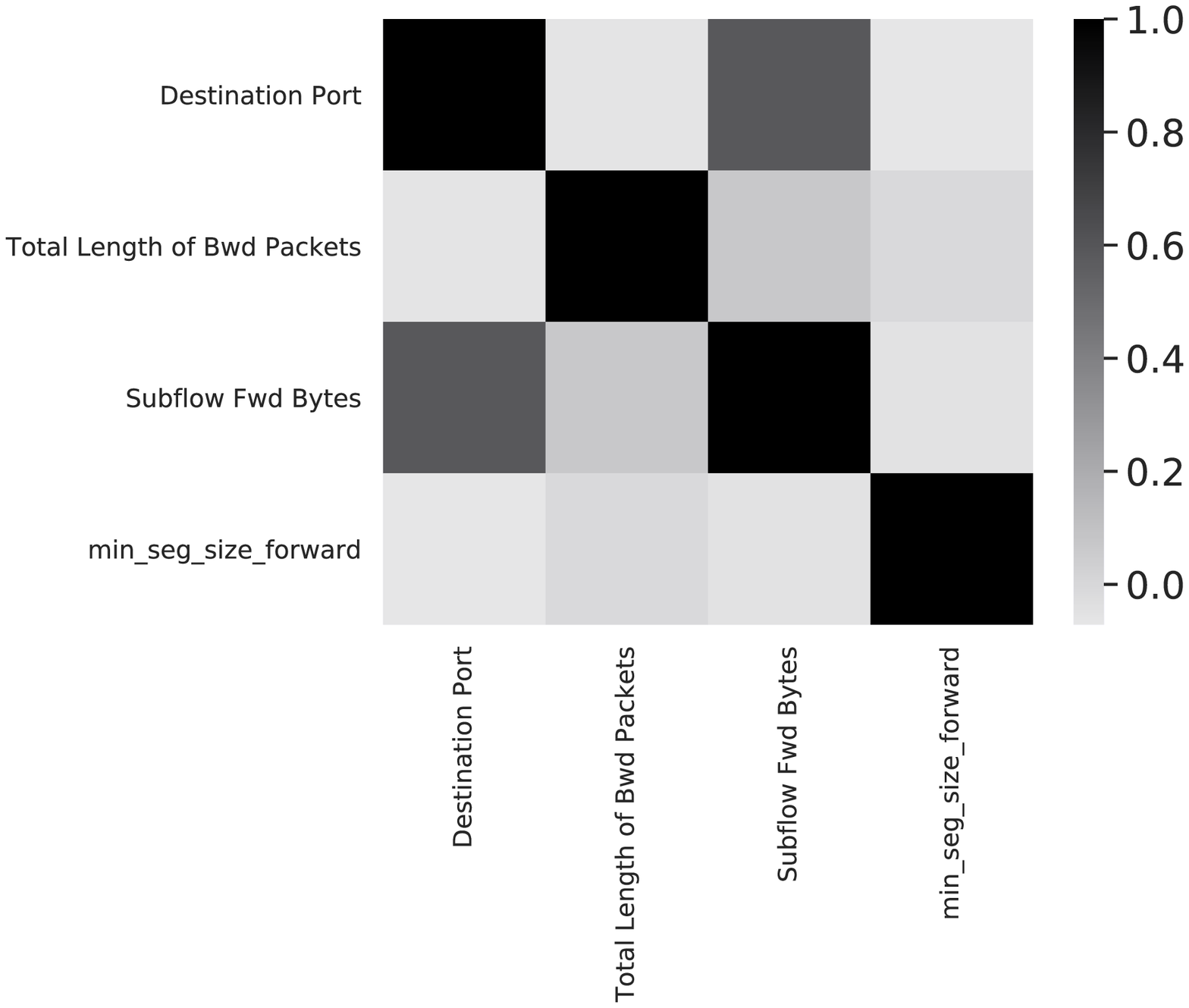}} \hfil
				\subfloat[MO-EA (5 fts)]{\includegraphics[width=2.1in]{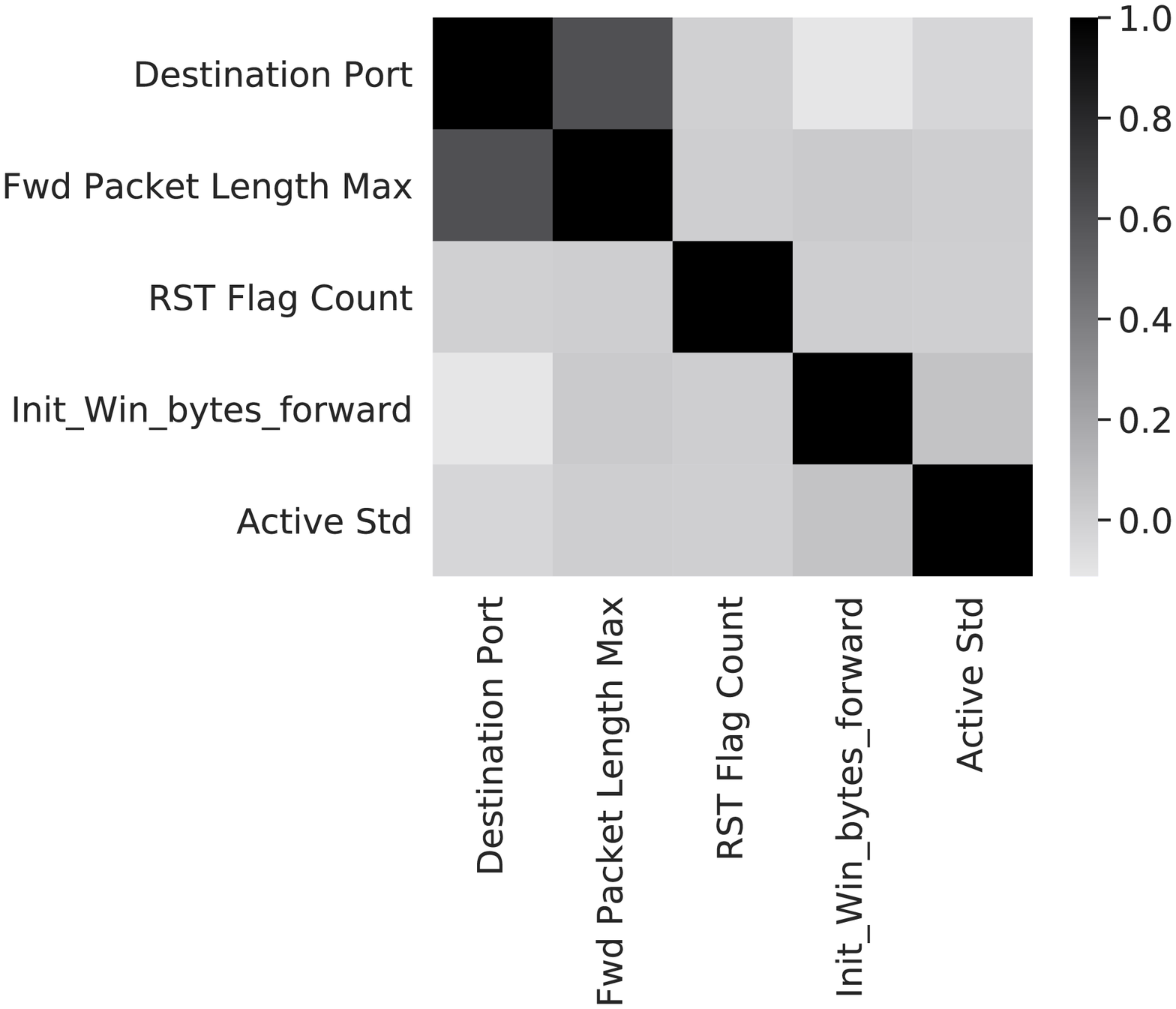}} \hspace{10mm} \\
				\subfloat[Ranking (10 fts)]{\includegraphics[width=2.3in]{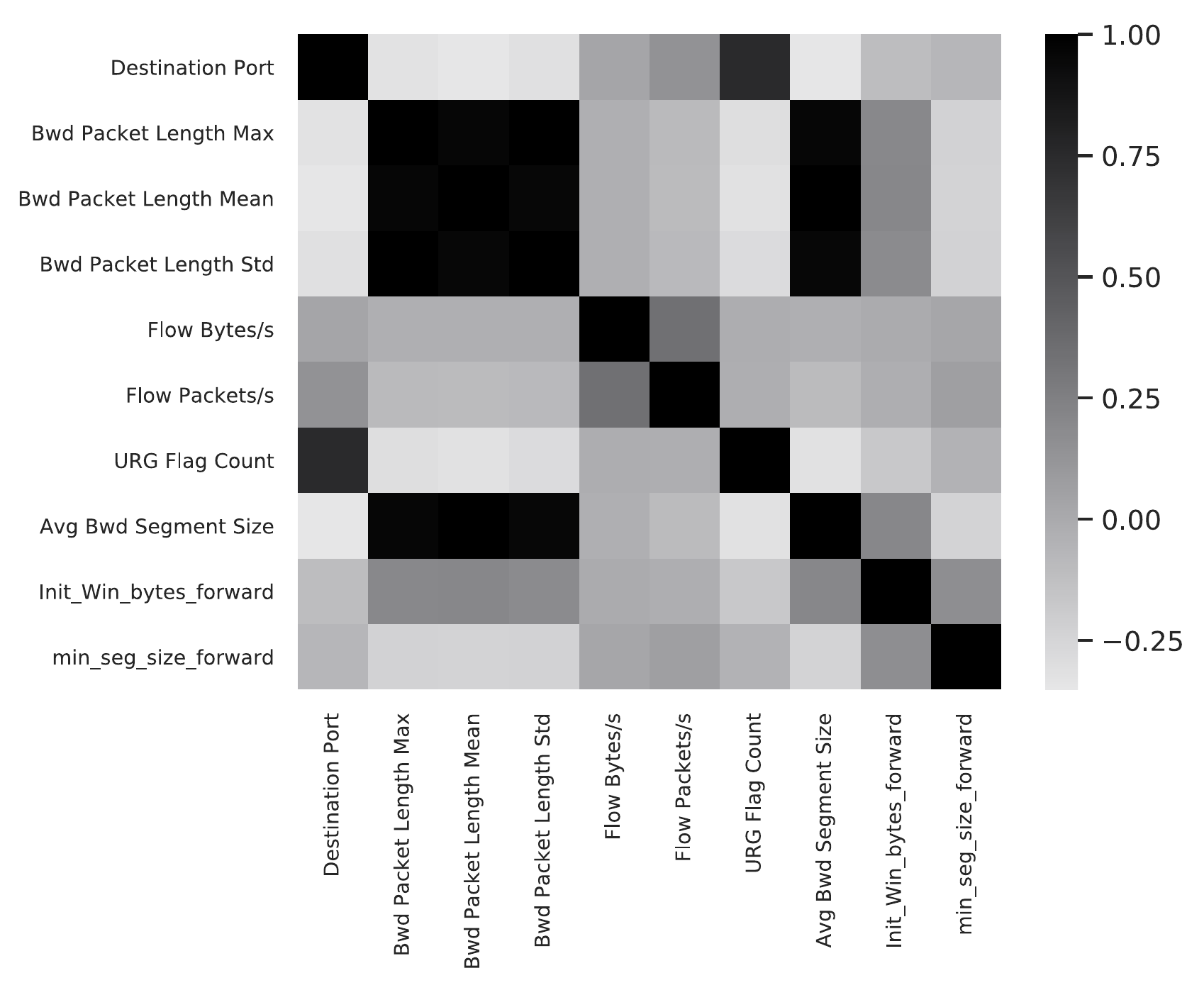}} \hfil 
				\subfloat[Cuckoo (7 fts)]{\includegraphics[width=2.3in]{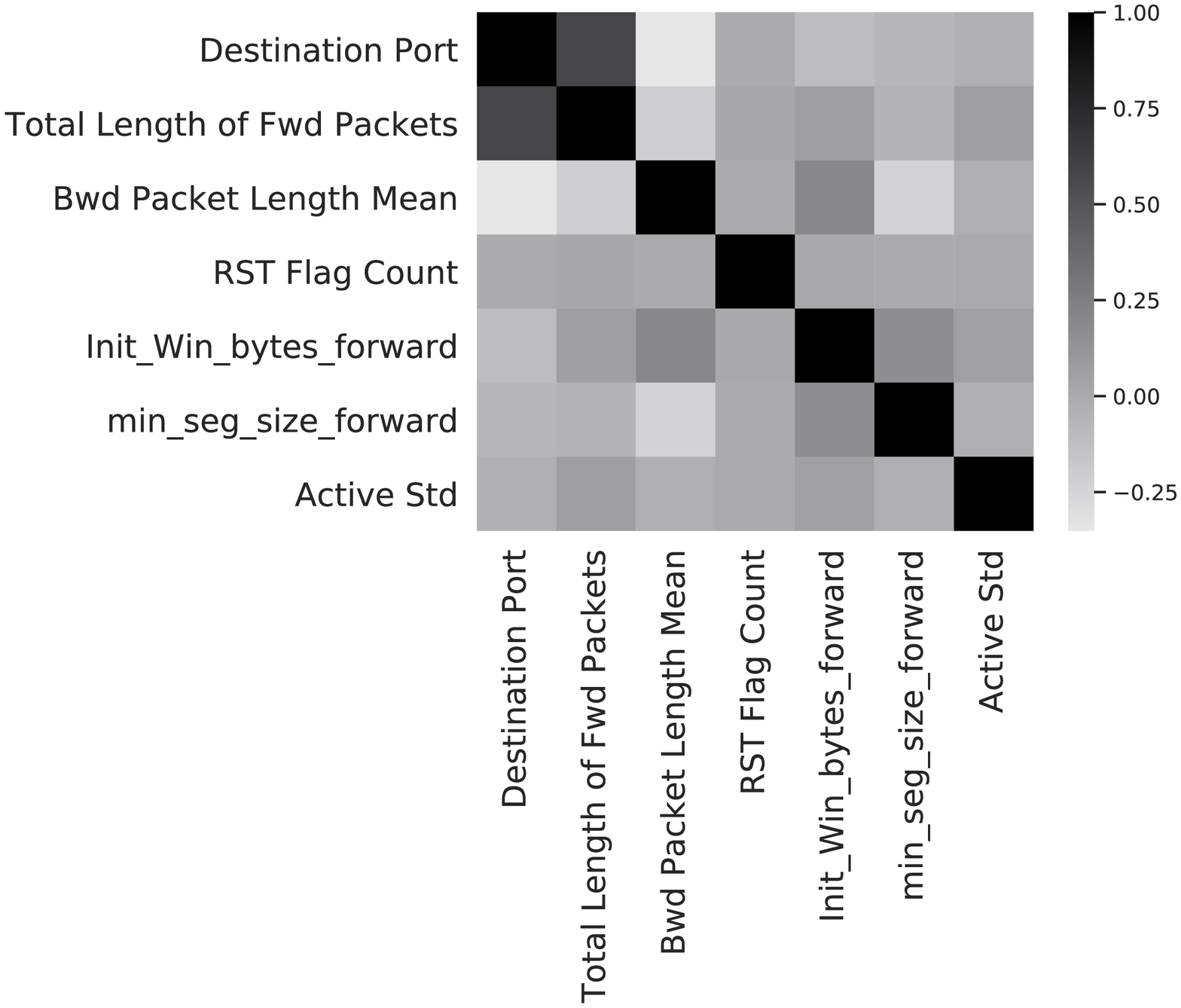}}  \hfil
				\subfloat[Tabu/LFS (6 fts)]{\includegraphics[width=2.1in]{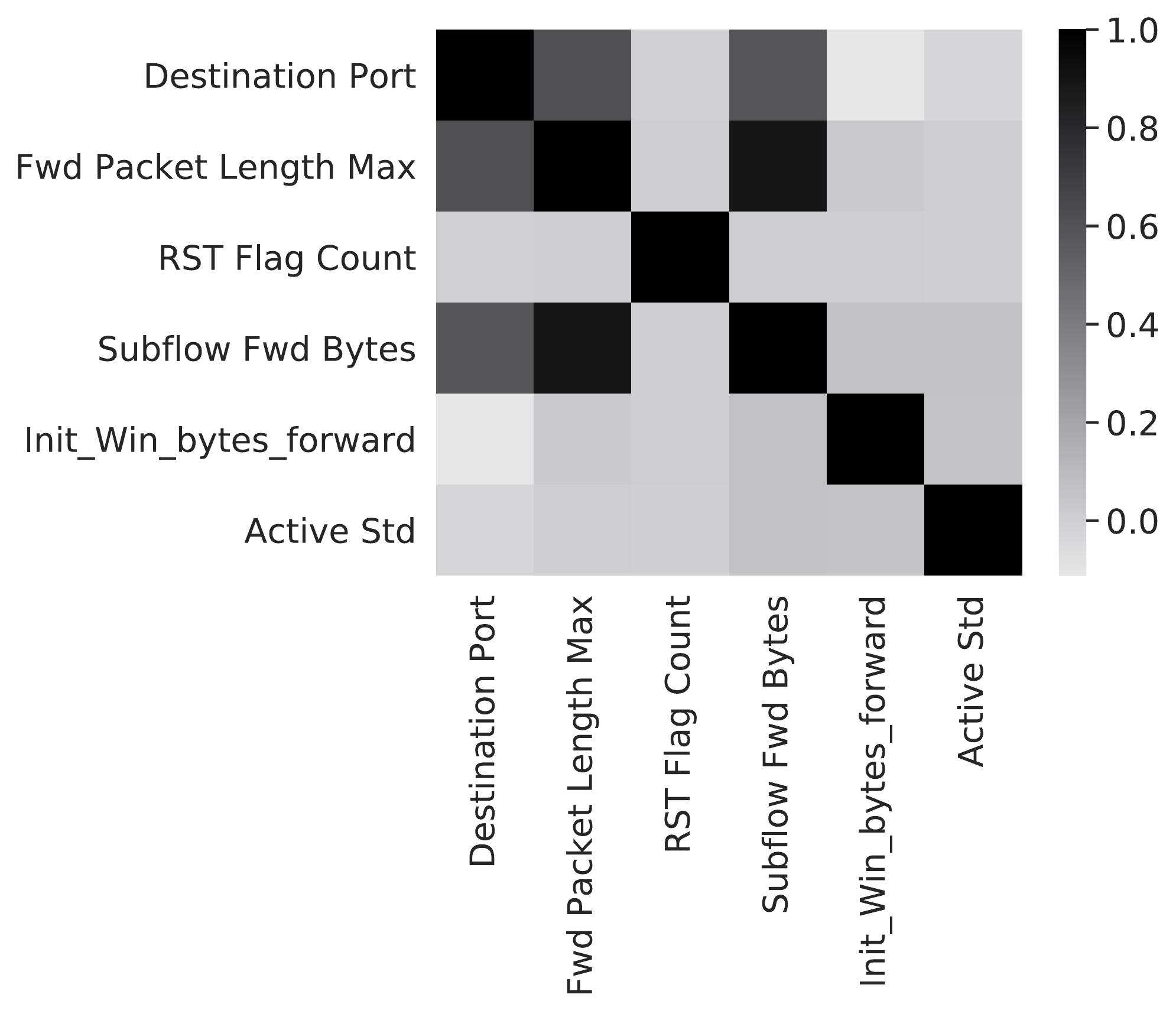}} \hfil \\
				\subfloat[Genetic (27 fts)]{\includegraphics[width=2.4in]{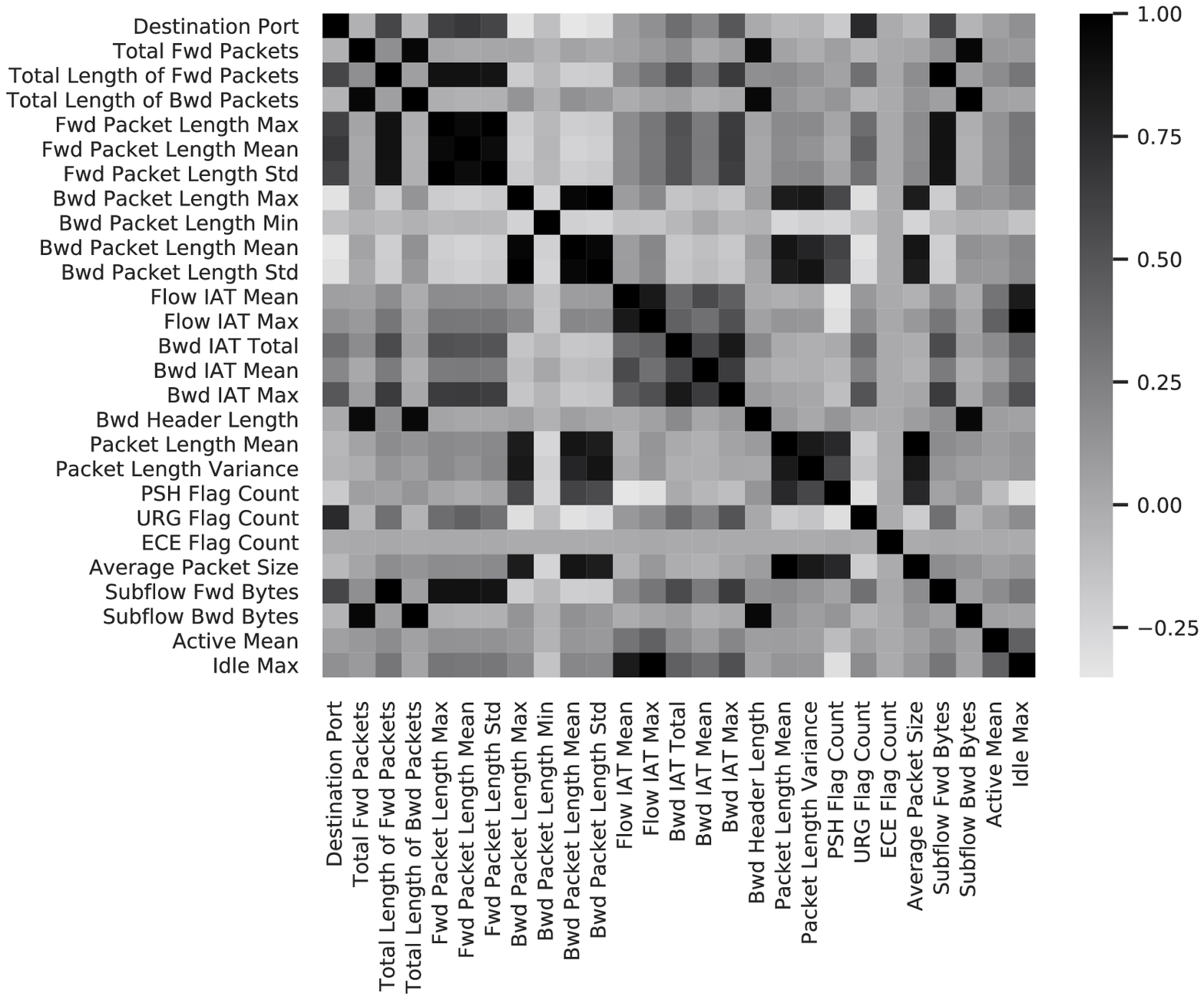}}   \hfil
				\subfloat[PSO (18 fts)]{\includegraphics[width=2.4in]{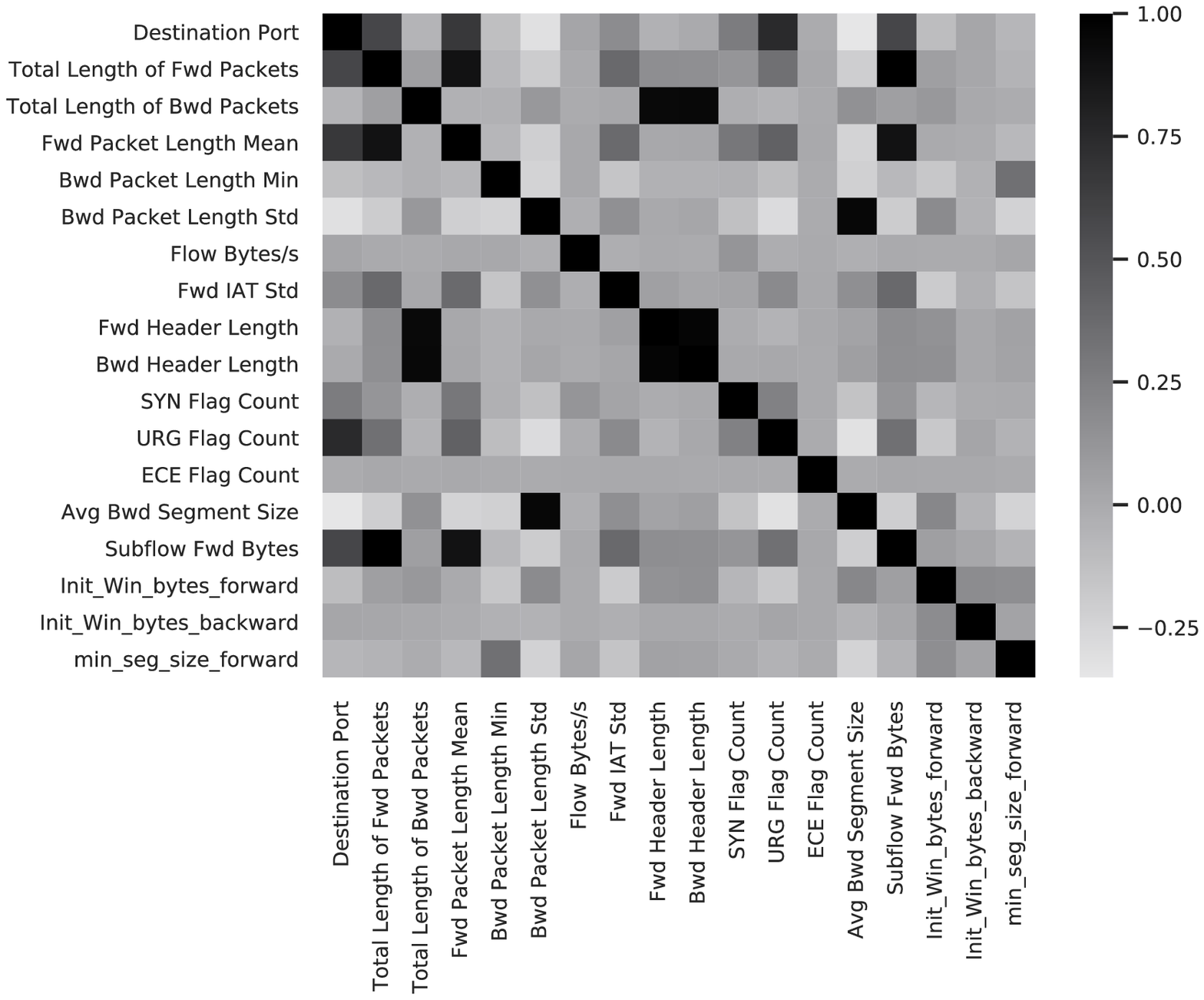}}  \hfil
	\end{tabular}
	\caption{Correlation maps for different algorithms - DDoS dataset. In parenthesis is reported the number of features surviving after the FS process.}
	\label{fig:corrmaps_ddos}
\end{figure*}

Herein we describe our analytical study, which required the development of a dedicated Python routine, to normalize/balance the datasets and to automate the comparison among FS algorithms. 
We have validated the scrutinized search algorithms using the Correlation-based Feature Selector (CFS) as objective function \cite{cfs1,cfs3}. 

The main idea behind CFS is that a good feature subset includes those features that are highly correlated with the class, while being strongly uncorrelated among them. A formal definition is offered in \cite{cfs5}: a feature $X_i$ is relevant {\em iff} there is some $x_i$ and $y$ for which $p(X_i=x_i)>0$ such that

\beq
p(Y=y|X_i=x_i)\neq p(Y=y).
\eeq
Namely, $X_i$ is relevant if $Y$ is conditionally dependent on $X_i$.
Thus, CFS is a filter algorithm that can rank feature subsets according to a correlation-based heuristic function. Precisely, given a subset $S$ including $k$ features, the heuristic {\em merit} $M_{S,k}$ is defined as:

\beq
M_{S,k}=\frac{k \overline{r_{fc}}}{\sqrt{k+k(k-1)\overline{r_{ff}}}},
\label{eq:merit}
\eeq
where $\overline{r_{fc}}$ is the average value of feature/class correlations, and $\overline{r_{ff}}$ is the average value of feature/feature correlations.  
The numerator of (\ref{eq:merit}) may be seen as an indicator of how far a set of features is predictive of a class; whereas, the denominator contains information about how much redundancy there is among features.

Our assessment is split into two parts: the first one concerns a \textit{single class} analysis, where we evaluate datasets exhibiting \textit{dichotomous} information (malign/benign); the second one is focused on \textit{multi class} problems, where we evaluate the effectiveness of FS in the presence of multiple classes.

\subsection{Single Class Analysis}

\begin{figure*}[pos=htp] 
	\centering
	\begin{tabular}{cc}
		\subfloat{\includegraphics[width=3.2in]{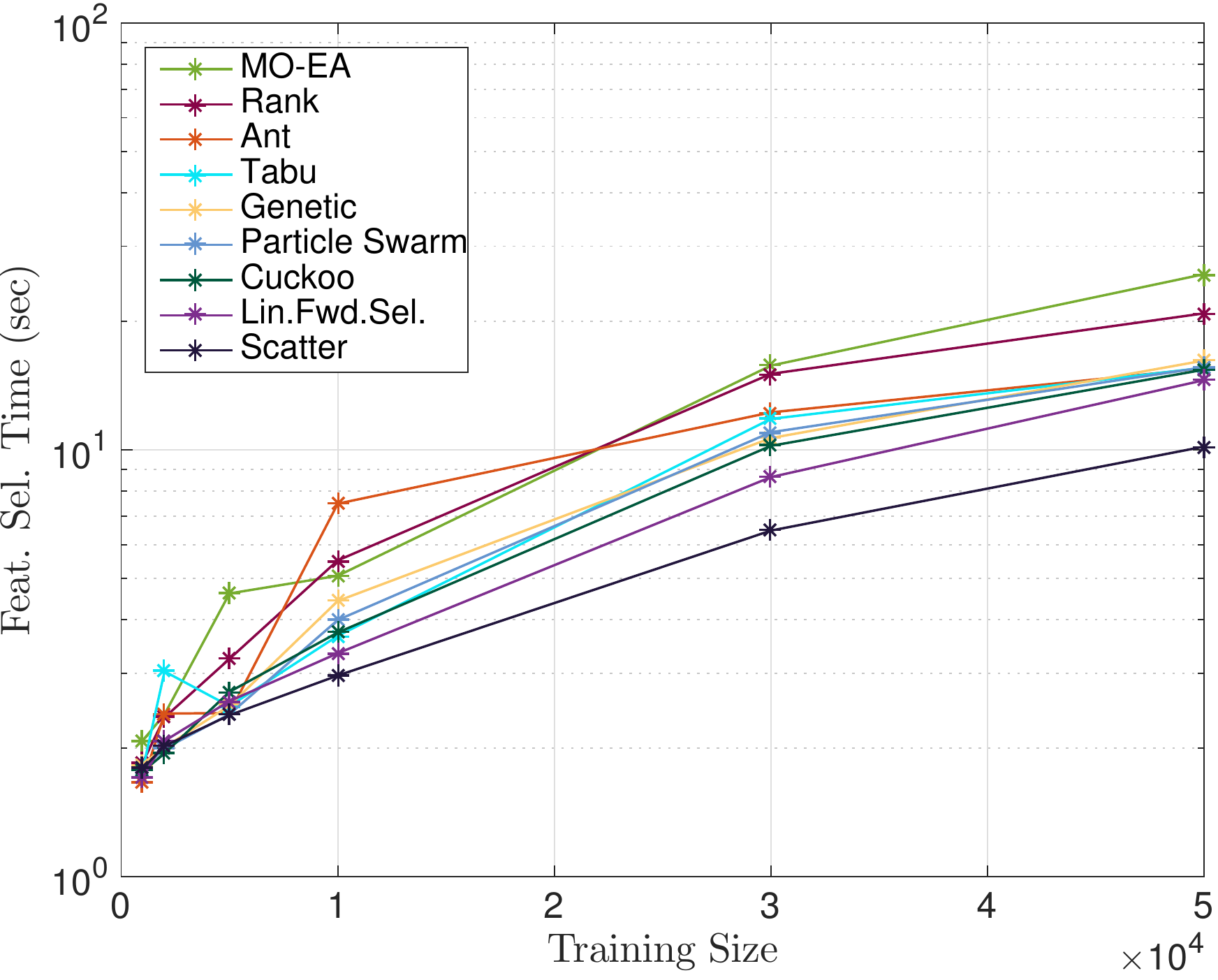}} \hspace{8mm}
		\subfloat{\includegraphics[width=3.2in]{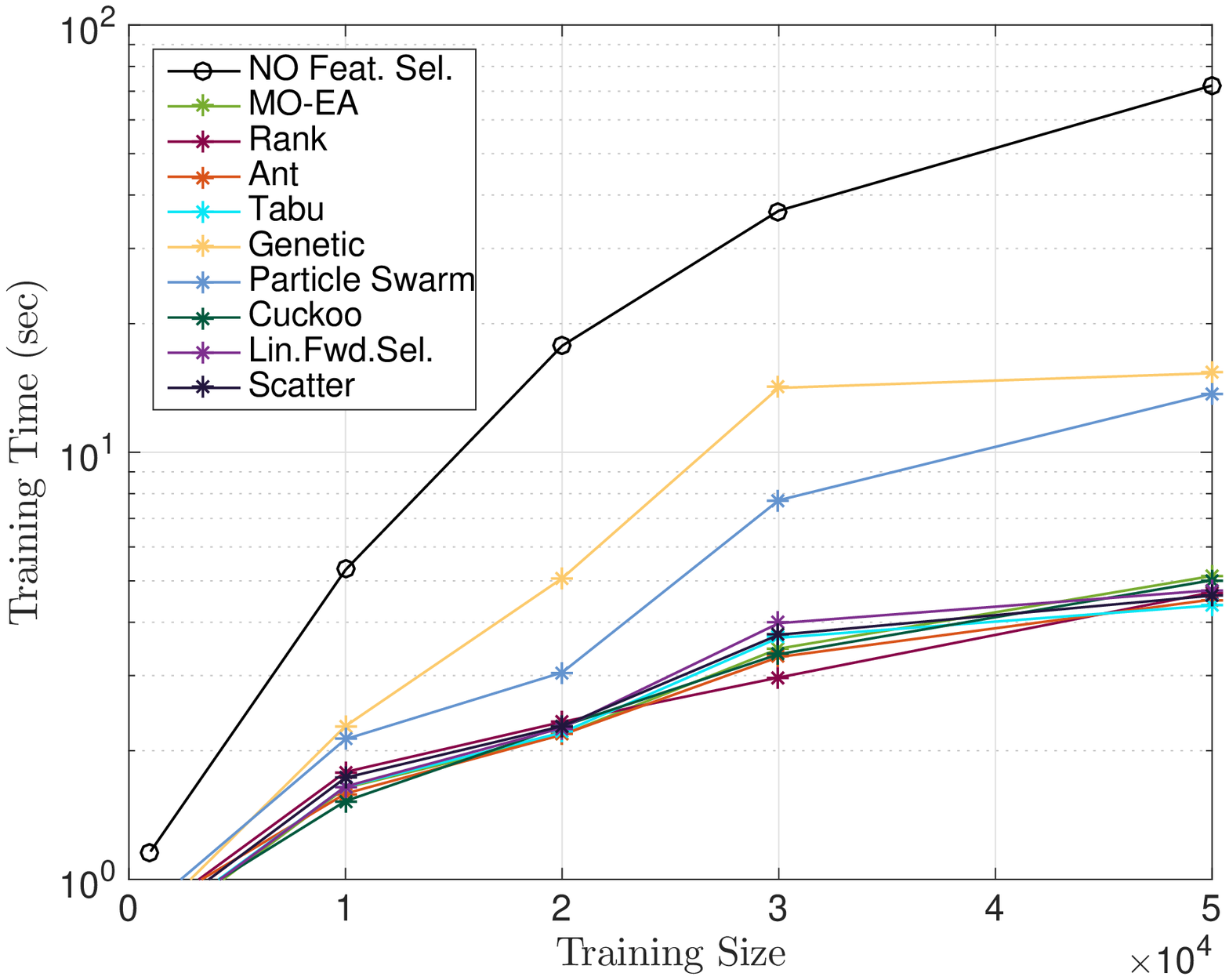}}
	\end{tabular}
	\caption{FS times - DDoS dataset (a); Training times - DDoS dataset (b).}
	\label{fig:times}
\end{figure*}

Let us consider the Distributed Denial of Service (DDoS) attack which, recently, is also affecting modern SDN-based networks \cite{jsac4,jsac5}. DDoS attacks are designed to overwhelm the target network resources by means of a \textit{botnet}, namely, a network composed of a large number of malicious nodes sending tiny packets towards the target, ultimately coordinated by a \textit{botmaster}. 

Let us now analyze the results obtained by pre-processing the DDoS dataset through the set of FS algorithms introduced above. In Fig. \ref{fig:corrmaps_ddos} we report, for each algorithm, the correlation map corresponding to a graphical representation of covariance matrices. This representation embeds three important pieces of information: \textit{i)} the number of features surviving after the FS processing step; \textit{ii)} the type of features; and \textit{iii)} the relationship existing among surviving features. The latter is taken into account by means of a gray scale, in which darker shades indicate higher levels of correlation. Thus, each $(i,j)$ ``pixel" gives the correlation level between feature $i$ and feature $j$. 
Accordingly, the pixels on the main diagonal are always black (maximum correlation, corr=$1$), due to the self-correlation.
As was to be expected, higher correlation are found among those features belonging to the same family (Time-based, Flow-based, etc.). 

Some interesting considerations about the various correlation maps arise. First, the number of features retained by different algorithms may significantly diverge, which is due to the specific approaches adopted by each algorithm. The Genetic algorithm is the one retaining the most features. This is to be ascribed to the particular strategy of this algorithm, which strives to escape local optima by applying the mutation operator, thus allowing to consider more paths, namely, more features. Second, some common features retained by all the algorithms can be recognized. For instance, the destination port feature is always present since, in a DDoS attack, a target victim is typically reached on a particular exposed TCP/UDP port. Moreover, since DDoS attacks are characterized by a large amount of small-size packets, features embodying information about packet lengths are retained. The difference is that, some algorithms (e.g. Scatter, MO-EA, Cuckoo, Tabu, LFS) just keep the essential features related to packet length (e.g. total packet length, total number of bytes sent in initial window); whereas, other algorithms (e.g. Ranking, Genetic, PSO, Ant) prefer to retain more features belonging to the same family. DDoS is also characterized by some kind of synchronization among the bots, which are coordinated to launch an almost-simultaneous attack. This means that time-related features will often provide useful information to detect DDoS. Interestingly, the Genetic algorithm retains $5$ features relating to the inter-arrival flow times, resulting in a dark gray cluster at the center of the correlation map (Fig. \ref{fig:corrmaps_ddos}(g)). 

\begin{figure*}[pos=ht] 
	\centering
	\begin{tabular}{cccc}
		\subfloat[]{\includegraphics[scale=0.48]{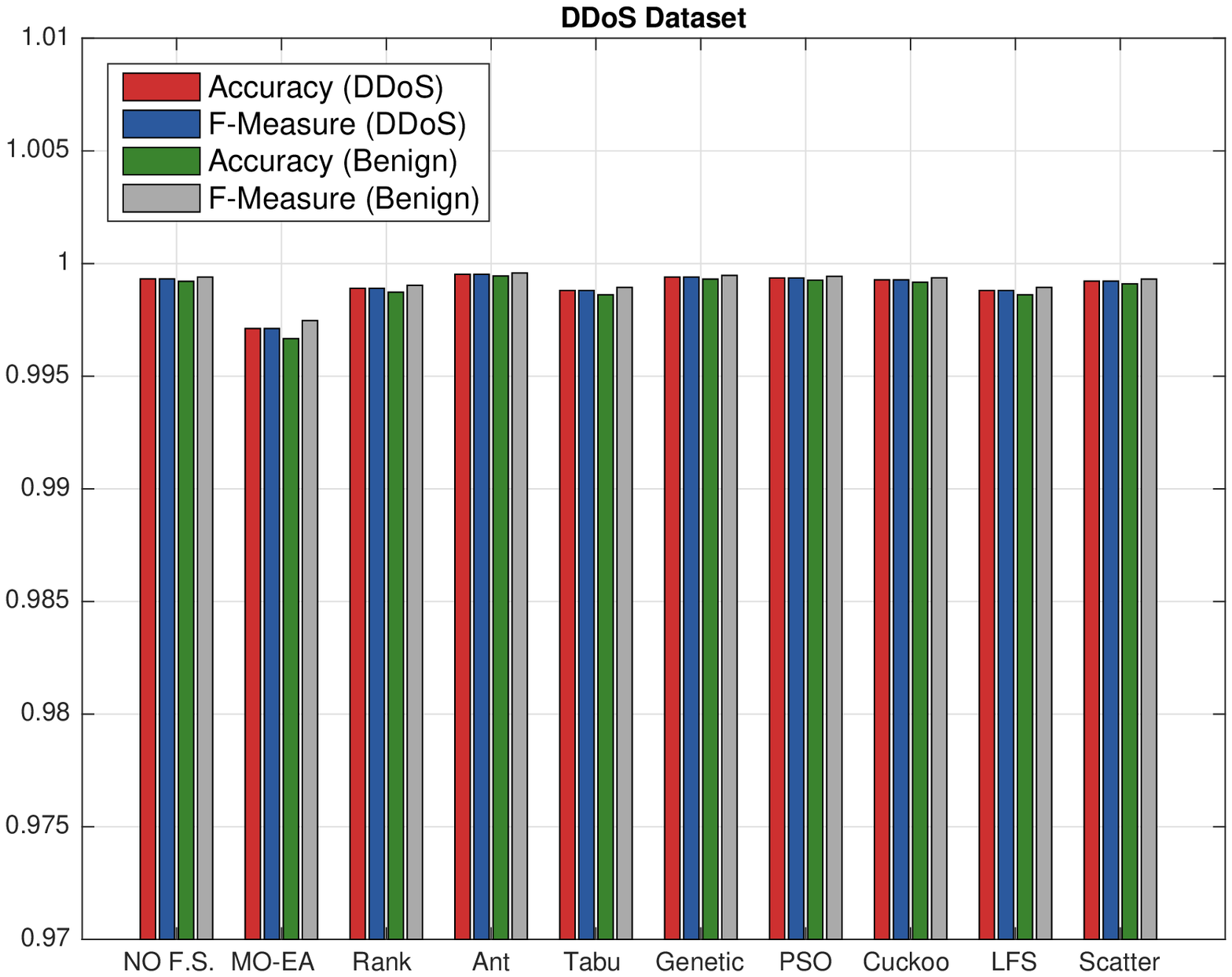}}  \hspace{8mm}
		\subfloat[]{\includegraphics[scale=0.48]{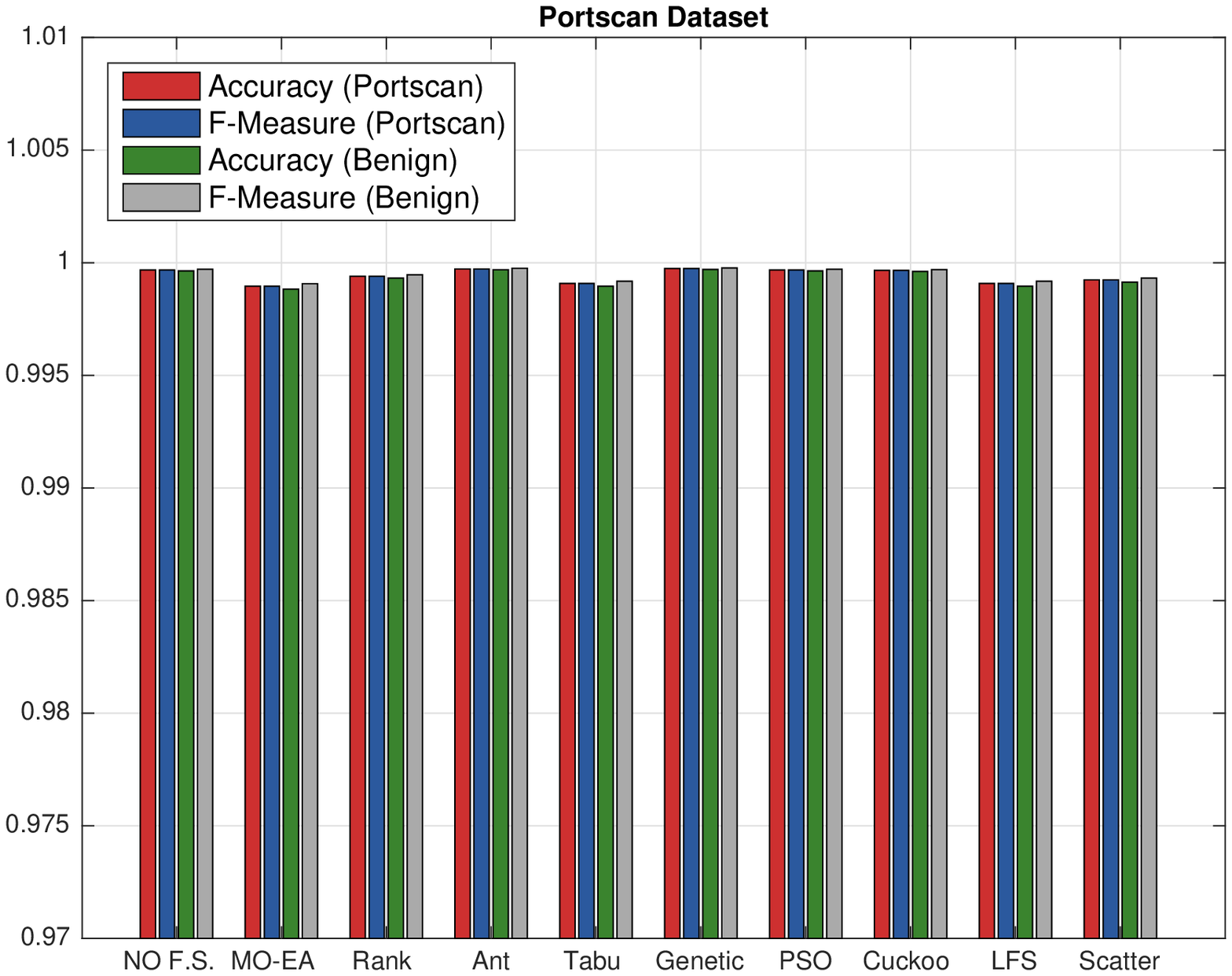}}  \hspace{8mm} \\
		\subfloat[]{\includegraphics[scale=0.48]{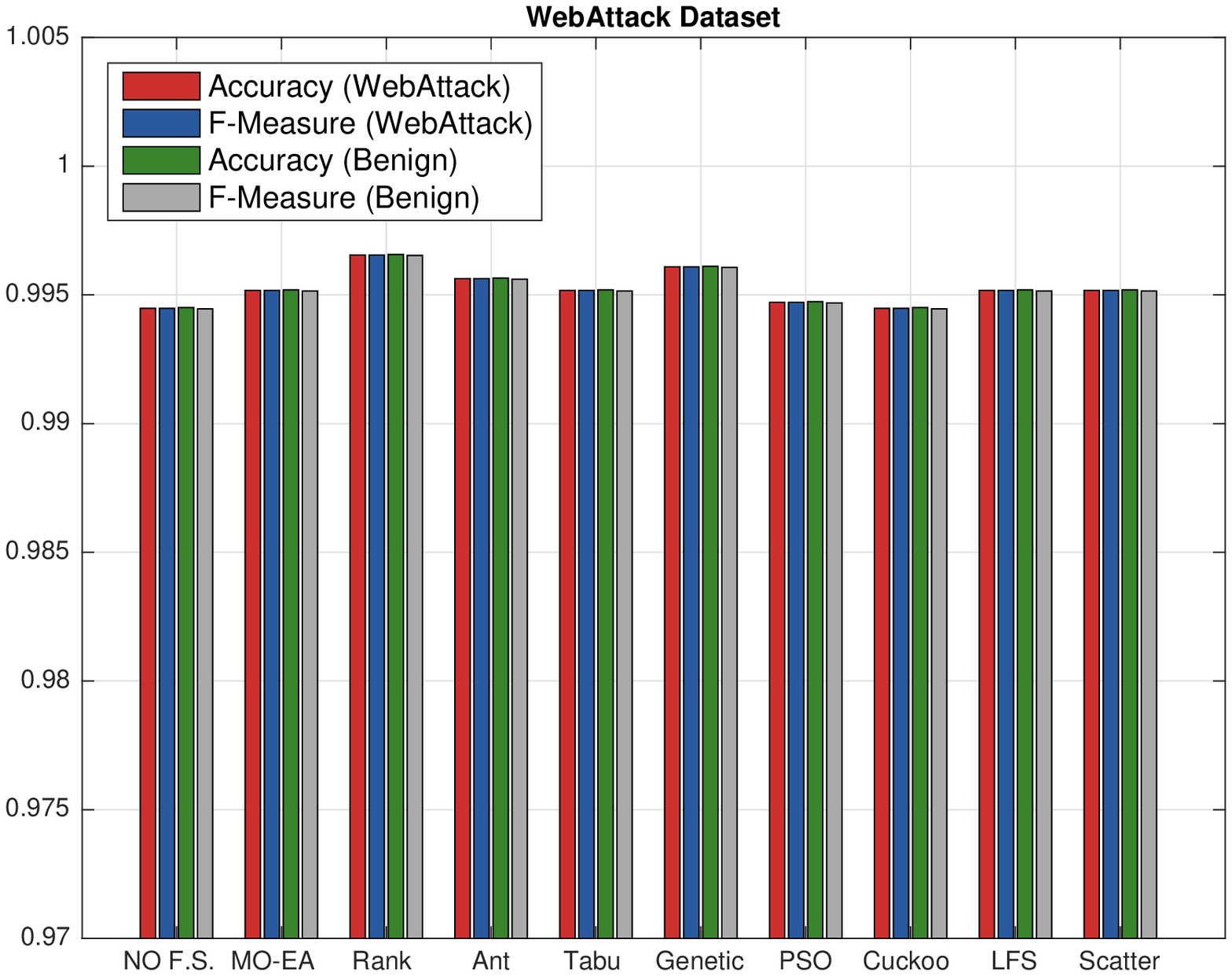}}  \hspace{8mm}
		\subfloat[]{\includegraphics[scale=0.48]{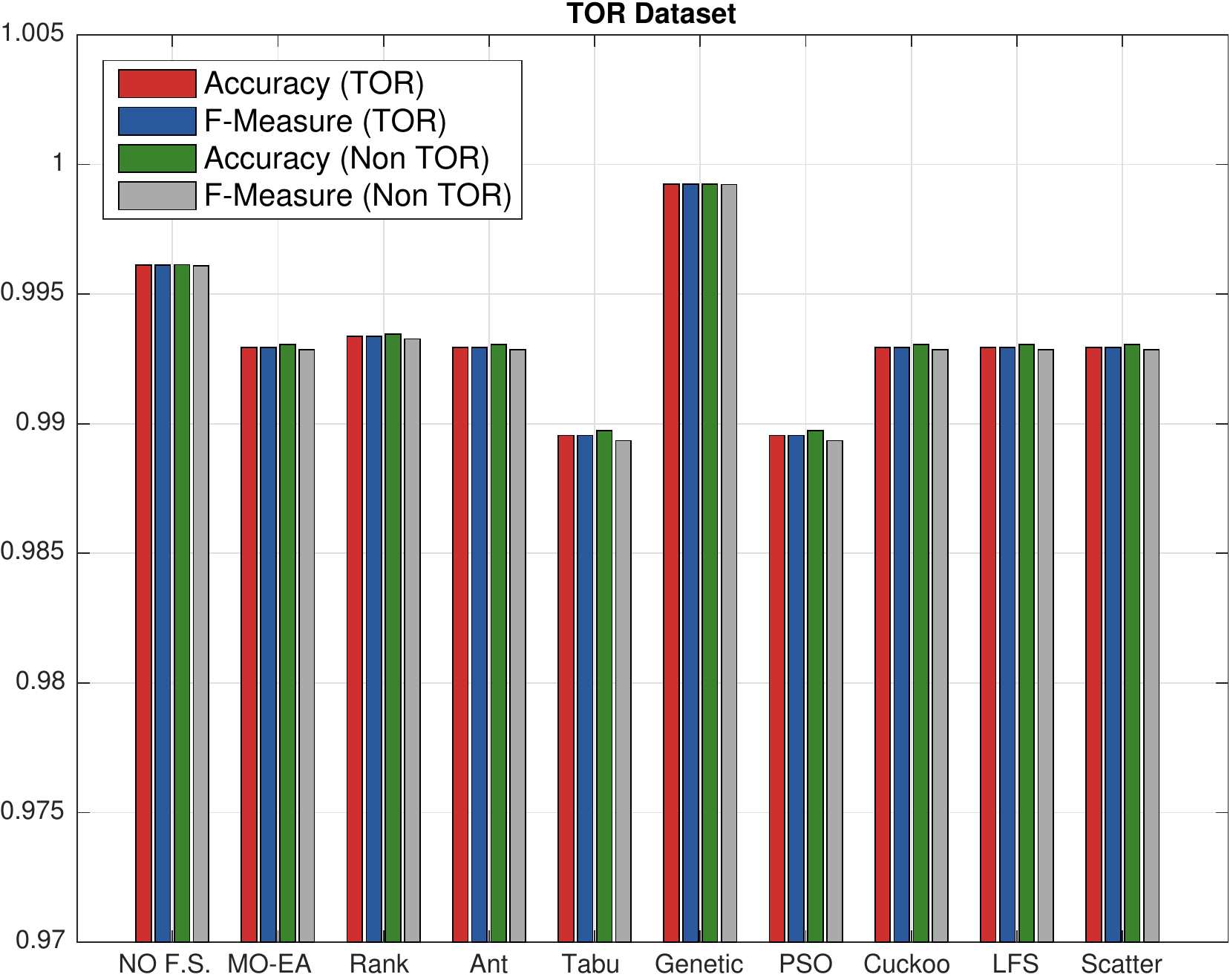}} \hspace{8mm}
	\end{tabular}
	\caption{Performance in terms of Accuracy/F-Measures for different single class datasets: DDoS (a), Portscan (b), WebAttack (c), TOR (d).}
	\label{fig:perf_single}
\end{figure*}

It is also possible for DDoS attacks to be even more effective through the modification of the IP flags (e.g. SYN/RST flooding). Accordingly, features embodying information about IP flags (e.g. RST-SYN-URG flag count) are retained by algorithms such as Ant (Fig. \ref{fig:corrmaps_ddos}(a)), MO-EA (Fig. \ref{fig:corrmaps_ddos}(c)), Cuckoo (Fig. \ref{fig:corrmaps_ddos}(e)), Genetic (Fig. \ref{fig:corrmaps_ddos}(g)), and PSO (Fig. \ref{fig:corrmaps_ddos}(h)). 
Let us note that many algorithms opt for selecting features that are uncorrelated among them (few dark gray or black clusters are present) since they convey more variegated information.

\begin{figure*}[pos=ht]
	\centering
	\begin{tabular}{cccc}
		\subfloat[Ant (22 fts)]{\includegraphics[width=2.4in]{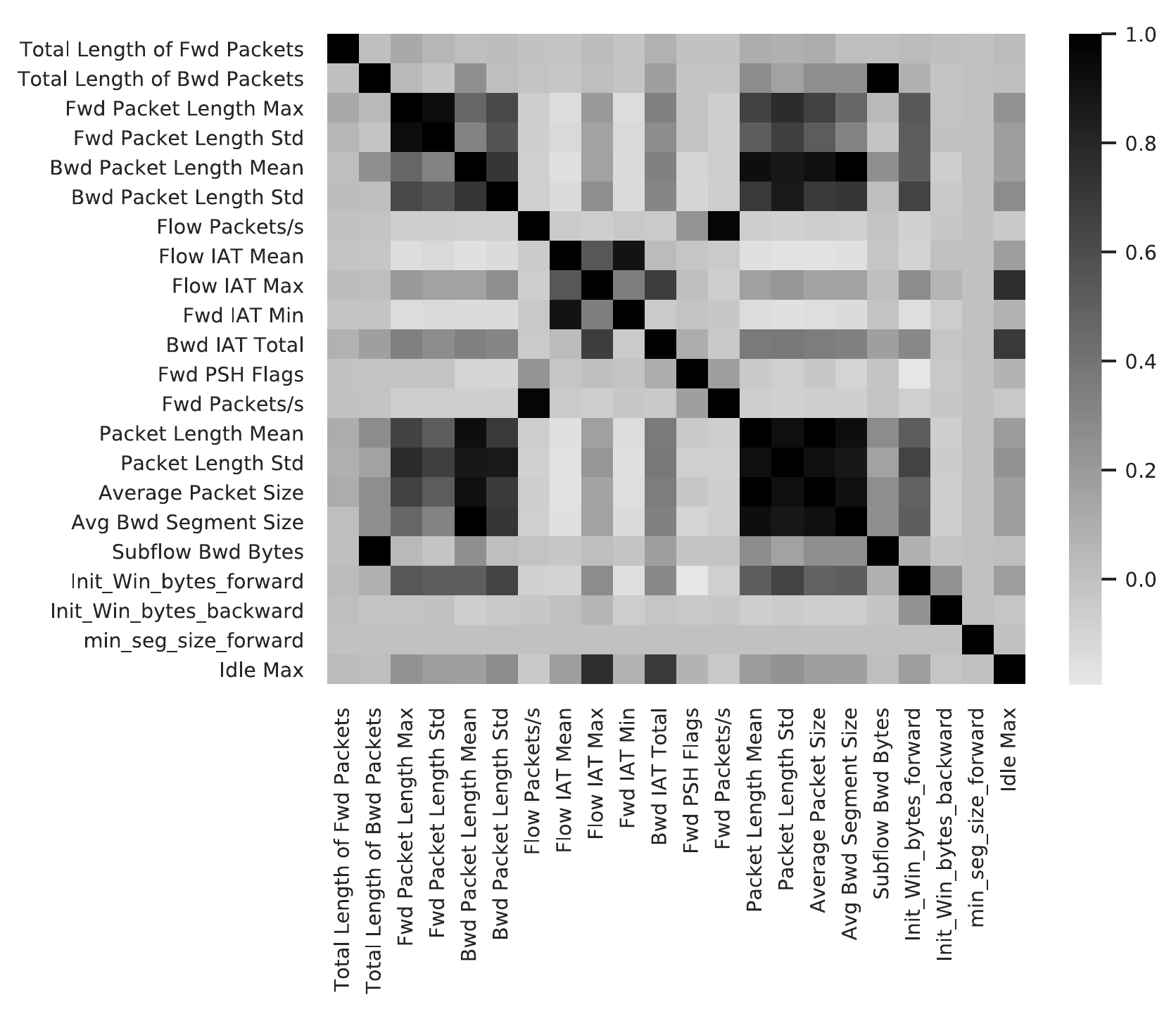}} \hfil
		\subfloat[Scatter/Tabu (9 fts)]{\includegraphics[width=2.4in]{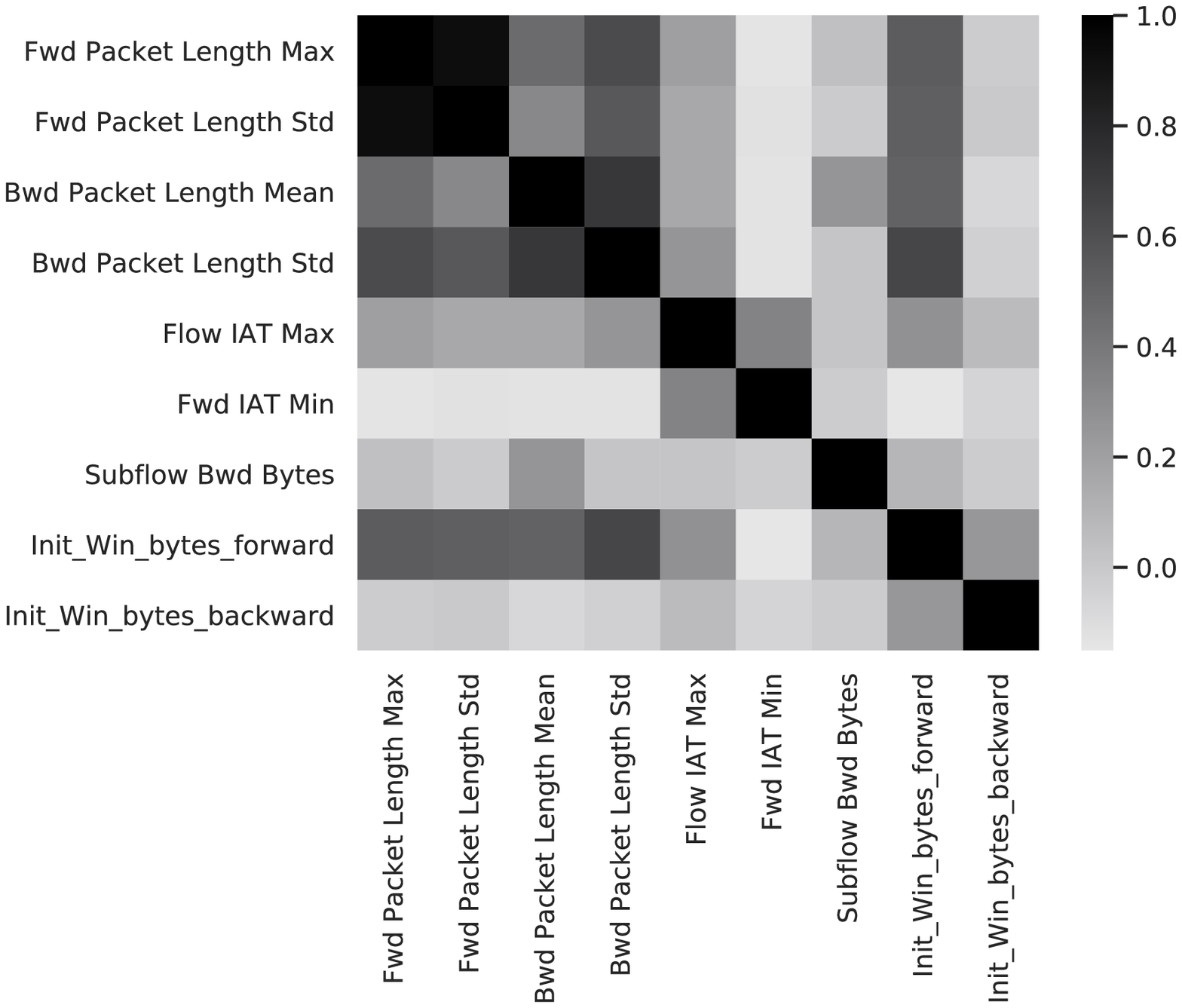}} \hfil
		\subfloat[MO-EA (6 fts)]{\includegraphics[width=2.2in]{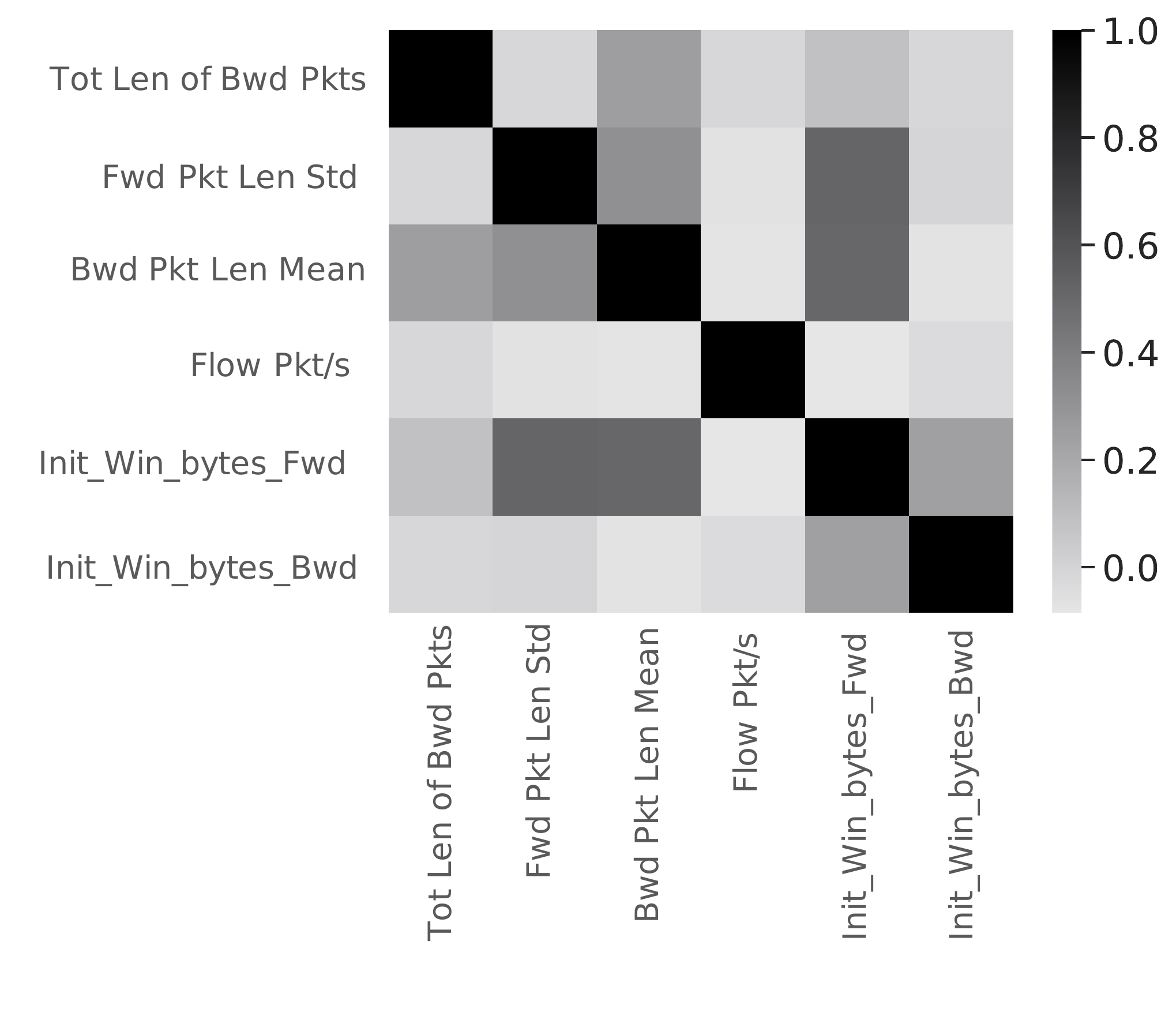}} \hspace{10mm} \\
		\subfloat[Ranking (28 fts)]{\includegraphics[width=2.3in]{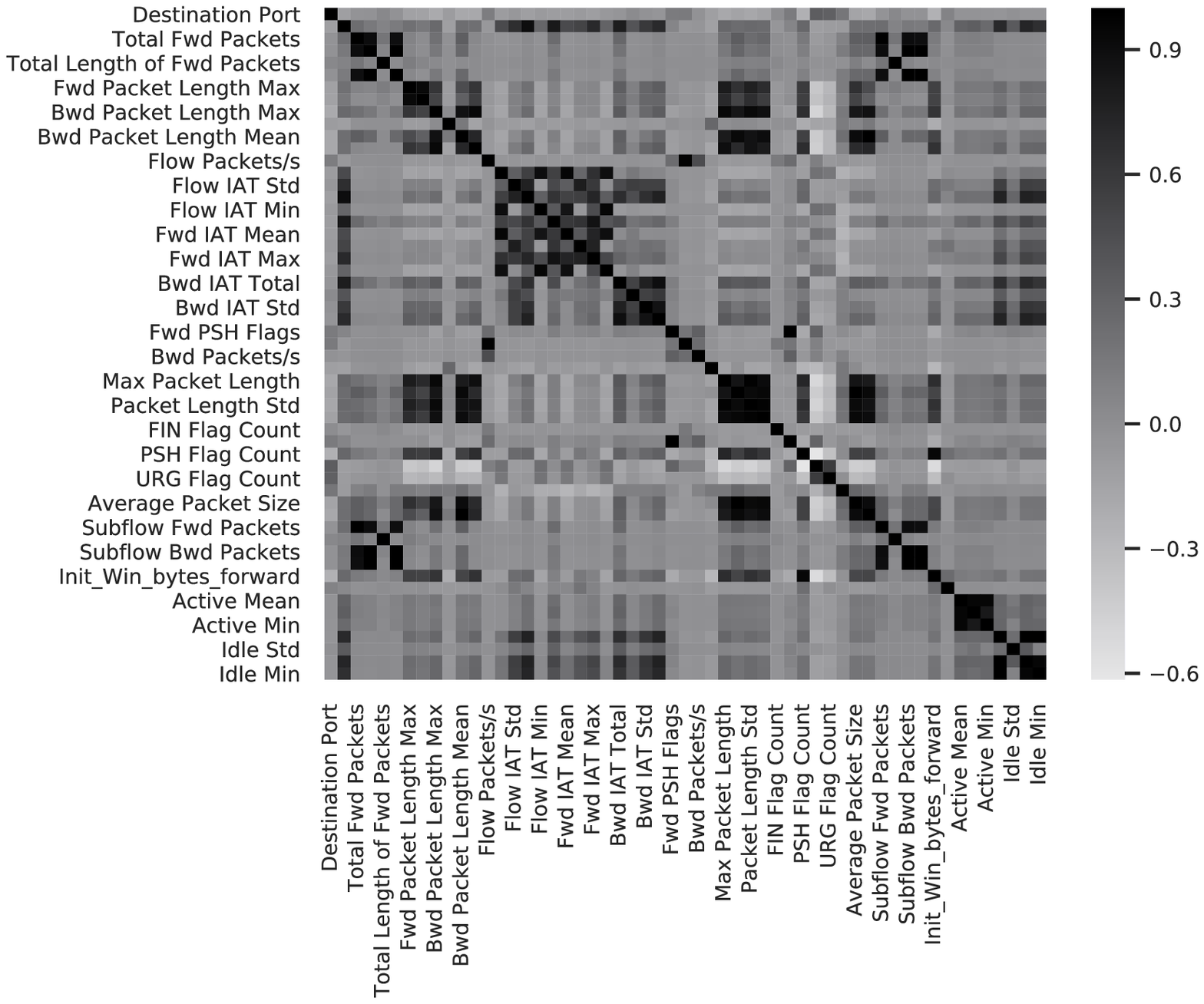}} \hfil 
		\subfloat[Cuckoo (17 fts)]{\includegraphics[width=2.3in]{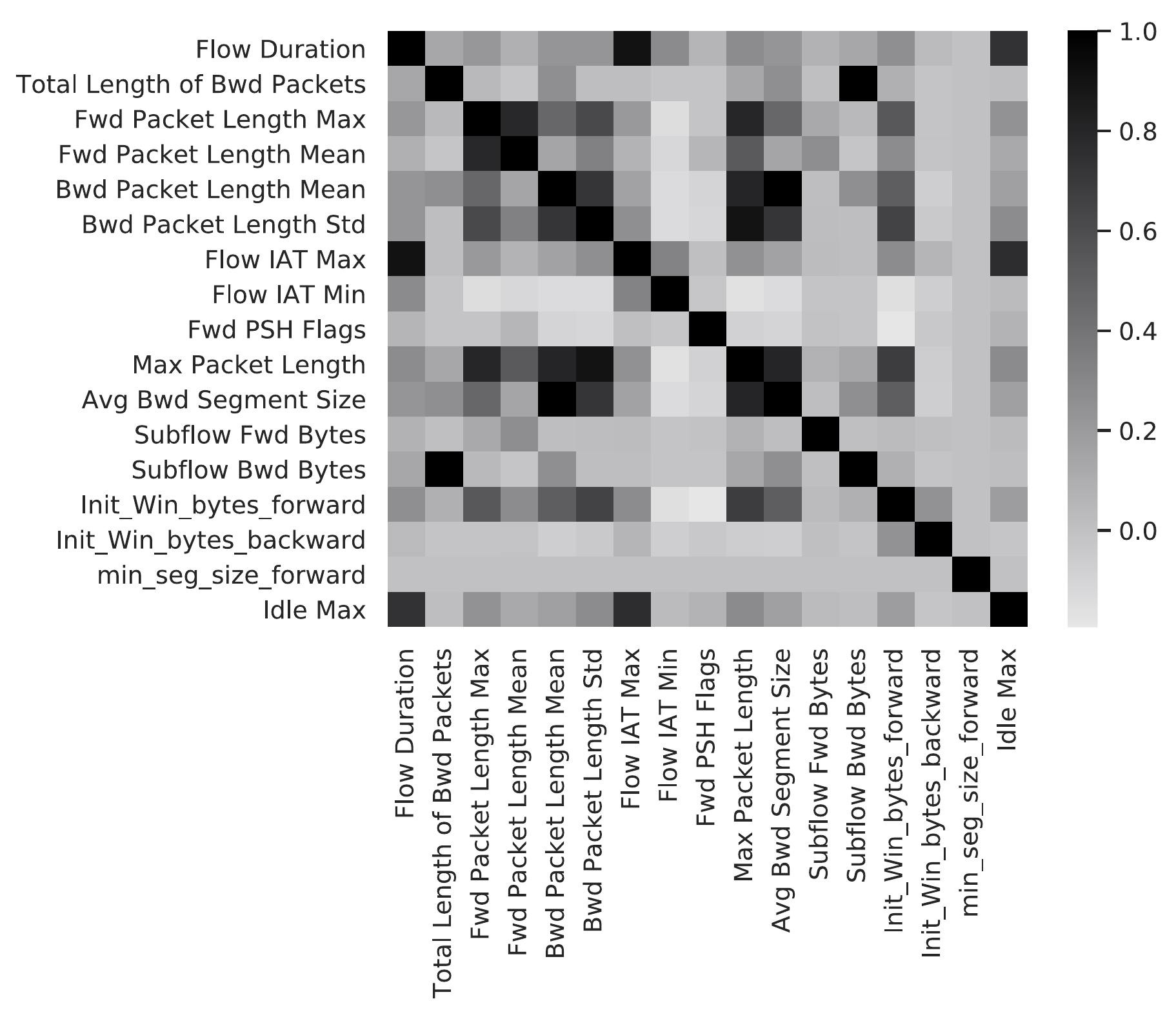}}  \hfil
		\subfloat[LFS (9 fts)]{\includegraphics[width=2.1in]{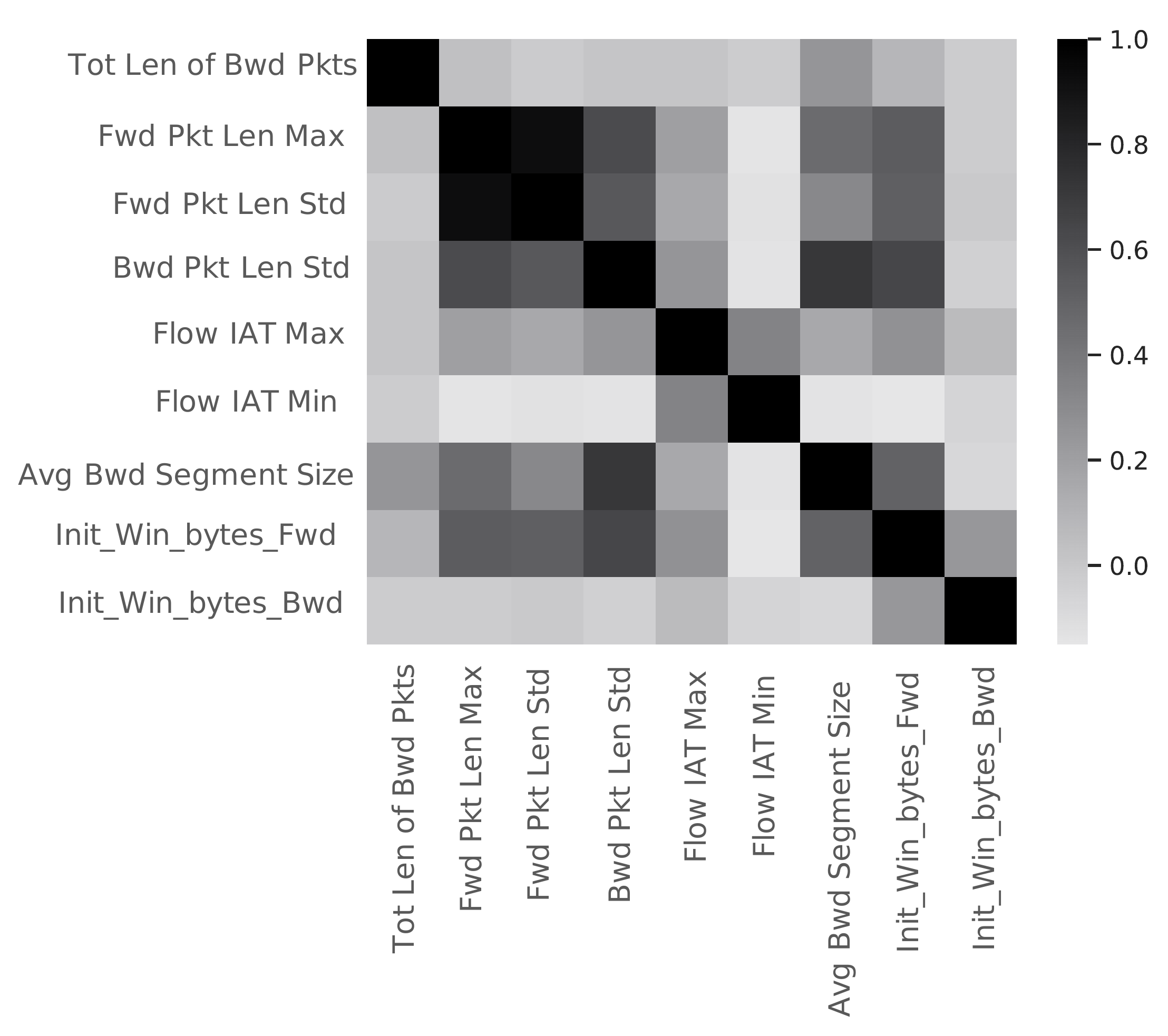}} \hfil \\
		\subfloat[Genetic (31 fts)]{\includegraphics[width=2.6in]{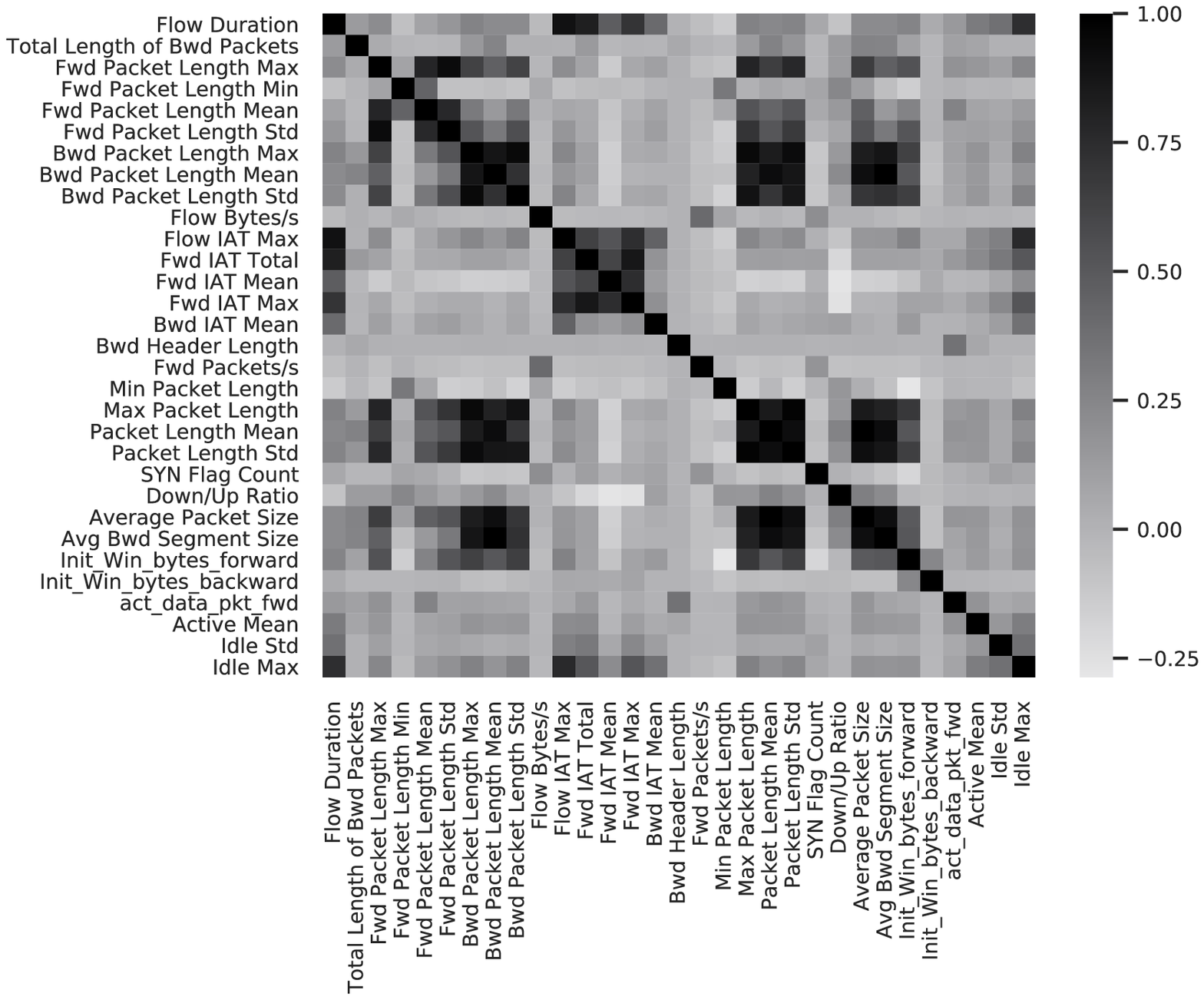}}   \hfil
		\subfloat[PSO (23 fts)]{\includegraphics[width=2.6in]{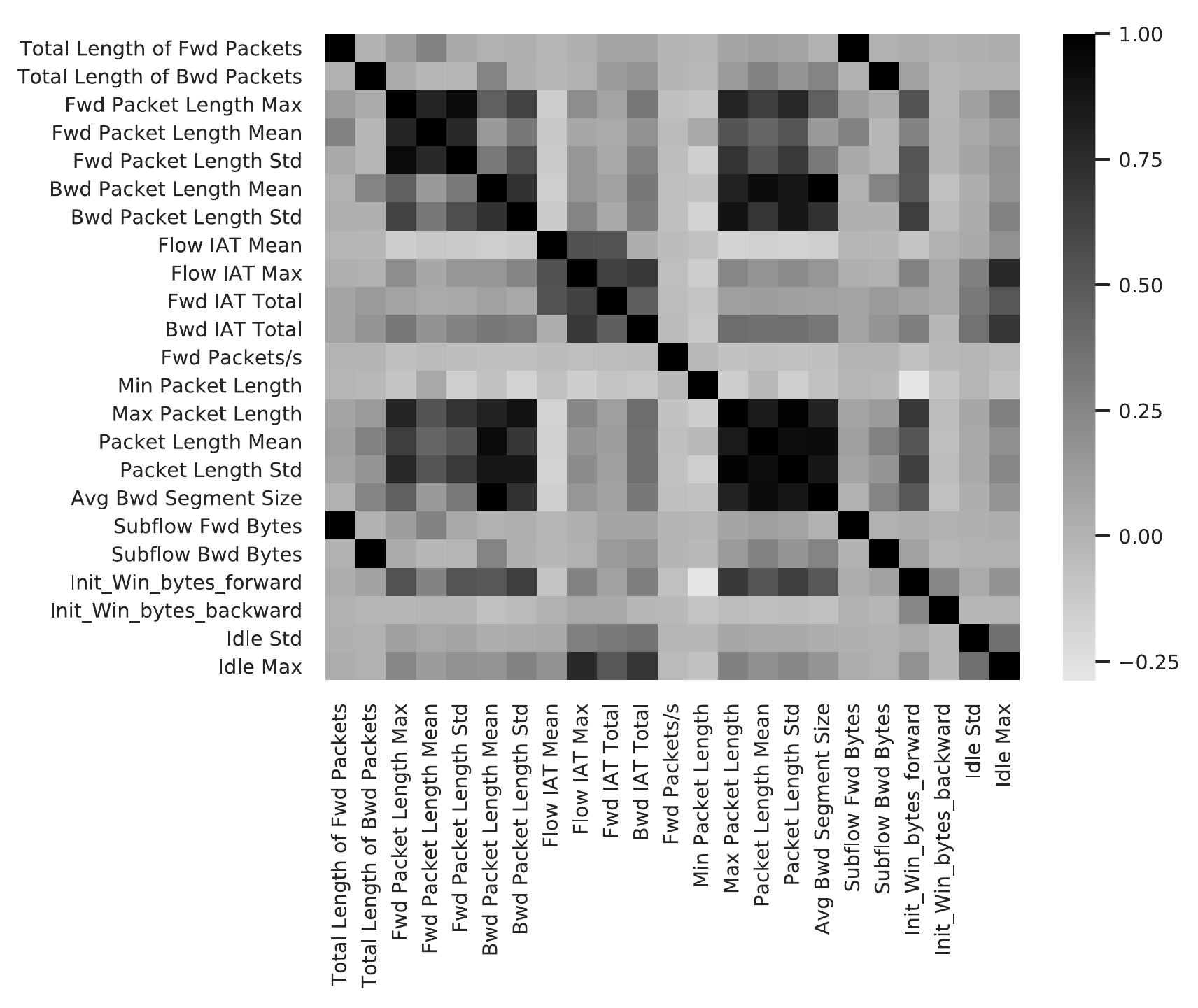}}  \hfil
	\end{tabular}
	\caption{Correlation maps - MultiAndroid dataset. In parenthesis is reported the number of features surviving after the FS process.}
	\label{fig:corrmapsmulti}
\end{figure*}

Let us now analyze some findings obtained from the time-complexity evaluation. To this aim, we use a PC equipped with Intel Core\textsuperscript{TM} i5-7200U CPU@ 2.50GHz CPU and 16 GB of RAM.
In Fig. \ref{fig:times}(a), we show how the FS time varies with training size, for the DDoS dataset. No dramatic differences are observed across the various algorithms, even more significantly as the training size grows. Considering a relatively large training size (with $5\cdot10^4$ training instances), FS times range from about $10$ seconds (Scatter algorithm) to almost $26$ seconds (MO-EA algorithm). Surprisingly, the FS times are rather uniform, in spite of the broad variation in number of retained features (by each of the algorithms). For instance, remaining in the case of $5\cdot10^4$ training instances, Scatter retains the minimum number of features (4), while Genetic retains the maximum number of features (27); yet FS times are comparable ($16.19$ and $10.18$ seconds, respectively). Although it is legitimate to expect that higher FS time could be justified to produce a more reduced feature space, the scarce correlation between such observables is due to the particular logic implemented in each FS algorithm.  

On the other hand, Fig. \ref{fig:times}(b) provides the training times obtained by applying the J48 benchmark algorithm, downstream of the FS processing step. Here, the black line (with empty circles) gives the training times obtained when no FS processing is employed. We can observe how FS leads to significant improvements, in terms of both times and trends. The black (benchmark) line grows rapidly to almost 80 seconds, while most algorithms peak to almost 5 seconds, with the exception of the Genetic algorithm (yellow line) and the Particle Swarm algorithm (light blue line) that take over 10 seconds to complete.     
This indicates that the FS process, on the whole, brings gains in the range of about one order of magnitude, which may become even more significant as the dataset grows.

 Let us now analyze the performance of the proposed FS algorithms in terms of Accuracy and F-Measure. These two metrics, widely used in the field of traffic classification \cite{accuracy1,accuracy2}, are defined as follows:
\begin{itemize}
\item \textbf{Accuracy}: the ratio of the correctly predicted observations to the total observations. This is the most intuitive indicator.
\item \textbf{F-Measure}: the weighted average of precision (ratio of correctly classified flows over all predicted flows in a class) and recall (ratio of correctly classified flows over all ground truth flows in a class). This is an indicator of a per-class performance.
\end{itemize}

\begin{figure*}[pos=ht] 
	\centering
	\begin{tabular}{cc}
		\subfloat[]{\includegraphics[width=3.2in]{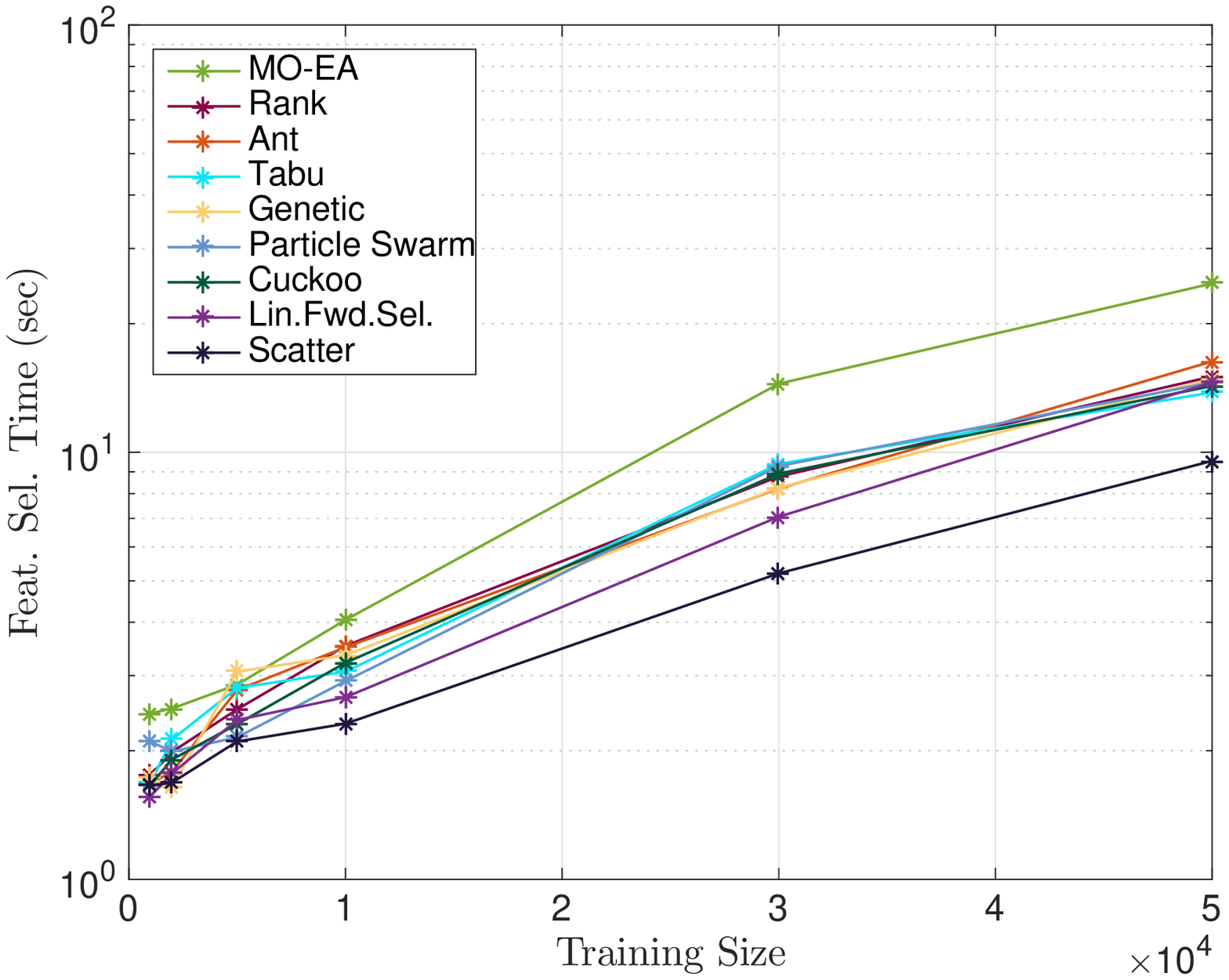}} \hspace{8mm}
		\subfloat[]{\includegraphics[width=3.2in]{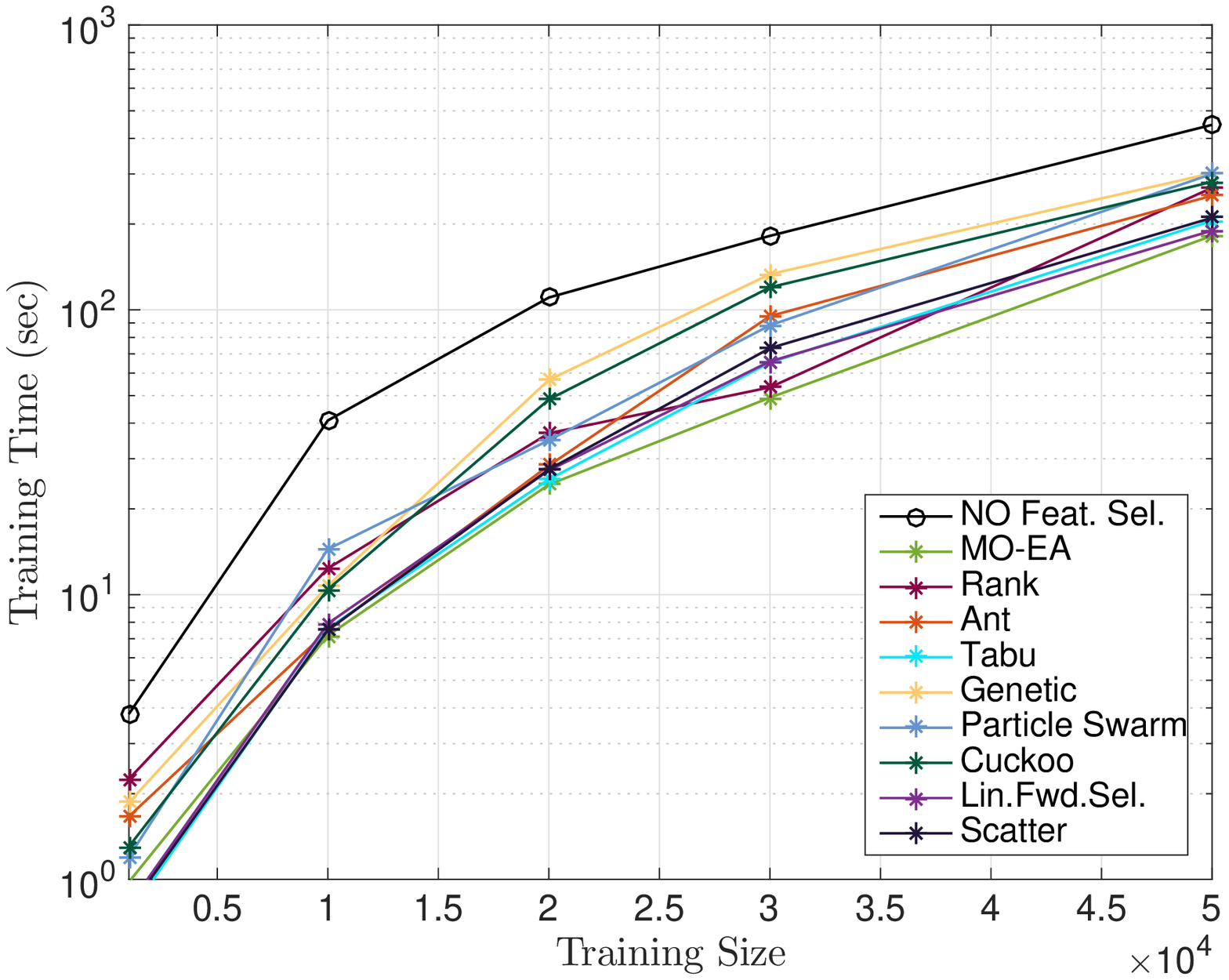}} 
	\end{tabular}
	\caption{FS times - MultiAndroid dataset (a); Training times - MultiAndroid dataset (b).}
	\label{fig:multitimes}
\end{figure*}

To verify that the effectiveness of the FS algorithms is not linked to specific datasets, we have considered the $4$ different datasets introduced in Sect. $5$ (DDoS, Portscan, WebAttack, and TOR), reporting our findings in Fig. \ref{fig:perf_single}.
Just like for the previous experiments, we have used the tree-based J$48$ algorithm as a benchmark. 
We have adopted a $10$-fold cross-validation which is typical in applied ML, and offers a good trade-off between training time and robustness. Noticeably, all FS algorithms perform satisfactorily (both in accuracy and F-measure) in comparison to the benchmark (first bars in all the histograms, labeled as ``NO F.S.'') for the four datasets. 

In some instances the FS algorithms performed even better than the benchmark (e.g., Rank and Genetic algorithms in the WebAttack dataset). This can be explained by a phenomenon that is well-known in ML, whereby models based on too many features may lead to biased classification. On the other hand, when FS manages to retain a sufficiently high number of meaningful features, there is a positive effect on accuracy. This is the case of the Genetic algorithm applied to the TOR dataset (Fig. \ref{fig:perf_single}(d)) that performs better than the other methods.

\subsection{Multi Class Analysis}

Another fruitful analysis is aimed at evaluating FS algorithms when multi-instance datasets are considered. This turns out to be particularly useful when it is not possible to discern different types of data traffic via some pre-processing filter (e.g. IP/Port-based filtering).
To assess this case, we consider two datasets: the \textit{MultiAndroid} dataset, containing benign traffic mixed up with five different types of Android-based threats; and the \textit{DDoS/Portscan} dataset, including a mix of DDoS, Portscan, and benign traffic.
The MultiAndroid dataset, includes the following types of malign traffic: 

\begin{itemize}
\item {\em FakeApp.AL}: a scareware hidden inside a fake {\em Minecraft} application, one of the most popular game applications;
\item {\em Android Defender}: a malware which, once accidentally downloaded and installed, raises some fake alerts;
\item {\em Gooligam}: an insidious malware that has already infected more than $1$ million Android-based devices, aimed at stealing Google accounts for Drive, Docs, Gmail, etc.;
 \item {\em Feiwo}: belonging to the adware family, it acts by showing advertisements in the system notification bar, and by sending device GPS coordinates to a remote server; 
 \item {\em Charger}: a ransomware hidden in some Google Play applications, which gains root privileges and steals contacts before asking for a ransom.
\end{itemize}

Let us analyze how FS algorithms impact on the MultiAndroid dataset in terms of feature correlation referring first to the panels of Fig. \ref{fig:corrmapsmulti}. 
Comparing these results with the ones of Fig. \ref{fig:corrmaps_ddos}, an interesting difference emerges: all FS algorithms retain more features w.r.t. the single-class case. This behavior is coherent with the fact that, to deal with different types of threats (ransomware, adware, malware) we need more features, to be able to capture this higher variability. This effect is even more evident in time-based features (mainly inter-arrival times) and in size-based features (mainly packet lengths). 

Looking at DDoS, we observe a difference between single- and multi-class analysis. In the latter, the destination port is not retained as a crucial feature. This is possibly because malwares exploit different mechanisms to create damage: rather than directly overwhelming a particular target port, they first act in the background (e.g. by stealing privacy data) and then produce malicious traffic in egress. On the other hand, DDoS attacks generate ingress traffic from the infected device. 

It is worth noticing that, when applied to multi-class problems, all algorithms have preserved their original logic. For instance, with 31 surviving features, the Genetic algorithm is still the algorithm that saves more features, thanks to the role played by the mutation operator. Another example is the MO-EA algorithm that, just like in the single-class experiment, retains the smallest number of features ($6$). This is mainly due to the diversity-preservation mechanism, which forces the selection of a representative subset of the whole Pareto front. It optimizes conflicting objective functions, thus few solutions survive.

The time-complexity evaluation is reported in Fig. \ref{fig:multitimes}, which evaluates the usual FS algorithms onto the MultiAndroid dataset. FS times exhibit the same order of magnitude as in single-class analysis (Figs.\ref{fig:times}(a)). For a training size amounting to $5\cdot10^4$ instances, the fastest algorithm is Scatter (FS time amounting to $9.541$ seconds); whereas the slowest one is MO-EA (FS time amounting to $24.827$ seconds). 

The situation changes dramatically when we consider training times for the J$48$ benchmark algorithm (Fig. \ref{fig:multitimes}(b)). Notably, multi-class algorithms are roughly one order of magnitude slower than their single-class counterpart. For instance, let us consider the Genetic algorithm (yellow curve). For a $10^3$ training size, Genetic FS reduces the training time to $1.861$ seconds, growing to the following (X;Y) points: ($10^4$; $10.731$); ($2*10^4$; $56.748$); ($3*10^4$; $133.346$); ($5*10^4$; $301.997$). The longer training times arise from the process of training multiple classes. Nevertheless, significant gains are still obtained by all FS algorithms compared to the ``NO F.S.'' benchmark, which peaks at $446.329$ secs. 

\begin{figure*}[pos=ht] 
	\centering
	\begin{tabular}{cccc}
		\subfloat[]{\includegraphics[scale=0.48]{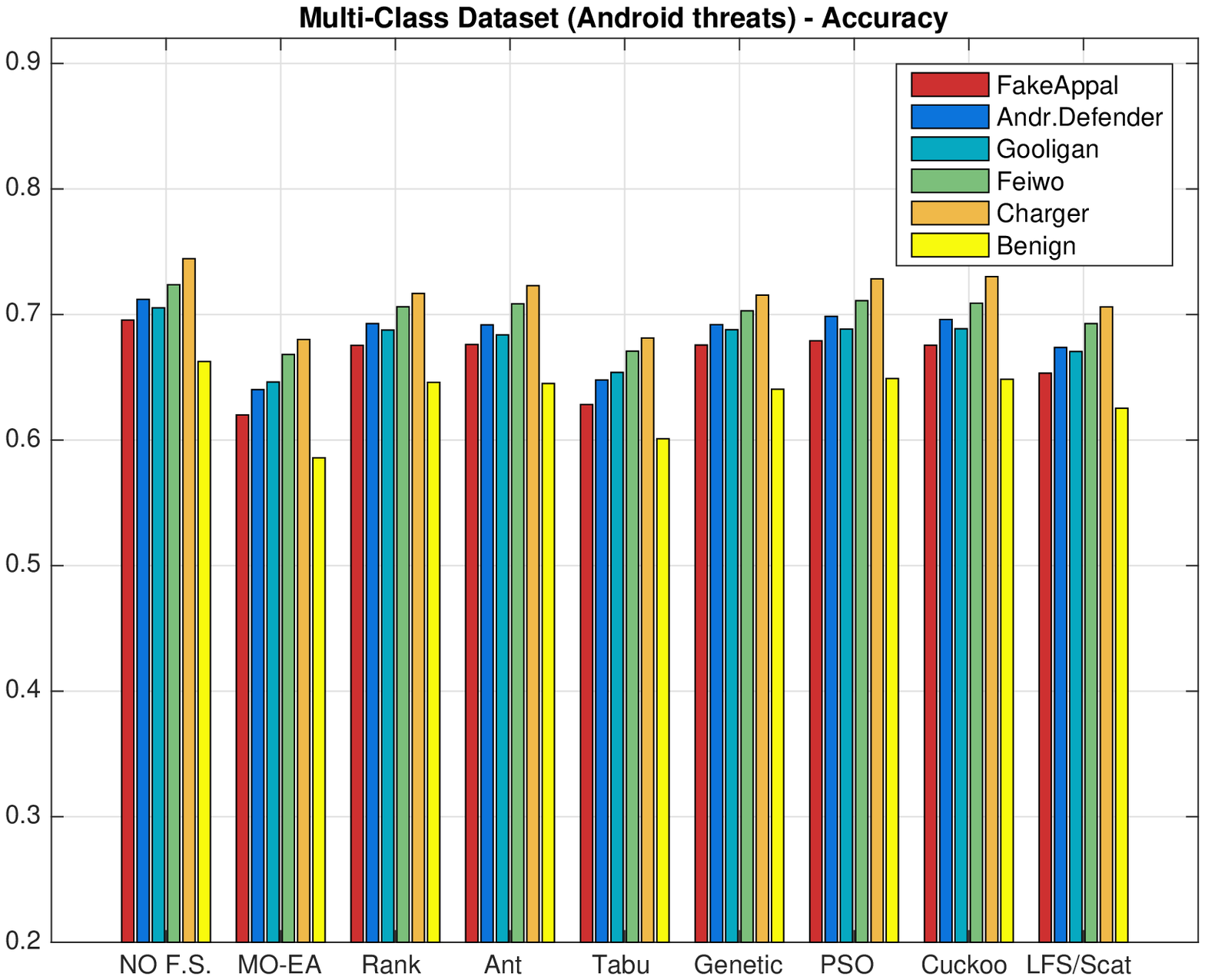}}  \hspace{8mm}
		\subfloat[]{\includegraphics[scale=0.48]{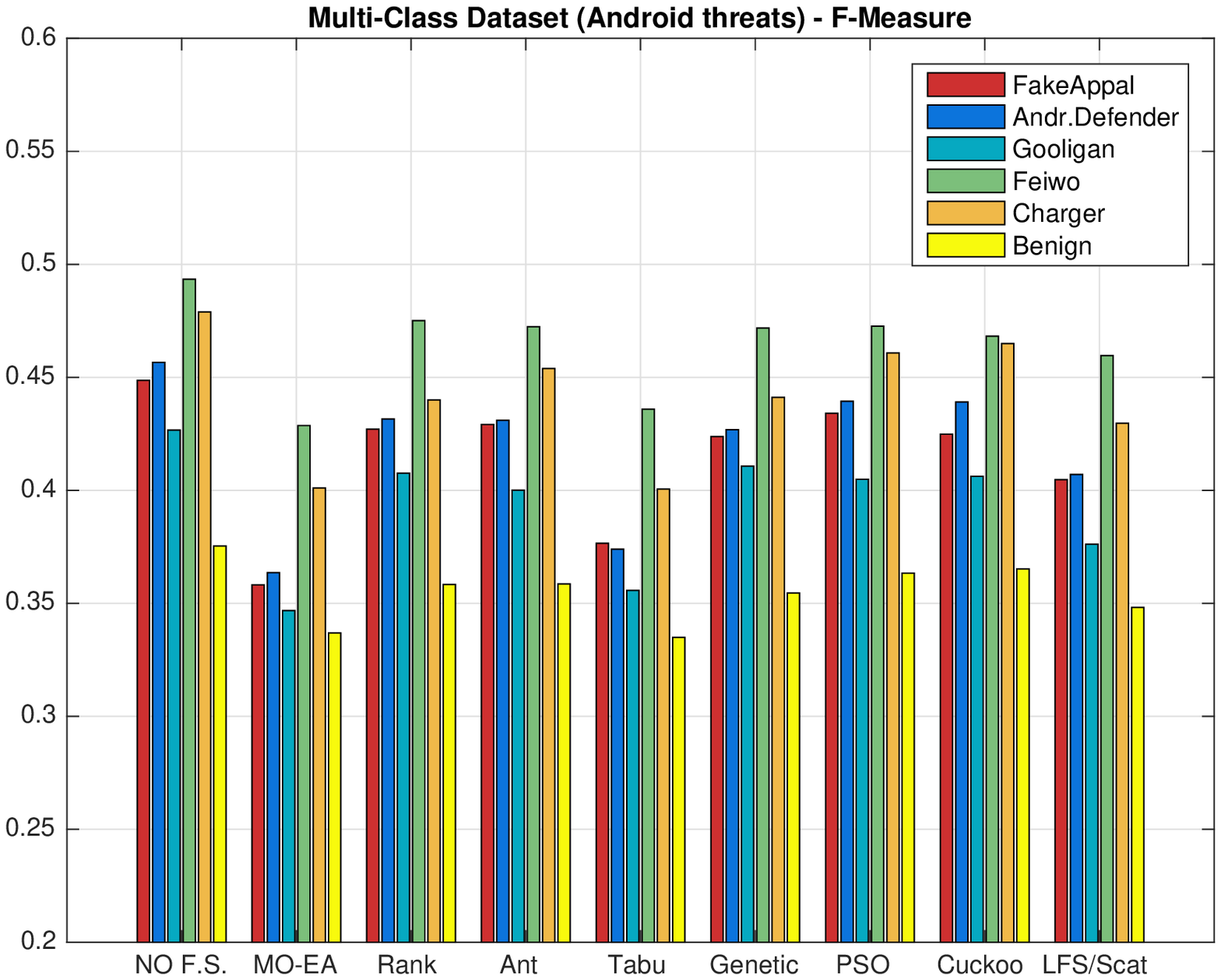}}  \hspace{8mm} \\
		\subfloat[]{\includegraphics[scale=0.48]{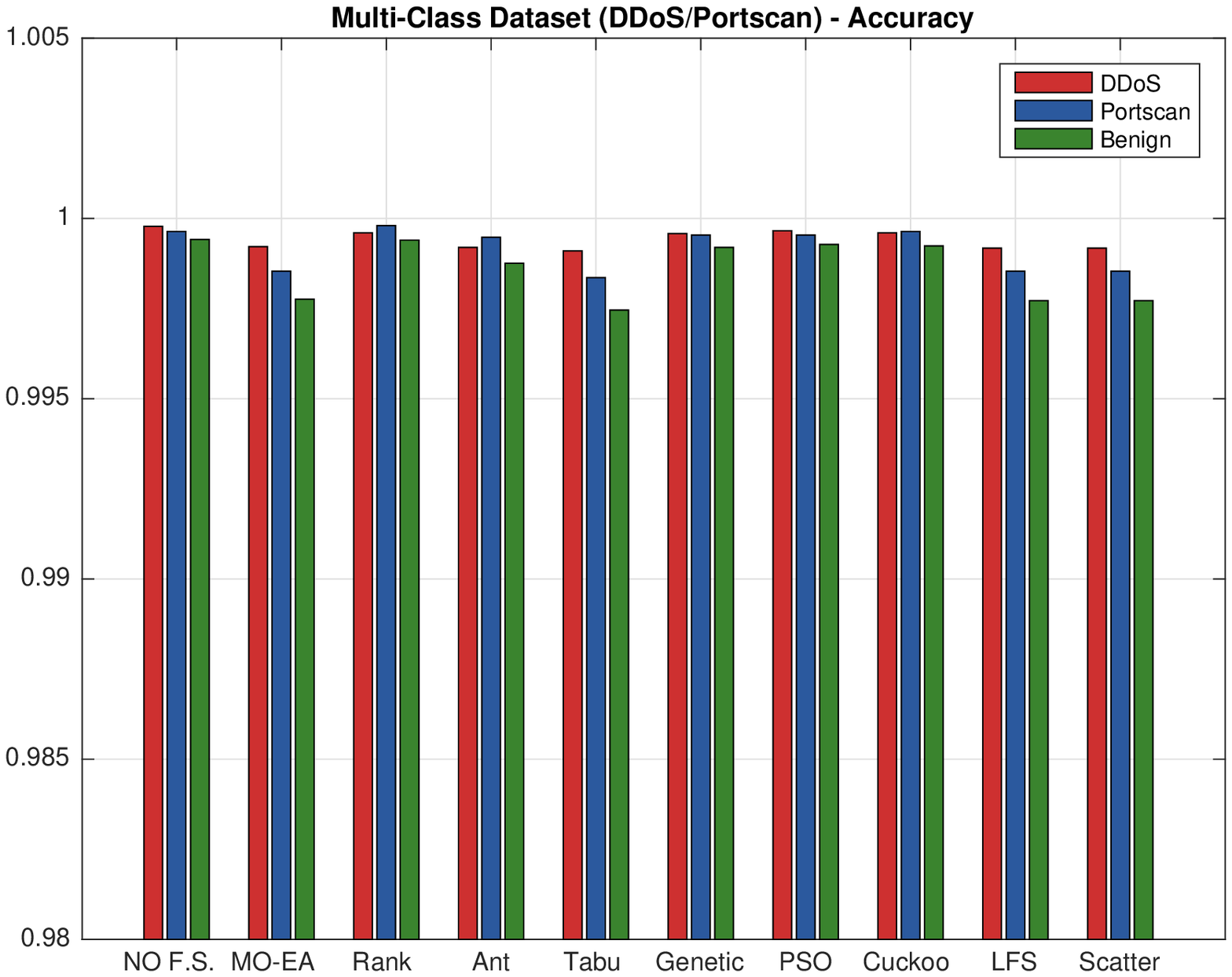}}  \hspace{8mm}
		\subfloat[]{\includegraphics[scale=0.48]{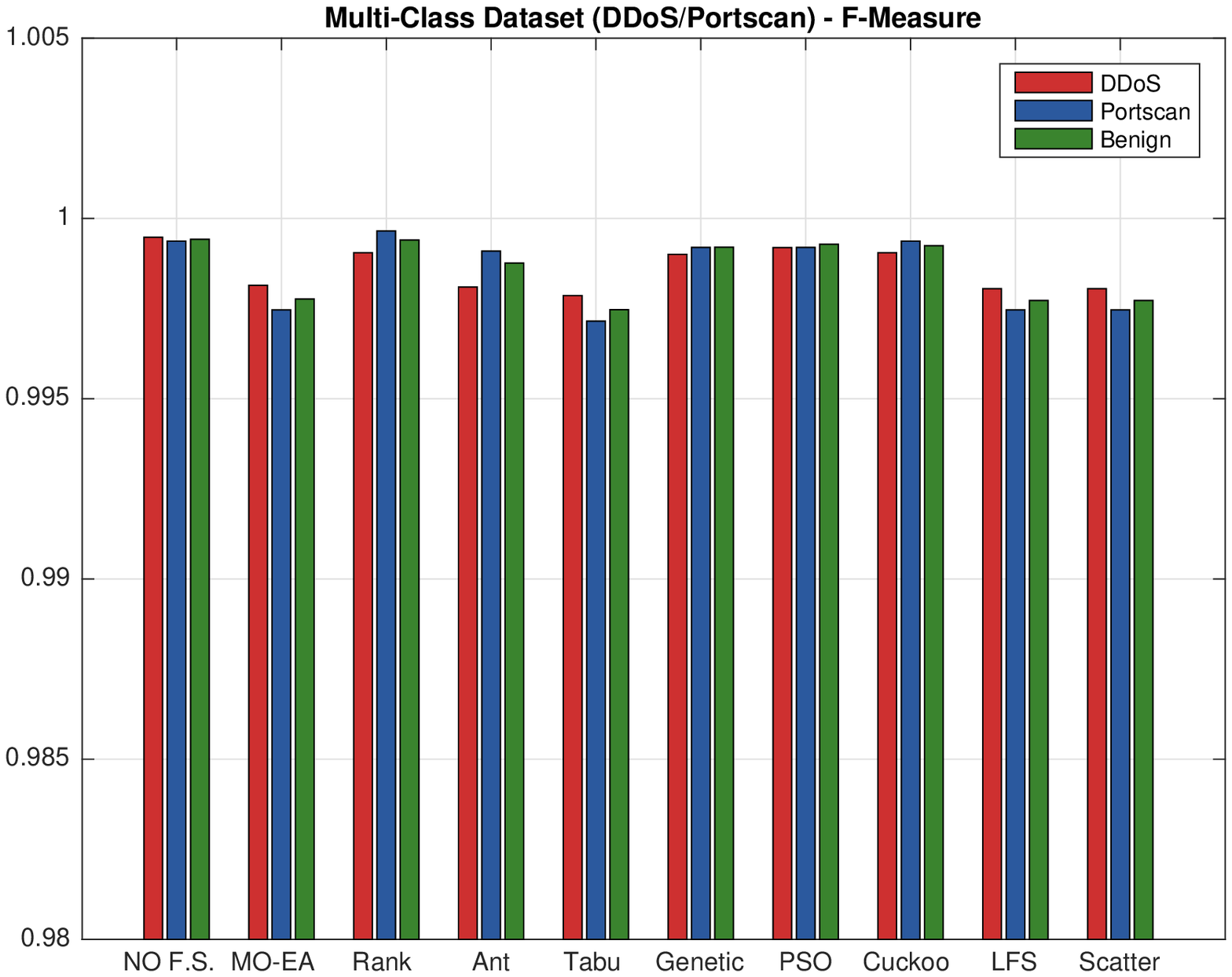}} \hspace{8mm}
	\end{tabular}
	\caption{MultiAndroid dataset: Accuracy (a), F-Measure (b); DDoS/Portscan dataset: Accuracy (c), F-Measure (d);}
	\label{fig:perf_multi}
\end{figure*}

Turning now to the performance analysis, in Fig. \ref{fig:perf_multi} we compare the two multi-class datasets, MultiAndroid and DDoS/Portscan, drawing some interesting considerations. It is comparably more difficult to detect Android threats than DDoS/Portscan attacks - MultiAndroid accuracy is below $0.7$ and F-Measure is below $0.5$. However, this issue is not generated by the FS processes, since the ``NO F.S.'' performance is poor too, particularly with the ``Benign'' class. This issue arises from two facts. First, mobile network attacks are often accompanied by activities that do not directly/immediately generate network anomalies. Examples are ransomware and malware, whereby the anomalies arise after the user has downloaded the malicious application. There is typically a lag between infection and anomalies, as the malicious program initially establishes a secret/silent communication with a remote server, and then gradually steals/sends private user data. Another example is adware, where those annoying banners actually incur very little data, thus making it hard to detect from the regular traffic. A second reason for the poor MultiAndroid performance is the strong similarity among different malign classes (e.g., scareware,  adware, ransomware). 
Similar considerations hold true in the case in which we consider a dataset including Webattack and TOR traffic (not reported for space constraints), whereby the high similarity between the two classes resulted in poor classification performance.
We should however stress that FS algorithms are still very beneficial, since the time-complexity benefits identified are achieved with no dramatic loss in accuracy. 

By contrast, the DDoS/Portscan multi-class case achieves outstanding performance (Figs.\ref{fig:perf_multi}(c) and (d)). This is because these types of attacks are radically distinct in the way they exploit network vulnerabilities: DDoS falls under the umbrella of volumetric attacks; whereas Portscan attacks employ monitoring strategies to unveil possible open ports. 
In other words, a peculiar {\em symptom} of a DDoS attack is the presence of an exceptionally large number of connections coming from different nodes and heading towards one network target’s port. Conversely, a {\em symptom} of Portscan attacks is the presence of just a single node (or a few nodes in case of simultaneous Portscans) opening a considerably large number of connections towards multiple ports of a certain network target.
Thus it is relatively easier to differentiate between these two attacks. 

\subsection{General Remarks}
Overall, we can observe that FS algorithms do lead to an effective reduction in feature space, ranging from $65 \%$ (Single Class, Genetic) to $95 \%$ (Single Class, Scatter) and from $60 \%$ (Multi Class, Genetic) to $92 \%$ (Multi Class, MO-EA). Such feature-space reduction translates into significant computational-time improvements, which become even more remarked as the training size grows. For instance, with a training set of $50k$ samples (single-class DDoS) the MO-EA algorithm takes $24.8$ secs to perform FS, while the training time compared to the benchmark drops from $72.2$ to $5.13$ secs. At the same time, performance is not significantly degraded by the feature reduction process - accuracy drops from $0.9993$ to $0.9971$. Similar considerations hold for all other algorithms. 

The performed assessment provides invaluable guidelines for network/security management practitioners dealing with traffic classification problems. Our evaluation framework aims at weighing the practical benefits of the various FS techniques in terms of time-complexity reduction and performance guarantees. For instance, if we aimed at minimizing the overall processing time (i.e., FS plus training times), the Scatter algorithm would be the best choice. This incurs a total processing time amounting to $14.338$ seconds for the single-class case (FS= $10.178$ secs plus training= $4.16$ secs), and to $219.963$ seconds for multi-class (FS= $9.541$ secs plus training= $210.422$ secs). Conversely, the Genetic method would be preferable to maximize performance. 

\section{Conclusion and Future Direction}
A prominent research direction for network intrusion detection is the adoption of machine learning methods, particularly for the detection of anomalous (and often malicious) network-traffic flows. Looking at the literature, we find ample examples of network classification problems. Yet, little attention has been turned towards feature selection, which is an essential classification pre-processing step. We argue that the main reason for this overlook is that most studies have been based on the obsolete KDD$99$ dataset, which includes few features, thus making FS irrelevant. 
On the other hand, we consider that modern network engines generate much richer features (in fact, hundreds of features), which allow more fine and granular network traffic analyses. However, this extra capability results into impractical ML training times, making it necessary to understand how FS may be realized effectively. 

To this end, herein we have carried out an experimental comparative evaluation of prominent methods, with the view to provide insights as to how the different FS algorithms perform in the peculiar context of network-traffic classification. Our assessment shows how few, relevant features are retained, but also that the FS reduction process is virtually lossless, with a significant acceleration of the overall training process. 

To sum up, the novelties of our work are: 

\textit{i)} we carry out an experimental-based review, considering recent datasets (including DDoS, Portscan, WebAttacks, and Android threats), as opposed to the obsolete KDD$99$ dataset adopted in most literature; 

\textit{ii)} we compare and contrast a representative number of alternative FS algorithm types, including classic rank-guided methods (LFS, Ranking), meta-heuristic methods (Particle Swarm, Tabu, Scatter), nature-inspired methods (Ant, Cuckoo), and evolutionary methods (Genetic, MO-EA); 

\textit{iii)} we provide actual experimental results, unveiling trade-offs between performance (Accuracy/F-Measure) and computational time, at different scales (training set size).  

Ultimately, our analysis shows the benefits linked to embedding the FS process into network analysis, providing a valuable tool for identifying the most useful features out of hundreds of possibilities. This will prove invaluable to the fields of network management, security management, intrusion detection and incident response. We should note that, the purpose of our comparative evaluation was not to claim the predominance of some FS algorithms over others but, rather, to suggest a methodical framework to work with FS. 

As a byproduct of our investigation, some interesting open research directions emerge:
\textit{i)} extending the present analysis to \textit{unsupervised} FS techniques, which would be useful to deal with datasets lacking class labels, or with new types of (unknown) malicious traffic - this is the case of so called {zero-day} attacks that have no prior information; \textit{ii)} considering the case of streamed data analysis, which is necessary when dealing with extremely time-variant streams, whereby the FS process should be repeated across time (e.g. by using a mobile time window), so as to periodically update the resulting dataset with the freshest features; 
\textit{iii)} designing routines to automatically manage the best FS strategies to be applied in accordance to specific criteria (e.g. accuracy target, latency needs, etc.).
Our investigation goes into the direction of the $6$G paradigm that, according to most network scientists, will be characterized by intelligent resource management, smart adjustments, and automatic service provisioning. 


\printcredits

\bibliographystyle{IEEEtran}
\bibliography{fs}

\begin{thebibliography}{100}
\providecommand{\url}[1]{#1}
\csname url@samestyle\endcsname
\providecommand{\newblock}{\relax}
\providecommand{\bibinfo}[2]{#2}
\providecommand{\BIBentrySTDinterwordspacing}{\spaceskip=0pt\relax}
\providecommand{\BIBentryALTinterwordstretchfactor}{4}
\providecommand{\BIBentryALTinterwordspacing}{\spaceskip=\fontdimen2\font plus
\BIBentryALTinterwordstretchfactor\fontdimen3\font minus
  \fontdimen4\font\relax}
\providecommand{\BIBforeignlanguage}[2]{{%
\expandafter\ifx\csname l@#1\endcsname\relax
\typeout{** WARNING: IEEEtran.bst: No hyphenation pattern has been}%
\typeout{** loaded for the language `#1'. Using the pattern for}%
\typeout{** the default language instead.}%
\else
\language=\csname l@#1\endcsname
\fi
#2}}
\providecommand{\BIBdecl}{\relax}
\BIBdecl

\bibitem{Camastra1}
F.~Camastra and A.~Staiano, ``Intrinsic dimension estimation: Advances and open
  problems,'' \emph{Information Sciences}, vol. 328, pp. 26 -- 41, 2016.

\bibitem{Camastra2}
F.~Camastra, ``Data dimensionality estimation methods: a survey,''
  \emph{Pattern Recognition}, vol.~36, no.~12, pp. 2945 -- 2954, 2003.

\bibitem{granville1}
I.~Possebon, A.~Santos~da Silva, L.~Zambenedetti~Granville, A.~Schaeffer-Filho,
  and A.~Marnerides, ``Improved network traffic classification using ensemble
  learning,'' in \emph{2019 IEEE Symposium on Computers and Communications
  (ISCC)}, 2019.

\bibitem{granville2}
F.~Grando, L.~Zambenedetti~Granville, and L.~Lamb, ``Machine learning in
  network centrality measures: Tutorial and outlook,'' \emph{ACM Comput.
  Surv.}, vol.~51, no.~5, pp. 102:1--102:32, 2018.

\bibitem{stadler}
R.~Stadler, R.~Pasquini, and V.~Fodor, ``Learning from network device
  statistics,'' \emph{Journal of Network and Systems Management}, vol.~25,
  no.~4, pp. 672--698, 2017.

\bibitem{boutaba}
A.~A. {Daya}, M.~A. {Salahuddin}, N.~{Limam}, and R.~{Boutaba}, ``A graph-based
  machine learning approach for bot detection,'' in \emph{2019 IFIP/IEEE
  Symposium on Integrated Network and Service Management (IM)}, 2019, pp.
  144--152.

\bibitem{bioinf0}
G.~Li, X.~Hu, X.~Shen, X.~Chen, and Z.~Li, ``A novel unsupervised feature
  selection method for bioinformatics data sets through feature clustering,''
  in \emph{2008 IEEE International Conference on Granular Computing}, 2008, pp.
  41--47.

\bibitem{bioinf1}
C.~Zheng, D.~Huang, L.~Zhang, and X.~Kong, ``Tumor clustering using nonnegative
  matrix factorization with gene selection,'' \emph{IEEE Transactions on
  Information Technology in Biomedicine}, vol.~13, no.~4, pp. 599--607, 2009.

\bibitem{bioinf2}
D.~{Huang} and H.~{Yu}, ``Normalized feature vectors: A novel alignment-free
  sequence comparison method based on the numbers of adjacent amino acids,''
  \emph{IEEE/ACM Transactions on Computational Biology and Bioinformatics},
  vol.~10, no.~2, pp. 457--467, 2013.

\bibitem{bioinf3}
H.~Abusamra, ``A comparative study of feature selection and classification
  methods for gene expression data of glioma,'' \emph{Procedia Computer
  Science}, vol.~23, pp. 5 -- 14, 2013.

\bibitem{image_recog1}
A.~Khotanzad and Y.~Hong, ``Rotation invariant image recognition using features
  selected via a systematic method,'' \emph{Pattern Recognition}, vol.~23,
  no.~10, pp. 1089 -- 1101, 1990.

\bibitem{image_recog2}
J.~Y. Choi, Y.~M. Ro, and K.~N. Plataniotis, ``Boosting color feature selection
  for color face recognition,'' \emph{IEEE Transactions on Image Processing},
  vol.~20, no.~5, pp. 1425--1434, 2011.

\bibitem{image_recog3}
A.~Goltsev and V.~Gritsenko, ``Investigation of efficient features for image
  recognition by neural networks,'' \emph{Neural Networks}, vol.~28, pp. 15 --
  23, 2012.

\bibitem{image_retr1}
D.~L. Swets and J.~J. Weng, ``Using discriminant eigenfeatures for image
  retrieval,'' \emph{IEEE Transactions on Pattern Analysis and Machine
  Intelligence}, vol.~18, no.~8, pp. 831--836, 1996.

\bibitem{image_retr2}
E.~Rashedi, H.~Nezamabadi-pour, and S.~Saryazdi, ``A simultaneous feature
  adaptation and feature selection method for content-based image retrieval
  systems,'' \emph{Knowledge-Based Systems}, vol.~39, pp. 85 -- 94, 2013.

\bibitem{image_recog4}
Y.~Liang, M.~Zhang, and W.~Browne, ``Image feature selection using genetic
  programming for figure-ground segmentation,'' \emph{Engineering Applications
  of Artificial Intelligence}, vol.~62, pp. 96--108, 2017.

\bibitem{image_recog5}
R.~Chatterjee, T.~Maitra, S.~{Hafizul Islam}, M.~{Mehedi Hassan}, A.~Alamri,
  and G.~Fortino, ``A novel machine learning based feature selection for motor
  imagery eeg signal classification in internet of medical things
  environment,'' \emph{Future Generation Computer Systems}, vol.~98, pp.
  419--434, 2019.

\bibitem{fault_diag1}
K.~Zhang, Y.~Li, P.~Scarf, and A.~Ball, ``Feature selection for
  high-dimensional machinery fault diagnosis data using multiple models and
  radial basis function networks,'' \emph{Neurocomputing}, vol.~74, no.~17, pp.
  2941 -- 2952, 2011.

\bibitem{fault_diag2}
T.~W. Rauber, F.~de~Assis~Boldt, and F.~M. Varejao, ``Heterogeneous feature
  models and feature selection applied to bearing fault diagnosis,'' \emph{IEEE
  Transactions on Industrial Electronics}, vol.~62, no.~1, pp. 637--646, 2015.

\bibitem{text_min1}
D.~Lewis, Y.~Yang, T.~Rose, and F.~Li, ``Rcv1: A new benchmark collection for
  text categorization research,'' \emph{Journal of Machine Learning Research},
  vol.~5, pp. 361--397, 2004.

\bibitem{text_min2}
S.~V. Landeghem, T.~Abeel, Y.~Saeys, and Y.~V. de~Peer, ``Discriminative and
  informative features for biomolecular text mining with ensemble feature
  selection,'' \emph{Bioinformatics}, vol.~26, no.~18, pp. 554--560, 2010.

\bibitem{text_min3}
M.~Labani, P.~Moradi, F.~Ahmadizar, and M.~Jalili, ``A novel multivariate
  filter method for feature selection in text classification problems,''
  \emph{Engineering Applications of Artificial Intelligence}, vol.~70, pp.
  25--37, 2018.

\bibitem{nids1}
M.~{Injadat}, A.~{Moubayed}, A.~B. {Nassif}, and A.~{Shami}, ``Multi-stage
  optimized machine learning framework for network intrusion detection,''
  \emph{IEEE Transactions on Network and Service Management}, pp. 1--1, 2020.

\bibitem{nids3}
C.~{Xu}, R.~{Zhang}, M.~{Xie}, and L.~{Yang}, ``Network intrusion detection
  system as a service in openstack cloud,'' in \emph{2020 International
  Conference on Computing, Networking and Communications (ICNC)}, 2020, pp.
  450--455.

\bibitem{nids4}
A.~Shahraki, M.~Abbasi, and O.~Haugen, ``Boosting algorithms for network
  intrusion detection: A comparative evaluation of real adaboost, gentle
  adaboost and modest adaboost,'' \emph{Engineering Applications of Artificial
  Intelligence}, vol.~94, 2020.

\bibitem{ava_1}
M.~{Di Mauro}, M.~{Longo}, F.~{Postiglione}, G.~{Carullo}, and M.~{Tambasco},
  ``Service function chaining deployed in an {NFV} environment: An availability
  modeling,'' in \emph{2017 IEEE Conference on Standards for Communications and
  Networking (CSCN)}, 2017, pp. 42--47.

\bibitem{ava_2}
M.~{Di Mauro}, M.~{Longo}, F.~{Postiglione}, and M.~{Tambasco}, ``Availability
  modeling and evaluation of a network service deployed via {NFV},'' in
  \emph{Digital Communication. Towards a Smart and Secure Future Internet},
  2017, pp. 31--44.

\bibitem{ava_3}
M.~{Di Mauro}, M.~{Longo}, and F.~{Postiglione}, ``Availability evaluation of
  multi-tenant service function chaining infrastructures by multidimensional
  universal generating function,'' \emph{IEEE Transactions on Services
  Computing}, pp. 1--1, 2018.

\bibitem{ava_4}
M.~{Di Mauro}, M.~{Longo}, F.~{Postiglione}, R.~{Restaino}, and M.~{Tambasco},
  ``Availability evaluation of the virtualized infrastructure manager in
  network function virtualization environments,'' in \emph{Proc. of the 26th
  European Safety and Reliability Conference, ESREL 2016}, 2017, pp.
  2591--2596.

\bibitem{esrel15}
M.~{Di Mauro}, M.~{Longo}, and F.~{Postiglione}, ``Reliability analysis of the
  controller architecture in software defined networks,'' in \emph{Proc. of the
  26th European Safety and Reliability Conference, ESREL 2015}, 2015, pp.
  1503--1510.

\bibitem{esrel17}
M.~{Di Mauro}, G.~{Galatro}, M.~{Longo}, F.~{Postiglione}, and M.~{Tambasco},
  ``Availability evaluation of a virtualized {IP} {M}ultimedia {S}ubsystem for
  5{G} network architectures,'' in \emph{Proc. of the 26th European Safety and
  Reliability Conference, ESREL 2017}, 2017, pp. 2203--2210.

\bibitem{tnsmava}
------, ``Comparative performability assessment of {SFC}s: The case of
  containerized {IP} {M}ultimedia {S}ubsystem,'' \emph{IEEE Trans. Netw.
  Service Manag.}, 2020.

\bibitem{nomsmdm1}
------, ``Performability management of softwarized {IP} {M}ultimedia
  {S}ubsystem,'' in \emph{IEEE/IFIP Network Operations and Management
  Symposium, 2020}, 2020, pp. 1--6.

\bibitem{nomsmdm2}
M.~{Di Mauro}, G.~{Galatro}, M.~{Longo}, A.~{Palma}, F.~{Postiglione}, and
  M.~{Tambasco}, ``Automated generation of availability models for {SFC}s: The
  case of virtualized {IP} {M}ultimedia {S}ubsystem,'' in \emph{IEEE/IFIP
  Network Operations and Management Symposium, 2020}, 2020, pp. 1--6.

\bibitem{ddos1}
V.~{Matta}, M.~{Di Mauro}, and M.~{Longo}, ``Botnet identification in
  randomized {DD}o{S} attacks,'' in \emph{Proceedings of the 24th European
  Signal Processing Conference}, 2016, pp. 2260--2264.

\bibitem{ddos2}
------, ``Botnet identification in multi-clustered {DD}o{S} attacks,'' in
  \emph{2017 25th European Signal Processing Conference (EUSIPCO)}, 2017, pp.
  2171--2175.

\bibitem{dimvoip}
P.~{Addesso}, M.~{Cirillo}, M.~{Di Mauro}, and V.~{Matta}, ``{ADV}o{IP}:
  Adversarial detection of encrypted and concealed voip,'' \emph{IEEE
  Transactions on Information Forensics and Security}, vol.~15, pp. 943--958,
  2020.

\bibitem{dimkend}
V.~{Matta}, M.~{Di Mauro}, M.~{Longo}, and A.~{Farina}, ``Cyber-threat
  mitigation exploiting the birth–death–immigration model,'' \emph{IEEE
  Transactions on Information Forensics and Security}, vol.~13, no.~12, pp.
  3137--3152, 2018.

\bibitem{cerroni1}
W.~Cerroni, G.~Moro, R.~Pasolini, and M.~Ramilli, ``Decentralized detection of
  network attacks through p2p data clustering of snmp data,'' \emph{Computers
  \& Security}, vol.~52, pp. 1 -- 16, 2015.

\bibitem{cerroni2}
------, ``Network attack detection based on peer-to-peer clustering of snmp
  data,'' in \emph{Lecture Notes of the Institute for Computer Sciences},
  vol.~22, 2009.

\bibitem{cerroni3}
W.~Cerroni, G.~Moro, T.~Pirini, and M.~Ramilli, ``Peer-to-peer data mining
  classifiers for decentralized detection of network attacks,'' in
  \emph{Proceedings of the Twenty-Fourth Australasian Database Conference -
  Volume 137}, 2013, pp. 101--107.

\bibitem{cicflowmeter}
``The {CSE-CIC-IDS2018} {D}ataset,''
  \url{https://github.com/alekzandr/flowmeter}, accessed: 2020-10-01.

\bibitem{survey_general_1}
G.~Chandrashekar and F.~Sahin, ``A survey on feature selection methods,''
  \emph{Computers \& Electrical Engineering}, vol.~40, no.~1, pp. 16 -- 28,
  2014.

\bibitem{survey_general_2}
S.~{Khalid}, T.~{Khalil}, and S.~{Nasreen}, ``A survey of feature selection and
  feature extraction techniques in machine learning,'' in \emph{2014 Science
  and Information Conference}, 2014, pp. 372--378.

\bibitem{survey_general_3}
L.~C. {Molina}, L.~{Belanche}, and A.~{Nebot}, ``Feature selection algorithms:
  a survey and experimental evaluation,'' in \emph{2002 IEEE International
  Conference on Data Mining, 2002. Proceedings.}, 2002, pp. 306--313.

\bibitem{relred}
L.~Yu and H.~Liu, ``Efficient feature selection via analysis of relevance and
  redundancy,'' \emph{The Journal of Machine Learning Research}, vol.~5, pp.
  1205--1224, 2004.

\bibitem{Camastra2008}
F.~Camastra and A.~Vinciarelli, \emph{Feature Extraction Methods and Manifold
  Learning Methods}.\hskip 1em plus 0.5em minus 0.4em\relax Springer London,
  2008.

\bibitem{blum}
A.~L. Blum and P.~Langley, ``Selection of relevant features and examples in
  machine learning,'' \emph{Artificial Intelligence}, vol.~97, no.~1, pp. 245
  -- 271, 1997.

\bibitem{dash}
M.~Dash and H.~Liu, ``Feature selection for classification,'' \emph{Intelligent
  Data Analysis}, vol.~1, no.~1, pp. 131 -- 156, 1997.

\bibitem{sel_vs_extr}
S.~Alelyani, J.~Tang, and H.~Liu, ``Feature selection for clustering: A
  review,'' in \emph{Data Clustering: Algorithms and Applications}, 2013.

\bibitem{sel_vs_extr2}
L.~Rendell and R.~Seshu, ``Learning hard concepts through constructive
  induction: Framework and rationale,'' in \emph{Proceedings of a Workshop on
  Computational Learning Theory and Natural Learning Systems (Vol. 1) :
  Constraints and Prospects: Constraints and Prospects}, 1994, pp. 83--141.

\bibitem{filters_wrappers}
L.~Talavera, ``An evaluation of filter and wrapper methods for feature
  selection in categorical clustering,'' in \emph{Advances in Intelligent Data
  Analysis VI}, 2005, pp. 440--451.

\bibitem{filters_wrappers_2}
M.~A. Hall and L.~A. Smith, ``Feature selection for machine learning: Comparing
  a correlation-based filter approach to the wrapper,'' in \emph{Proceedings of
  the Twelfth International Florida Artificial Intelligence Research Society
  Conference}, 1999, pp. 235--239.

\bibitem{filters_wrappers_3}
R.~Kohavi and G.~H. John, ``Wrappers for feature subset selection,''
  \emph{Artificial Intelligence}, vol.~97, no.~1, pp. 273 -- 324, 1997.

\bibitem{filters_wrappers_4}
I.~Guyon, ``An introduction to variable and feature selection,'' \emph{Journal
  of Machine Learning Research}, vol.~3, pp. 1157--1182, 2003.

\bibitem{taxonomy}
J.~C. {Ang}, A.~{Mirzal}, H.~{Haron}, and H.~N.~A. {Hamed}, ``Supervised,
  unsupervised, and semi-supervised feature selection: A review on gene
  selection,'' \emph{IEEE/ACM Transactions on Computational Biology and
  Bioinformatics}, vol.~13, no.~5, pp. 971--989, 2016.

\bibitem{tnsm1}
J.~{Dromard}, G.~{Roudière}, and P.~{Owezarski}, ``Online and scalable
  unsupervised network anomaly detection method,'' \emph{IEEE Trans. Netw.
  Service Manag.}, vol.~14, no.~1, pp. 34--47, 2017.

\bibitem{table_wang}
W.~{Wang}, Y.~{He}, J.~{Liu}, and S.~{Gombault}, ``Constructing important
  features from massive network traffic for lightweight intrusion detection,''
  \emph{IET Information Security}, vol.~9, no.~6, pp. 374--379, 2015.

\bibitem{table_jana}
T.~{Janarthanan} and S.~{Zargari}, ``Feature selection in unsw-nb15 and
  kddcup'99 datasets,'' in \emph{2017 IEEE 26th International Symposium on
  Industrial Electronics (ISIE)}, 2017, pp. 1881--1886.

\bibitem{table_khatib}
K.~{El-Khatib}, ``Impact of feature reduction on the efficiency of wireless
  intrusion detection systems,'' \emph{IEEE Transactions on Parallel and
  Distributed Systems}, vol.~21, no.~8, pp. 1143--1149, 2010.

\bibitem{table_chen}
Y.~Chen, Y.~Li, X.~Cheng, and L.~Guo, ``Survey and taxonomy of feature
  selection algorithms in intrusion detection system,'' in \emph{Proceedings of
  the Second SKLOIS Conference on Information Security and Cryptology}, 2006,
  pp. 153--167.

\bibitem{table_nisioti}
A.~{Nisioti}, A.~{Mylonas}, P.~D. {Yoo}, and V.~{Katos}, ``From intrusion
  detection to attacker attribution: A comprehensive survey of unsupervised
  methods,'' \emph{IEEE Communications Surveys Tutorials}, vol.~20, no.~4, pp.
  3369--3388, 2018.

\bibitem{table_iglesias}
F.~Iglesias and T.~Zseby, ``Analysis of network traffic features for anomaly
  detection,'' \emph{Machine Learning}, vol. 101, no.~1, pp. 59--84, 2015.

\bibitem{table_singh}
R.~{Singh}, H.~{Kumar}, and R.~K. {Singla}, ``Analysis of feature selection
  techniques for network traffic dataset,'' in \emph{2013 International
  Conference on Machine Intelligence and Research Advancement}, 2013, pp.
  42--46.

\bibitem{table_bahrololum}
M.~{Bahrololum}, E.~{Salahi}, and M.~{Khaleghi}, ``Machine learning techniques
  for feature reduction in intrusion detection systems: A comparison,'' in
  \emph{Progress in Computing, Analytics and Networking. Advances in
  Intelligent Systems and Computing}, 2009, pp. 1091--1095.

\bibitem{table_dhote}
Y.~{Dhote}, S.~{Agrawal}, and A.~J. {Deen}, ``A survey on feature selection
  techniques for internet traffic classification,'' in \emph{2015 International
  Conference on Computational Intelligence and Communication Networks (CICN)},
  2015, pp. 1375--1380.

\bibitem{classif_buczak}
A.~L. {Buczak} and E.~{Guven}, ``A survey of data mining and machine learning
  methods for cyber security intrusion detection,'' \emph{IEEE Communications
  Surveys Tutorials}, vol.~18, no.~2, pp. 1153--1176, 2016.

\bibitem{classif_mishra}
P.~{Mishra}, V.~{Varadharajan}, U.~{Tupakula}, and E.~S. {Pilli}, ``A detailed
  investigation and analysis of using machine learning techniques for intrusion
  detection,'' \emph{IEEE Communications Surveys Tutorials}, vol.~21, no.~1,
  pp. 686--728, 2019.

\bibitem{classif_dimauro}
M.~{Di Mauro} and C.~{Di Sarno}, ``Improving {SIEM} capabilities through an
  enhanced probe for encrypted skype traffic detection,'' \emph{Journal of
  Information Security and Applications}, vol.~38, pp. 85--95, 2018.

\bibitem{classif_dimauro2}
M.~{Di Mauro} and M.~{Longo}, ``Skype traffic detection: A decision theory
  based tool,'' in \emph{2014 International Carnahan Conference on Security
  Technology (ICCST)}, 2014, pp. 1--6.

\bibitem{classif_dimauro3}
M.~{Di Mauro} and C.~{Di Sarno}, ``A framework for internet data real-time
  processing: A machine-learning approach,'' in \emph{2014 International
  Carnahan Conference on Security Technology (ICCST)}, 2014, pp. 1--6.

\bibitem{classif_liotta}
F.~Cauteruccio, G.~Fortino, A.~Guerrieri, A.~Liotta, D.~Mocanu, C.~Perra,
  G.~Terracina, and M.~Torres~Vega, ``Short-long term anomaly detection in
  wireless sensor networks based on machine learning and multi-parameterized
  edit distance,'' \emph{Information Fusion}, vol.~52, pp. 13 -- 30, 2019.

\bibitem{classif_nn}
M.~{Di Mauro}, G.~{Galatro}, and A.~{Liotta}, ``Experimental review of
  neural-based approaches for network intrusion management,'' \emph{IEEE
  Transactions on Network and Service Management}, pp. 1--1, 2020.

\bibitem{classif_dimauro4}
M.~{Di Mauro} and M.~{Longo}, ``Revealing encrypted webrtc traffic via machine
  learning tools,'' in \emph{2015 12th International Joint Conference on
  e-Business and Telecommunications (ICETE)}, vol.~04, 2015, pp. 259--266.

\bibitem{ids_18_1}
H.~Benaddi, K.~Ibrahimi, and A.~Benslimane, ``Improving the intrusion detection
  system for nsl-kdd dataset based on pca-fuzzy clustering-knn,'' in \emph{2018
  6th International Conference on Wireless Networks and Mobile Communications
  (WINCOM)}, 2018, pp. 1--6.

\bibitem{ids_18_2}
W.~{Wang}, X.~{Du}, and N.~{Wang}, ``Building a cloud ids using an efficient
  feature selection method and svm,'' \emph{IEEE Access}, vol.~7, pp.
  1345--1354, 2019.

\bibitem{ids_19}
S.~M. Kasongo and Y.~Sun, ``A deep learning method with filter based feature
  engineering for wireless intrusion detection system,'' \emph{IEEE Access},
  vol.~7, pp. 38\,597--38\,607, 2019.

\bibitem{ids_18_3}
K.~Wu, Z.~Chen, and W.~Li, ``A novel intrusion detection model for a massive
  network using convolutional neural networks,'' \emph{IEEE Access}, vol.~6,
  pp. 50\,850--50\,859, 2018.

\bibitem{ids_16}
M.~A. {Ambusaidi}, X.~{He}, P.~{Nanda}, and Z.~{Tan}, ``Building an intrusion
  detection system using a filter-based feature selection algorithm,''
  \emph{IEEE Transactions on Computers}, vol.~65, no.~10, pp. 2986--2998, 2016.

\bibitem{ids_ann_19}
K.~A. {Taher}, B.~{Mohammed Yasin Jisan}, and M.~M. {Rahman}, ``Network
  intrusion detection using supervised machine learning technique with feature
  selection,'' in \emph{2019 International Conference on Robotics,Electrical
  and Signal Processing Techniques (ICREST)}, 2019, pp. 643--646.

\bibitem{ids_dnn_19}
J.~{Woo}, J.~{Song}, and Y.~{Choi}, ``Performance enhancement of deep neural
  network using feature selection and preprocessing for intrusion detection,''
  in \emph{2019 International Conference on Artificial Intelligence in
  Information and Communication (ICAIIC)}, 2019, pp. 415--417.

\bibitem{ids_2011}
F.~Amiri, M.~R. Yousefi, C.~Lucas, A.~Shakery, and N.~Yazdani, ``Mutual
  information-based feature selection for intrusion detection systems,''
  \emph{Journal of Network and Computer Applications}, vol.~34, no.~4, pp. 1184
  -- 1199, 2011.

\bibitem{nsl-kdd}
M.~{Tavallaee}, E.~{Bagheri}, W.~{Lu}, and A.~A. {Ghorbani}, ``A detailed
  analysis of the kdd cup 99 data set,'' in \emph{2009 IEEE Symposium on
  Computational Intelligence for Security and Defense Applications}, 2009, pp.
  1--6.

\bibitem{unsw1}
N.~{Moustafa} and J.~{Slay}, ``Unsw-nb15: a comprehensive data set for network
  intrusion detection systems (unsw-nb15 network data set),'' in \emph{2015
  Military Communications and Information Systems Conference (MilCIS)}, 2015,
  pp. 1--6.

\bibitem{unsw2}
N.~{Moustafa}, J.~{Slay}, and G.~{Creech}, ``Novel geometric area analysis
  technique for anomaly detection using trapezoidal area estimation on
  large-scale networks,'' \emph{IEEE Transactions on Big Data}, vol.~5, no.~4,
  pp. 481--494, 2019.

\bibitem{unsw3}
{Doreswamy}, M.~K. {Hooshmand}, and I.~{Gad}, ``Feature selection approach
  using ensemble learning for network anomaly detection,'' \emph{CAAI
  Transactions on Intelligence Technology}, vol.~5, no.~4, pp. 283--293, 2020.

\bibitem{unsw4}
A.~{Binbusayyis} and T.~{Vaiyapuri}, ``Identifying and benchmarking key
  features for cyber intrusion detection: An ensemble approach,'' \emph{IEEE
  Access}, vol.~7, pp. 106\,495--106\,513, 2019.

\bibitem{cic}
``\color{black} canadian institute for cybersecurity,''
  \url{https://www.unb.ca/cic/}, accessed: 2020-10-01.

\bibitem{rank}
M.~A. {Hall} and G.~{Holmes}, ``Benchmarking attribute selection techniques for
  discrete class data mining,'' \emph{IEEE Transactions on Knowledge and Data
  Engineering}, vol.~15, no.~6, pp. 1437--1447, 2003.

\bibitem{rank_fuzzy}
R.~{B} and G.~{S}, ``An intelligent fuzzy rule based feature selection for
  effective intrusion detection,'' in \emph{2018 International Conference on
  Recent Trends in Advance Computing (ICRTAC)}, 2018, pp. 206--211.

\bibitem{rank_svm}
V.~S. {Takkellapati} and G.~V. {Prasad}, ``Network intrusion detection system
  based on feature selection and triangle area support vector machine,''
  \emph{International Journal of Engineering Trends and Technology}, vol.~3,
  no.~4, pp. 466--470, 2012.

\bibitem{rank_c45}
S.~{Ganapathy}, K.~{Kulothungan}, P.~{Yogesh}, and A.~{Kannan}, ``An
  intelligent intrusion detection system for ad hoc networks,'' in \emph{IET
  Chennai 3rd International on Sustainable Energy and Intelligent Systems
  (SEISCON 2012)}, 2012, pp. 1--5.

\bibitem{rank_supervised}
J.~M. {HernÃ¡ndez JimÃ©nez} and K.~{Goseva-Popstojanova}, ``The effect on
  network flows-based features and training set size on malware detection,'' in
  \emph{2018 IEEE 17th International Symposium on Network Computing and
  Applications (NCA)}, 2018, pp. 1--9.

\bibitem{rank_knn}
P.~{Singh} and A.~{Tiwari}, ``An efficient approach for intrusion detection in
  reduced features of kdd99 using id3 and classification with knnga,'' in
  \emph{2015 Second International Conference on Advances in Computing and
  Communication Engineering}, 2015, pp. 445--452.

\bibitem{lfs}
M.~Gutlein, E.~Frank, M.~Hall, and A.~Karwath, ``Large-scale attribute
  selection using wrappers,'' in \emph{2009 IEEE Symposium on Computational
  Intelligence and Data Mining}, 2009, pp. 332--339.

\bibitem{lfs_android}
W.~{Wang}, X.~{Wang}, D.~{Feng}, J.~{Liu}, Z.~{Han}, and X.~{Zhang},
  ``Exploring permission-induced risk in android applications for malicious
  application detection,'' \emph{IEEE Transactions on Information Forensics and
  Security}, vol.~9, no.~11, pp. 1869--1882, 2014.

\bibitem{lfs_zero}
I.~Finizio, C.~Mazzariello, and C.~Sansone, ``Combining genetic-based misuse
  and anomaly detection for reliably detecting intrusions in computer
  networks,'' in \emph{Proceedings of the 13th International Conference on
  Image Analysis and Processing}, 2005, pp. 66--74.

\bibitem{lfs_ids}
Y.~{Liu}, Z.~{Xu}, J.~{Yang}, L.~{Wang}, C.~{Song}, and K.~{Chen}, ``A novel
  meta-heuristic-based sequential forward feature selection approach for
  anomaly detection systems,'' in \emph{2016 International Conference on
  Network and Information Systems for Computers (ICNISC)}, 2016, pp. 218--227.

\bibitem{glover_article}
F.~Glover, ``Future paths for integer programming and links to artificial
  intelligence,'' \emph{Computers \& Operations Research}, vol.~13, no.~5, pp.
  533 -- 549, 1986.

\bibitem{glover_book}
G.~F.W. and L.~M., \emph{Tabu Search}.\hskip 1em plus 0.5em minus 0.4em\relax
  New York, USA: Springer, 1997.

\bibitem{rego_book}
C.~Rego and B.~Alidaee, \emph{Metaheuristic Optimization via Memory and
  Evolution: Tabu Search and Scatter Search (Operations Research/Computer
  Science Interfaces Series)}.\hskip 1em plus 0.5em minus 0.4em\relax Berlin,
  Heidelberg: Springer-Verlag, 2005.

\bibitem{tabu}
A.-R. Hedar, J.~Wang, and M.~Fukushima, ``Tabu search for attribute reduction
  in rough set theory,'' \emph{Soft Computing}, vol.~12, no.~9, pp. 909--918,
  2008.

\bibitem{tabu_fuzzy}
H.~{Mohamadi}, J.~{Habibi}, and H.~{Saadi}, ``Intrusion detection in computer
  networks using tabu search based fuzzy system,'' in \emph{2008 7th IEEE
  International Conference on Cybernetic Intelligent Systems}, 2008, pp. 1--6.

\bibitem{tabu_knn}
W.~{Jian-guang}, T.~{Ran}, and L.~{Zhi-Yong}, ``An improving tabu search
  algorithm for intrusion detection,'' in \emph{2011 Third International
  Conference on Measuring Technology and Mechatronics Automation}, vol.~1,
  2011, pp. 435--439.

\bibitem{tabu_c45}
Y.~{Chen}, L.~{Dai}, and X.~{Cheng}, ``Gats-c4.5: An algorithm for optimizing
  features in flow classification,'' in \emph{2008 5th IEEE Consumer
  Communications and Networking Conference}, 2008, pp. 466--470.

\bibitem{tabu_genetic}
K.~{Bakour}, G.~S. {Das}, and H.~M. {Unver}, ``An intrusion detection system
  based on a hybrid tabu-genetic algorithm,'' in \emph{2017 International
  Conference on Computer Science and Engineering (UBMK)}, 2017, pp. 215--220.

\bibitem{tabu_genetic2}
{Xiaocong Z.}, {Dongling L.}, and {Yang Y.}, ``Improved incremental support
  vector machine with hybrid feature selection for network intrusion
  detection,'' in \emph{2013 International Conference on Information and
  Network Security (ICINS 2013)}, 2013, pp. 1--6.

\bibitem{scatter_source}
F.~Glover, ``Heuristics for integer programming using surrogate constraints,''
  \emph{Decision Sciences}, vol.~8, pp. 156--166, 1977.

\bibitem{scatter_nlp}
Z.~Ugray, L.~Lasdon, J.~Plummer, F.~Glover, J.~Kelly, and R.~Marti, ``Scatter
  search and local nlp solvers: a multistart framework for global
  optimization,'' \emph{Informs Journal on Computing}, vol.~19, no.~3, pp. 328
  -- 340, 2007.

\bibitem{scatter_roughset}
J.~Wang, A.~R. Hedar, S.~Wang, and J.~Ma, ``Rough set and scatter search
  metaheuristic based feature selection for credit scoring,'' \emph{Expert
  Systems with Applications}, vol.~39, no.~6, pp. 6123 -- 6128, 2012.

\bibitem{scatter_parallel}
F.~G. Lopez, M.~G. Torres, B.~M. Batista, J.~A.~M. Perez, and J.~M.
  Moreno-Vega, ``Solving feature subset selection problem by a parallel scatter
  search,'' \emph{European Journal of Operational Research}, vol. 169, no.~2,
  pp. 477 -- 489, 2006.

\bibitem{scatter_credit}
E.~Duman and M.~H. Ozcelik, ``Detecting credit card fraud by genetic algorithm
  and scatter search,'' \emph{Expert Systems with Applications}, vol.~38,
  no.~10, pp. 13\,057 -- 13\,063, 2011.

\bibitem{scatter_sec_sw}
D.~{Byers} and N.~{Shahmehri}, ``Prioritisation and selection of software
  security activities,'' in \emph{2009 International Conference on
  Availability, Reliability and Security}, 2009, pp. 201--207.

\bibitem{pso_scource}
J.~{Kennedy} and R.~{Eberhart}, ``Particle swarm optimization,'' in
  \emph{Proceedings of ICNN'95 - International Conference on Neural Networks},
  vol.~4, 1995, pp. 1942--1948 vol.4.

\bibitem{pso_svm}
Q.~{Yao}, J.~{Cai}, and J.~{Zhang}, ``Simultaneous feature selection and ls-svm
  parameters optimization algorithm based on {PSO},'' in \emph{2009 WRI World
  Congress on Computer Science and Information Engineering}, vol.~5, 2009, pp.
  723--727.

\bibitem{pso_svm2}
W.~{Hu}, J.~{Gao}, Y.~{Wang}, O.~{Wu}, and S.~{Maybank}, ``Online
  adaboost-based parameterized methods for dynamic distributed network
  intrusion detection,'' \emph{IEEE Transactions on Cybernetics}, vol.~44,
  no.~1, pp. 66--82, 2014.

\bibitem{pso_rf}
H.~{Li}, W.~{Guo}, G.~{Wu}, and Y.~{Li}, ``A {RF-PSO} based hybrid feature
  selection model in intrusion detection system,'' in \emph{2018 IEEE Third
  International Conference on Data Science in Cyberspace (DSC)}, 2018, pp.
  795--802.

\bibitem{pso_bigdata}
S.~{Fong}, R.~{Wong}, and A.~V. {Vasilakos}, ``Accelerated {PSO} swarm search
  feature selection for data stream mining big data,'' \emph{IEEE Transactions
  on Services Computing}, vol.~9, no.~1, pp. 33--45, 2016.

\bibitem{ant}
M.~{Dorigo}, V.~{Maniezzo}, and A.~{Colorni}, ``Ant system: optimization by a
  colony of cooperating agents,'' \emph{IEEE Transactions on Systems, Man, and
  Cybernetics, Part B (Cybernetics)}, vol.~26, no.~1, pp. 29--41, 1996.

\bibitem{ant_svm}
T.~{Mehmood} and H.~B.~M. {Rais}, ``Svm for network anomaly detection using aco
  feature subset,'' in \emph{2015 International Symposium on Mathematical
  Sciences and Computing Research (iSMSC)}, 2015, pp. 121--126.

\bibitem{ant_stream}
S.~{Harde} and V.~{Sahare}, ``Design and implementation of aco feature
  selection algorithm for data stream mining,'' in \emph{2016 International
  Conference on Automatic Control and Dynamic Optimization Techniques
  (ICACDOT)}, 2016, pp. 1047--1051.

\bibitem{ant_faco}
H.~{Peng}, C.~{Ying}, S.~{Tan}, B.~{Hu}, and Z.~{Sun}, ``An improved feature
  selection algorithm based on ant colony optimization,'' \emph{IEEE Access},
  vol.~6, pp. 69\,203--69\,209, 2018.

\bibitem{cuckoo_book}
X.~Yang, \emph{Cuckoo Search and Firefly Algorithm: Theory and
  Applications}.\hskip 1em plus 0.5em minus 0.4em\relax London, UK: Springer,
  2013.

\bibitem{cuckoo_pca}
Z.~{Li}, Y.~{Su}, and Q.~{Han}, ``Intrusion detection based on pca and fuzzy
  clustering optimized by cs,'' in \emph{2017 Chinese Automation Congress
  (CAC)}, 2017, pp. 6334--6339.

\bibitem{cuckoo_ids}
E.~J.~L. D.~Asir Antony Gnana~Singh, R.~Priyadharshini, ``Cuckoo optimisation
  based intrusion detection system for cloud computing,'' \emph{Internation
  Journal of Computer Network and Information Security}, vol.~11, pp. 42--49,
  2018.

\bibitem{cuckoo_ann}
K.~Rithesh, ``Anomaly-based nids using artificial neural networks optimised
  with cuckoo search optimizer,'' in \emph{Emerging Research in Electronics,
  Computer Science and Technology}, 2019, pp. 23--35.

\bibitem{cuckoo_svm}
W.~{Niu}, X.~{Zhang}, G.~{Yang}, Z.~{Ma}, and Z.~{Zhuo}, ``Phishing emails
  detection using cs-svm,'' in \emph{2017 IEEE International Symposium on
  Parallel and Distributed Processing with Applications and 2017 IEEE
  International Conference on Ubiquitous Computing and Communications
  (ISPA/IUCC)}, 2017, pp. 1054--1059.

\bibitem{cuckoo_sentiment}
M.~{Redmond}, S.~{Salesi}, and G.~{Cosma}, ``A novel approach based on an
  extended cuckoo search algorithm for the classification of tweets which
  contain emoticon and emoji,'' in \emph{2017 2nd International Conference on
  Knowledge Engineering and Applications (ICKEA)}, 2017, pp. 13--19.

\bibitem{cuckoo_sdn}
I.~H. {Abdulqadder}, D.~{Zou}, I.~T. {Aziz}, B.~{Yuan}, and W.~{Li},
  ``Secsdn-cloud: Defeating vulnerable attacks through secure software-defined
  networks,'' \emph{IEEE Access}, vol.~6, pp. 8292--8301, 2018.

\bibitem{weise}
T.~Weise, ``Global optimization algorithms -- theory and application,'' 2009.

\bibitem{genetic1}
W.~W. Bledsoe and I.~Browning, ``Pattern recognition and reading by machine,''
  in \emph{Papers Presented at the December 1-3, 1959, Eastern Joint
  IRE-AIEE-ACM Computer Conference}, 1959, pp. 225--232.

\bibitem{genetic2}
H.~J. Bremermann, ``Optimization through evolution and recombination,'' in
  \emph{Self-Organizing Systems}, M.~C. Yovits, G.~T. Jacobi, , and G.~D.
  Goldstein, Eds.\hskip 1em plus 0.5em minus 0.4em\relax Spartan Books, 1962.

\bibitem{holland_1962}
J.~Holland, ``Outline for a logical theory of adaptive systems,'' \emph{Journal
  of ACM}, vol.~9, no.~3, pp. 297--314, 1962.

\bibitem{holland_1992}
------, \emph{Adaptation in Natural and Artificial Systems: An Introductory
  Analysis with Applications to Biology, Control and Artificial
  Intelligence}.\hskip 1em plus 0.5em minus 0.4em\relax Cambridge, MA, USA: MIT
  Press, 1992.

\bibitem{genetic_survey}
C.~Coello, ``An updated survey of ga-based multiobjective optimization
  techniques,'' \emph{ACM Computing Surveys}, vol.~32, no.~2, pp. 109--143,
  2000.

\bibitem{Goldberg:1989:GAS:534133}
D.~Goldberg, \emph{Genetic Algorithms in Search, Optimization and Machine
  Learning}, 1st~ed.\hskip 1em plus 0.5em minus 0.4em\relax Boston, MA, USA:
  Addison-Wesley Longman Publishing Co., Inc., 1989.

\bibitem{ga_ann}
S.~{Guha}, S.~S. {Yau}, and A.~B. {Buduru}, ``Attack detection in cloud
  infrastructures using artificial neural network with genetic feature
  selection,'' in \emph{2016 IEEE 14th Intl Conf on Dependable, Autonomic and
  Secure Computing}, 2016, pp. 414--419.

\bibitem{ga_j48}
B.~{Senthilnayaki}, K.~{Venkatalakshmi}, and A.~{Kannan}, ``An intelligent
  intrusion detection system using genetic based feature selection and modified
  {J}48 decision tree classifier,'' in \emph{2013 Fifth International
  Conference on Advanced Computing (ICoAC)}, 2013, pp. 1--7.

\bibitem{ga_svm}
------, ``Intrusion detection using optimal genetic feature selection and svm
  based classifier,'' in \emph{2015 3rd International Conference on Signal
  Processing, Communication and Networking (ICSCN)}, 2015, pp. 1--4.

\bibitem{ga_svm2}
H.~{Gharaee} and H.~{Hosseinvand}, ``A new feature selection ids based on
  genetic algorithm and svm,'' in \emph{2016 8th International Symposium on
  Telecommunications (IST)}, 2016, pp. 139--144.

\bibitem{ga_svm3}
P.~{Tao}, Z.~{Sun}, and Z.~{Sun}, ``An improved intrusion detection algorithm
  based on {GA} and {SVM},'' \emph{IEEE Access}, vol.~6, pp. 13\,624--13\,631,
  2018.

\bibitem{moea_source}
C.~M. {Fonseca} and P.~J. {Fleming}, ``An overview of evolutionary algorithms
  in multiobjective optimization,'' \emph{Evolutionary Computation}, vol.~3,
  no.~1, pp. 1--16, 1995.

\bibitem{moea}
F.~Jimenez, G.~Sanchez, J.~Garcia, G.~Sciavicco, and L.~Miralles,
  ``Multi-objective evolutionary feature selection for online sales
  forecasting,'' \emph{Neurocomputing}, vol. 234, pp. 75 -- 92, 2017.

\bibitem{moea_ensemb}
M.~S. {Aliakbarian} and A.~{Fanian}, ``Internet traffic classification using
  moea and online refinement in voting on ensemble methods,'' in \emph{2013
  21st Iranian Conference on Electrical Engineering (ICEE)}, 2013, pp. 1--6.

\bibitem{moea_nsga1}
Y.~Zhu, J.~Liang, J.~Chen, and Z.~Ming, ``An improved nsga-iii algorithm for
  feature selection used in intrusion detection,'' \emph{Knowledge-Based
  Systems}, vol. 116, pp. 74--85, 2017.

\bibitem{moea_nsga2}
``Building an effective intrusion detection system using unsupervised feature
  selection in multi-objective optimization framework,''
  \url{arxiv.org/pdf/1905.06562.pdf}, accessed: 2020-10-01.

\bibitem{moea_fuzzy}
P.~{Ducange}, G.~{Mannara}, F.~{Marcelloni}, R.~{Pecori}, and M.~{Vecchio}, ``A
  novel approach for internet traffic classification based on multi-objective
  evolutionary fuzzy classifiers,'' in \emph{2017 IEEE International Conference
  on Fuzzy Systems (FUZZ-IEEE)}, 2017, pp. 1--6.

\bibitem{kdd99}
``The {KDD99} {D}ataset,''
  \url{http://kdd.ics.uci.edu/databases/kddcup99\\/kddcup99.html}, accessed:
  2020-10-01.

\bibitem{icissp-dataset}
G.~Draper-Gil, A.~H. Lashkari, M.~S.~I. Mamun, and A.~A. Ghorbani,
  ``Characterization of encrypted and vpn traffic using time-related
  features,'' in \emph{International Conference on Information Systems Security
  and Privacy}, 2016.

\bibitem{icissp-dataset2}
``Cicflowmeter network analyzer,''
  \url{https://www.unb.ca/cic/datasets/ids-2018.html}, accessed: 2020-10-01.

\bibitem{cfs1}
``Correlation-based feature selection for machine learning,''
  \url{cs.waikato.ac.nz/~mhall/thesis.pdf}, accessed: 2020-10-01.

\bibitem{cfs3}
F.~Fabris, A.~A. Freitas, and J.~M.~A. Tullet, ``An extensive empirical
  comparison of probabilistic hierarchical classifiers in datasets of
  ageing-related genes,'' \emph{IEEE/ACM Transactions on Computational Biology
  and Bioinformatics}, vol.~13, no.~6, pp. 1045--1058, 2016.

\bibitem{cfs5}
R.~Kohavi and G.~H. John, ``Wrappers for feature subset selection,''
  \emph{Artificial Intelligence}, vol.~97, no.~1, pp. 273 -- 324, 1997.

\bibitem{jsac4}
L.~{Zhu}, X.~{Tang}, M.~{Shen}, X.~{Du}, and M.~{Guizani}, ``Privacy-preserving
  ddos attack detection using cross-domain traffic in software defined
  networks,'' \emph{IEEE Journal on Selected Areas in Communications}, vol.~36,
  no.~3, pp. 628--643, 2018.

\bibitem{jsac5}
K.~{Kalkan}, L.~{Altay}, G.~{Gür}, and F.~{Alagöz}, ``Jess: Joint
  entropy-based ddos defense scheme in sdn,'' \emph{IEEE Journal on Selected
  Areas in Communications}, vol.~36, no.~10, pp. 2358--2372, 2018.

\bibitem{accuracy1}
H.~Kim, K.~Claffy, M.~Fomenkov, D.~Barman, M.~Faloutsos, and K.~Lee, ``Internet
  traffic classification demystified: Myths, caveats, and the best practices,''
  in \emph{Proceedings of the 2008 ACM CoNEXT Conference}, 2008, pp.
  11:1--11:12.

\bibitem{accuracy2}
J.~{Zhang}, Y.~{Xiang}, Y.~{Wang}, W.~{Zhou}, Y.~{Xiang}, and Y.~{Guan},
  ``Network traffic classification using correlation information,'' \emph{IEEE
  Transactions on Parallel and Distributed Systems}, vol.~24, no.~1, pp.
  104--117, 2013.

\end{thebibliography}



\end{document}